\documentclass[12pt, a4paper]{article}
\usepackage[T1]{fontenc}
\usepackage[utf8]{inputenc}
\usepackage[absolute]{textpos}
\usepackage[top=3cm,bottom=3cm,left=2cm,right=2cm]{geometry}
\usepackage{amsmath,amssymb,mathtools,dsfont,emptypage,graphicx,dcolumn,bm,booktabs,colortbl,collcell,enumerate,float,verbatim,multirow,xparse,slashed,mathrsfs,color}
\usepackage{cite}
\usepackage{braket}
\usepackage{tabularx}
\usepackage{multirow}
\usepackage{graphicx}
\usepackage{listings}
\usepackage{subcaption}
\usepackage{appendix}
\usepackage{booktabs} % Enhance the tables
\usepackage{colortbl}
\usepackage{collcell}
\usepackage{xparse}
\usepackage[labelfont=bf, labelsep=period]{caption}
\usepackage{sectsty}
\usepackage{pifont}
\usepackage{relsize}
\usepackage{pifont}
\usepackage{bbold}
\usepackage{fontawesome}
\sectionfont{\fontsize{15}{12}\selectfont}
\subsectionfont{\fontsize{13}{12}\selectfont}

\usepackage[compat=1.1.0]{tikz-feynman}
\usepackage{contour}%FeynamnDiagrams e contorni

\newcommand{\virg}[1]{``#1''}
\newcommand{\mnulight}{m_{\nu}^{\rm lightest}}
\DeclareFixedFont{\ttm}{T1}{txtt}{m}{n}{11} % for normal

\def\Mav{\bar{M}}

\def\sph{\text{sph}}
\def\LNC{\text{LNC}}
\def\LNV{\text{LNV}}

\definecolor{myred}{cmyk}{0,1,1,0.55}
\definecolor{mygold}{rgb}{0.8, 0.54, 0.08}
\definecolor{mygray}{gray}{.95}
\usepackage[linktoc = page, colorlinks=true,urlcolor=mygold,linkcolor=mygold,citecolor=mygold]{hyperref}

\definecolor{darkgreen}{RGB}{0, 171, 148}
\definecolor{carnelian}{rgb}{0.7, 0.11, 0.11}

\newcommand\pythonstyle{\lstset{
language=Python,
basicstyle=\ttm,
otherkeywords={self},    % Add keywords here
keywordstyle=\ttb\color{deepblue},
emph={MyClass,__init__, uls},   % Custom highlighting
emphstyle=\ttb\color{deepred}, % Custom highlighting style
stringstyle=\color{deepgreen},
frame=tb,       % Any extra options here
showstringspaces=false,   % 
commentstyle=\tti,
morecomment=[s]{"""}{"""},
}}
\newcommand\bashstyle{\lstset{
language=bash,
basicstyle=\ttm,
otherkeywords={self},    % Add keywords here
keywordstyle=\ttb\color{deepblue},
emph={uls-calc,uls-scan,uls-nest},   % Custom highlighting
emphstyle=\ttb\color{deepred}, % Custom highlighting style
stringstyle=\color{deepgreen},
frame=tb,       % Any extra options here
showstringspaces=true,   % 
commentstyle=\tti,
}}

\lstnewenvironment{python}[1][]
{
\pythonstyle
\lstset{#1}
}
{}

\lstnewenvironment{bash}[1][]
{
\bashstyle
\lstset{#1}
}
{}

\begin{document}

\vspace*{-15mm}
\begin{flushright}
IPPP/26/27, IFT-UAM/CSIC-26-64
\end{flushright}
\vspace*{0.7cm}

\begin{center}
{\bf \large ULYSSES the Third:\\
An Odyssey Towards a Unified Python Toolkit for Leptogenesis}\\
[5mm]
 \renewcommand*{\thefootnote}{\fnsymbol{footnote}}
Alessandro Granelli\footnote{\href{mailto:alessandro.granelli@ific.uv.es}{alessandro.granelli@ific.uv.es}}$^{a}$,
Juraj Klari\'c\footnote{\href{mailto:jklaric@phy.hr}{jklaric@phy.hr}}$^{b,c,d}$,
Dhruv Pasari\footnote{\href{mailto:dhruv.pasari@durham.ac.uk}{dhruv.pasari@durham.ac.uk}}$^{e}$,
Yuber F.\ Perez‑Gonzalez\footnote{\href{mailto:yuber.perez@uam.es}{yuber.perez@uam.es}}$^{f}$ and 
Jessica Turner\footnote{\href{mailto:jessica.turner@durham.ac.uk}{jessica.turner@durham.ac.uk}}$^{e}$\\
\vspace{2mm}
$^{a}${\it Instituto de Física Corpuscular (IFIC), CSIC-Universitat de València, Parc Científic UV,\\ c/ Catedrático José Beltrán 2, E-46980 Paterna, Spain}\\
$^{b}${\it Institute for Theoretical Physics Amsterdam and Delta Institute for Theoretical Physics,
University of Amsterdam, Science Park 904, 1098 XH Amsterdam, The Netherlands}\\
$^{c}${\it Nikhef, Theory Group, Science Park 105, 1098 XG, Amsterdam, The Netherlands}\\
$^{d}${\it University of Zagreb, Faculty of Science, Department of Physics, 10000 Zagreb, Croatia}\\
$^{e}${\it Institute for Particle Physics Phenomenology, Department of Physics, Durham University, Durham DH1 3LE, U.K.}\\
$^{f}${\it Departamento de Física Teórica and Instituto de Física Teórica UAM/CSIC,\\ Universidad Autónoma de Madrid, Cantoblanco, 28049 Madrid, Spain}

\end{center}

%%%%%%%%%%%%%%%%ABSTRACT%%%%%%%%%%%%%%%%%%%55
\begin{center}
 {\bf Abstract}
\end{center}
 We present the third release of \texttt{ULYSSES}, a Python package for the numerical evaluation of the baryon asymmetry generated through leptogenesis. This version includes code implementing state-of-the-art density matrix equations for low-scale leptogenesis with three quasi-degenerate right-handed neutrinos. We extend the validity of the code in this scenario beyond the 100 GeV right-handed neutrino mass scale, into the regime of resonant leptogenesis, by including neutrino production rates valid in both the relativistic and non-relativistic regimes. In addition, in the high-scale vanilla scenario, we provide routines for computing $\Delta L = 1$ scattering processes, enabling full phase-space evolution of the right-handed neutrino and lepton asymmetry. A new \texttt{-{}-extended} parameter interface allows users to pass model-specific inputs beyond the standard leptogenesis runcard without modifying the core infrastructure and demonstrate its use with a toy module that simultaneously solves the vanilla leptogenesis equations and the freeze-in production of dark matter. On top of these improvements, we introduce an alternative parametrisation of the Casas-Ibarra matrix, update the default neutrino oscillation parameters and report cross-checks of the new low-scale leptogenesis module against published benchmarks and independent codes. \texttt{ULYSSES} is publicly available on GitHub \href{https://github.com/earlyuniverse/ulysses}{\faGithub} and pip-installable from PyPI.

\newpage

\noindent{\bf PROGRAM SUMMARY}

\noindent \underline{Program title}: ULYSSES

% \noindent CPC Library link to program files: (to be added by Technical Editor)

\noindent \underline{Developer's repository link}: \url{https://github.com/earlyuniverse/ulysses}

\noindent \underline{Licensing provisions}: MIT License

\noindent \underline{Programming language}: Python 3

\noindent \underline{Journal reference of previous version}: Comput. Phys. Commun. 291 (2023) 108834

\noindent \underline{Does the new version supersede the previous version?}: Yes

\noindent \underline{Reasons for the new version}:
Extension of the code capabilities to include low-scale leptogenesis scenarios, complete $\Delta L = 1$ scattering processes for high-scale leptogenesis, improved user-defined model support, and enhanced usability through a Python API and Jupyter notebook interface.

\noindent \underline{Summary of revisions}:
Implementation of a unified treatment of low-scale leptogenesis with three right-handed neutrinos; inclusion of complete $\Delta L = 1$ scattering processes in the leptogenesis evolution equations; addition of alternative Casas--Ibarra parametrisations; support for user-specified initial heavy-neutrino abundances; extension of the framework to accommodate custom leptogenesis models and parameters; introduction of a Python API and Jupyter notebook workflow.

\noindent \underline{Nature of problem}:
Numerical solution of coupled Boltzmann and Density Matrix equations governing thermal, resonant, and low-scale leptogenesis, including the evaluation of computationally demanding thermal integrals and reaction rates.

\noindent \underline{Solution method}:
The evolution equations are solved using the \texttt{odeint} and \texttt{solve\_ivp} integrators available in Python. Integrals are evaluated using a Numba-accelerated wrapper around the \texttt{cquadpack} integration library.

% \noindent Additional comments including restrictions and unusual features:
% The package includes a Numba wrapper for \texttt{cquadpack}, providing significant performance improvements for repeated numerical integrations required in leptogenesis calculations.
%\\
% \vspace{1.5em}

\renewcommand*{\thefootnote}{\arabic{footnote}}
\setcounter{footnote}{0}

%%%%%%%%%%%%%%%%%%%%%%%%%%%%%%%%%%%%%%%%%%
\tableofcontents

\section{Introduction}

Since its initial proposal~\cite{Fukugita:1986hr}, leptogenesis has been one of the most compelling
mechanisms for explaining the observed matter-antimatter asymmetry of the
Universe. The most well-studied realisation of leptogenesis is that within the type-I seesaw extension of the Standard Model (SM), which introduces right-handed neutrinos (RHNs) and also accounts for the generation of the masses and mixing of light neutrinos via the seesaw mechanism~\cite{Minkowski:1977sc, Yanagida:1979as, GellMann:1980vs, Gell-Mann:1979ijt, Glashow:1979nm, Mohapatra:1979ia}. In this scenario, a lepton asymmetry can be generated in the early Universe through decays, inverse decays, scatterings and CP-violating
oscillations involving RHNs, and later translated into an asymmetry in baryons by sphalerons \cite{Kuzmin:1985mm, DOnofrio:2014rug}. Different regimes of leptogenesis within the type-I seesaw framework can be realised,
depending on the mass scale and mass hierarchy associated to the RHNs \cite{Pilaftsis:1997jf, Pilaftsis:2003gt, Akhmedov:1998qx, Asaka:2005pn, Racker:2012vw}. Although leptogenesis could, in principle, be described by a unique set of kinetic equations valid across all regimes, depending on which processes dominate the generation of the asymmetry, each regime can be studied with distinct approximated forms of the full evolution equations.

\texttt{ULYSSES}
\cite{Granelli:2020pim, Granelli:2023vcm} is a \textit{Python} package that solves the
semi-classical Boltzmann equations (BEs) and density matrix equations (DMEs) for
leptogenesis within the type-I seesaw framework, covering different leptogenesis regimes and a wide range of RHN masses.\footnote{No upper limit is imposed by the code itself; however, one must note that $\Delta L=2$ processes mediated by off-shell RHNs, which scale as $M_N{m_\nu}^2$ where $m_\nu\lesssim 1\,\text{eV}$ ($M_N\gg m_\nu$) is the light neutrino (RHN) mass scale, can only be neglected at $M_N\ll 10^{14}$ GeV. The Primordial Black Hole module \cite{Granelli:2023vcm} takes this correction into account, while the thermal high scale modules do not and hence one must take care in interpreting results in the regime where such scattering processes are relevant.} \texttt{ULYSSES} version~1~\cite{Granelli:2020pim} provided code for solving the
momentum-averaged BEs relevant to high-scale thermal leptogenesis based on the out-of-equilibrium
decays of RHNs, covering both resonant and non-resonant
regimes. Effects such as lepton flavour, scatterings and spectator processes were also
included. \texttt{ULYSSES} version~2~\cite{Granelli:2023vcm} extended this with the
\virg{complete} set of thermal leptogenesis BEs that properly accounts for quantum statistics
without assuming kinetic equilibrium for the RHNs, state-of-the-art BEs for low-scale
leptogenesis via oscillations with two RHNs~\cite{Asaka:2005pn,Akhmedov:1998qx} and primordial black hole-induced leptogenesis \cite{Perez-Gonzalez:2020vnz,Bernal:2022pue,Gunn:2024xaq}.

In this third release, we present the following main new features:
\begin{itemize}
 \item State-of-the-art density matrix equations for low-scale leptogenesis with three quasi-degenerate-in-mass
 RHNs, extended to the non-relativistic limit and including contributions
 from both lepton-number-conserving and lepton-number-violating processes to the
 Hamiltonian and rates, thereby extending their validity up to the RHN mass scale of $\sim 100 \,\text{TeV}$. We have also included speed-up routines for the numerical integration of the DMEs in this scenario.

 \item Routines implementing $\Delta L = 1$ scattering processes  in the vanilla scenario of high-scale leptogenesis with one decaying RHN, that complete the modules introduced in the second version. This implementation enables a more comprehensive phase-space evolution of the RHN and lepton asymmetry beyond the momentum-averaged approximation.
 
 \item An \texttt{-{}-extended} parameter interface that allows users to pass model-specific
 parameters beyond the standard leptogenesis runcard, enabling custom model implementations
 without modifying the core infrastructure.

 \item An alternative parameterisation of the Casas-Ibarra matrix~\cite{Drewes:2021nqr},
 expressed in terms of a single complex angle and four real angles.

 \item Updated default neutrino oscillation parameters, adopting the best-fit values
 from the \href{http://www.nu-fit.org/?q=node/309}{\texttt{NuFit-6.1}} global analysis~\cite{Esteban:2024eli}. 
\end{itemize}

A key motivation for this release is the rapidly approaching experimental era for 
testing low-scale leptogenesis. The
leptogenesis via oscillation mechanism
 requires RHNs with masses at or below the electroweak scale and 
small but non-negligible mixing angles with the active neutrinos; precisely the 
parameter space targeted by a new generation of intensity-frontier experiments searching 
for heavy neutral leptons (HNLs). Most notably, the \textsc{SHiP} experiment~\cite{Alekhin:2015byh} at the CERN SPS has been approved and funded and is designed to probe HNL mixing 
angles many orders of magnitude below current bounds for masses in the $\mathcal{O}
(\text{GeV})$ range. Complementary sensitivity will be provided by experiments including 
\textsc{FASER2}~\cite{Kling:2018wct, Feng:2017uoz}, MATHUSLA \cite{Curtin:2018mvb} as well as future 
colliders~\cite{Mekala:2022cmm,Blondel:2022qqo,Wang:2019xvx,Boyarsky:2022epg}. Together, these experiments will probe a significant portion of the leptogenesis 
parameter space consistent with the observed baryon asymmetry $\eta_B$~\cite{Drewes:2024bla
}. The density matrix equations implemented in \texttt{ULYSSES}~version 3 for three RHNs 
provide the theoretical framework needed to map this experimental reach onto the 
leptogenesis parameter space in a systematic way, making 
\texttt{ULYSSES} a timely tool for confronting low-scale leptogenesis with data.
Due to the experimental relevance, many groups have been working on predicting the baryon asymmetry precisely from
low-scale leptogenesis via oscillations. 
A significant challenge in the current literature is the lack of a unified benchmark between the various groups. In this work, we address this gap by providing what is, to the best of our knowledge, the first and only systematic cross-comparison of results from several independent collaborations. By reproducing benchmarks from \cite{Hernandez:2022ivz, Abada:2018oly} and checking with the public \texttt{C++} code \texttt{amiqs}, as well as the private one used, e.g., in \cite{Drewes:2021nqr}, we offer a bridge between different numerical implementations. 

We also verify the compatibility between the low-scale leptogenesis module, here equipped with density matrix equations valid up to the $\mathcal{O}(100\,\mathrm{TeV})$ scale, and the module for resonant leptogenesis, demonstrating their consistency in the regime where both descriptions are expected to overlap \cite{Klaric:2020phc, Klaric:2021cpi}. This consistency check underpins the design philosophy of \texttt{ULYSSES} as a universal infrastructure for leptogenesis. In particular, \texttt{ULYSSES}, with its suite of modules adapted to different regimes of temperatures,
provides a robust architecture for the construction of a unified set of leptogenesis equations applicable across all regimes.  Finally, the motivation for the \texttt{-{}-extended} parameter interface is to enhance the versatility of \texttt{ULYSSES} to study a broader class of non-standard leptogenesis models, such as primordial black hole-induced leptogenesis module in version 2. Along these lines, we provide an example module that simultaneously solve equations for vanilla leptogenesis and the
production of dark matter.

As in the previous versions, the emphasis of the code is on user flexibility and rapid
evaluation.  \texttt{ULYSSES} is publicly available at
\href{https://github.com/earlyuniverse/ulysses}{https://github.com/earlyuniverse/ulysses}; the release accompanying this paper is tagged \texttt{v3.0.0}, and all figures shown here can be reproduced from the runcards shipped in the repository under that tag (see the \virg{Code Availability} paragraph at the end of this work for details).
For convenience, we briefly recall the naming convention used for the
predefined \texttt{ULYSSES} modules. In modules denoted
\texttt{\(n\)BE\(m\)F}, \(n\) is the number of dynamically evolved RHNs,
\texttt{BE} denotes Boltzmann equations, and \(m\) is the number of resolved
charged-lepton flavours. Similarly, \texttt{\(n\)DME} denotes density-matrix
equations with \(n\) dynamical RHNs. Additional suffixes identify specific
ingredients, such as \texttt{sct} for scattering processes, \texttt{sp} for
spectator effects, and \texttt{sf} for evolution in terms of the scale
factor. For conventional decay leptogenesis, the one-flavour treatment is normally
appropriate when the dominant asymmetry production occurs at
\(T\gtrsim10^{12}\,\mathrm{GeV}\), the two-flavour treatment for
\(10^{9}\,\mathrm{GeV}\lesssim T\lesssim10^{12}\,\mathrm{GeV}\), and the
three-flavour treatment for \(T\lesssim10^{9}\,\mathrm{GeV}\). These
boundaries are approximate, and a density-matrix module should be used near
the flavour transitions or whenever flavour coherences are important \cite{Moffat:2018smo, Granelli:2021fyc}.
Specialised resonant and low-scale modules are subject to the additional
mass-degeneracy and oscillation-temperature requirements discussed in their
respective sections. A complete and version-specific table of the available modules, including a
short description of the physical scenario implemented by each one, is
maintained in the repository documentation. The same catalogue can be
displayed for the locally installed version by running
\texttt{uls-models}. An overview of the modules introduced in the previous
releases is also given in Table~1 of \texttt{ULYSSESv1}~\cite{Granelli:2020pim}.

\texttt{ULYSSES} version~3 applies all of the same conventions (Yukawa matrix
parametrisation, Higgs vacuum expectation value, normalisation of number densities) as
the previous versions and we refer the reader to Refs.~\cite{Granelli:2020pim,
Granelli:2023vcm} for a discussion of these matters. We refrain from reviewing the
different regimes and subtleties of the leptogenesis mechanism and instead refer the
reader to recent reviews (see, $e.g.$, Ref.~\cite{Bodeker:2020ghk} and references therein) on various
aspects of thermal, resonant and low-scale leptogenesis. For installation instructions and basic usage, see Ref.~\cite{Granelli:2020pim} and the repository README. A self-contained Jupyter notebook (\texttt{ULYSSES\_intro.ipynb}) is available in the repository.

The paper is organised as follows. In Section~\ref{sec:setup} we describe updates to
the general set-up of \texttt{ULYSSES}, including the new Casas-Ibarra parameterisation
and the treatment of a decoupled heavy neutrino. Section~\ref{sec:lowlep} presents the new
density matrix equations for low-scale leptogenesis with three RHNs, together with
numerical speed-up strategies and benchmark cross-checks. In
Section~\ref{sec:scat} we describe the implementation of $\Delta L = 1$ scattering
processes for the full phase-space evolution. Section~\ref{sec:DM} details the use of the new \texttt{-{}-extended} parameter through a toy dark matter module. Thermally averaged expressions for the rates appearing in the density
matrix equations are collected in Appendix~\ref{app:Rates}. We conclude with a summary in Sec.~\ref{sec:conclusions}.

\section{Overview and updates of the \texttt{ULYSSES}'s general set-up}\label{sec:setup}
\subsection{\texttt{ULYSSES}'s notation for the type-I seesaw framework}
The type-I seesaw extension of the SM~\cite{Minkowski:1977sc,Yanagida:1979as,GellMann:1980vs,Glashow:1979nm,Mohapatra:1979ia} augments the SM by introducing $n_R$
RHNs. Being singlets under the SM gauge group, these neutrinos can naturally acquire Majorana mass terms. They also couple to the SM left-handed lepton doublets and the Higgs doublet through Yukawa interactions. \texttt{ULYSSES} is designed to work with $n_R = 2$ or $3$, 
\footnote{Parameters in \texttt{ULYSSES} are always input as in the $n_R=3$ scenario, but can be chosen such that the heaviest neutrino effectively decouples, allowing the user to study scenarios with only two RHNs. See Sec.~\ref{ssec:RHNdecouple} for details.} in the basis where both the charged lepton Yukawa matrix and the RHN mass matrix are diagonal. The type-I seesaw Lagrangian is defined as:
\begin{equation}
{\cal L}_{\text{Type-I}} ={\cal L}_{\rm SM} + \frac{1}{2} \,\overline{N_{j}}(i\slashed{\partial}-M_j)N_{j} 
-\left[
 \,Y_{\alpha j} \overline{\psi_{\alpha L}}\,i\sigma_2\,\Phi^*P_R\,N_{j}
 + \hbox{h.c.}\right], 
\end{equation}
where $\mathcal{L}_{\rm SM}$ denotes the Standard Model Lagrangian; $Y_{\alpha j}$ are the components of the Yukawa coupling matrix $Y$, with $\alpha = e,\,\mu,\,\tau$ labelling lepton flavour and $j = 1, \ldots, n_R$ % j$
indexing the generations of RHNs in the chosen basis; the symbol $\sigma_2$ refers to the second Pauli matrix and $P_R$ is the right-chiral projection operator; $\psi_{\alpha L} = (\nu_{\alpha L}^T\;\alpha_L^T)^T$ and $\Phi = (\Phi^+\;\Phi^{(0)})^T$ denote, respectively, the SM left-handed lepton doublet of flavour $\alpha$ and the Higgs doublet; $\nu_{\alpha L}$ and $\alpha_L$ correspond to the left-handed active neutrino and charged lepton 
of flavour $\alpha$; $N_j$
is a \textit{heavy Majorana neutrino} (i.e.~heavier than the sub-eV scale of the light neutrinos) 
with mass $M_j$.

\subsection{An alternative parameterisation for the Casas-Ibarra matrix}
The \texttt{ULYSSES} code allows the user to either manually set the Yukawa matrix $Y$, or to adopt the Casas-Ibarra (CI) parameterisation with the inclusion of one-loop radiative corrections \cite{Casas:2001sr, Lopez-Pavon:2015cga}:
%%%%%%%%%%%%%%%%%%%%%%%%%%%%%%%%%%%%%
\begin{equation}
Y =
 \pm \, \dfrac{i}{v}\,U\, \sqrt{\hat{m}_\nu}\,R^T\sqrt{f^{-1}_{\rm loop}(M)}\,,
\label{eq:Casas-Ibarra}
\end{equation}
%%%%%%%%%%%%%%%%%%%%%%%%%%%%%%%%%%%%%%
%
where $v = 174\,\text{GeV}$ is the vacuum expectation value of the neutral component of the Higgs doublet; $U$ is the Pontecorvo-Maki-Nakagawa-Sakata (PMNS) neutrino mixing matrix; $\hat{m}_\nu = \text{diag}(m_1, m_2,m_3)$ is the light neutrino mass matrix, $m_a$ being the masses of the active neutrinos, $a=1,2,3$; $R$ is a $3\times 3$ complex orthogonal matrix, satisfying $R^T\,R = R\,R^T = \mathbb{1}$. The loop corrections are encoded in $f_{\rm loop}(M) = \text{diag}(f(M_1), f(M_2), f(M_3))$, with \cite{Pilaftsis_1992, Grimus_2002, Aristizabal_Sierra_2011, Lopez_Pavon_2013}
\begin{equation}
f(M_j) \equiv \frac{1}{M_j}\left\{1 - \frac{M_j^2}{16\pi^2v^2}\left[\frac{\log{(M_j^2/m_H^2)}}{M_j^2/m_H^2 - 1}+3\frac{\log{(M_j^2/m_Z^2)}}{M_j^2/m_Z^2 - 1}\right]\right\},
~j=1,2,3\,,
\label{eq:Mloop}
\end{equation}
$m_H = 125$ GeV and 
$m_Z = 91.2$ GeV being respectively the Higgs and $Z$ boson 
masses. Based on the user's discretion, the code allows the user to switch on or off the loop corrections,
see Ref.~\cite{Granelli:2020pim} for instructions. 
The PMNS matrix is defined in the code according to the following parameterisation \cite{Tanabashi:2018oca}:
%%%%%%%%%%%%%%%%%%%%%%%%%%%%
\begin{equation}
\label{PMNS}
U = \begin{pmatrix}
c_{12}c_{13}&s_{12}c_{13}&s_{13}\text{e}^{-i\delta}\\
-s_{12}c_{23}-c_{12}s_{23}s_{13}\text{e}^{i\delta}&c_{12}c_{23}-s_{12}s_{23}s_{13}\text{e}^{i\delta}&s_{23}c_{13}\\
s_{12}s_{23}-c_{12}c_{23}s_{13}\text{e}^{i\delta}&-c_{12}s_{23}-s_{12}c_{23}s_{13}\text{e}^{i\delta}&c_{23}c_{13}
\end{pmatrix}
\begin{pmatrix}
1&0&0\\
0&\text{e}^{i\alpha_{21}/2}&0\\
0&0& \text{e}^{i\alpha_{31}/2}
\end{pmatrix},
\end{equation}
where $c_{kl} \equiv \cos\theta_{kl}$, $s_{kl} \equiv \sin\theta_{kl}$, $k,\,l = 1,2,3$ and $k\neq l$;
$0\leq \delta < 2\pi$ is the Dirac phase,
while $0 \leq \alpha_{21}, \alpha_{31} < 2\pi$ are the two Majorana phases \cite{Bilenky:1980cx}. Also, the standard numbering of the light neutrinos, which can have either a spectrum with normal ordering (NO) $\mnulight \equiv m_1 < m_2 < m_3$ or inverted ordering (IO) $\mnulight \equiv m_3<m_1<m_2$, is assumed. For this new version of \texttt{ULYSSES} we adopt as default input of the PMNS angles and squared mass differences $\Delta m_{21}^2 \equiv m_2^2-m_1^2$ and $\Delta m_{31(32)}^2 \equiv m_3^2-m_{1(2)}^2$ the best-fit values provided by the \texttt{NuFit-6.1} 
global analysis \cite{%nufit, 
Esteban:2024eli}.

In the previous versions of the \texttt{ULYSSES} code, the only available parameterisation of the CI matrix $R$ was the Euler one based on three complex rotations, that is
\begin{equation}\label{eq:Euler_O}
R = R^{(23)}_1R^{(13)}_2R^{(12)}_3,
\end{equation}
% %%%%%%%%%%%%%%%%%%%%%%%%%%%%%%%%%%
% %
where $R^{(jk)}_l\equiv R^{(jk)}(z_l)$, $j,k,l = 1,2,3$, $j\neq k$, $j\neq l$, $k\neq l$, $R^{(jk)}$ being a $3\times 3$ complex orthogonal rotation matrix in the $j-k$ plane, namely
\begin{equation}\label{eq:complex_rotations}
R^{(23)}(z_1)=\begin{pmatrix}
 1&0&0\\
 0&c_1&s_1\\
 0&-s_1&c_1
 \end{pmatrix},\,
R^{(13)}(z_2)= \begin{pmatrix}
 c_2&0&s_2\\
 0&1&0\\
 -s_2&0&c_2
 \end{pmatrix},\,
 R^{(12)}(z_3)=\begin{pmatrix}
 c_3&s_3&0\\
 -s_3&c_3&0\\
 0&0&1
 \end{pmatrix},
 \end{equation}
 and $c_l \equiv \cos z_l$ and $s_l \equiv \sin z_l$, $z_l = x_l+i y_l$, $l=1,2,3$, with $x_1$, $x_2$, $x_3$, $y_1$, $y_2$ and $y_3$ being six free real parameters.
 In this new release, we introduce the alternative parameterisation \cite{Drewes:2021nqr}
%%%%%%%%%%%%%%%%%%%%%%%%%%%5
\begin{equation}
\label{eq:alt_O}
R^T = R^{(23)}(x_1^\nu)R^{(13)}(x_2^\nu)R^{(12)}(z)R^{(13)}(x_2^N)R^{(23)}(x_1^N) ,
\end{equation}
%%%%%%%%%%%%%%%%%%%%%%%%%
%
in terms of a single complex angle $z = x + i y$ and four real angles $x_1^\nu$, $x_2^\nu$, $x_1^N$, $x_2^N$. This parameterisation is particularly convenient to deal with relatively large Yukawa couplings, since only one single imaginary parameter $iy$ controls their overall size rather than the three $iy_1$, $iy_2$ and $iy_3$ in the Euler parameterisation. Moreover, the remaining real angles, $x^\nu_1$, $x_2^\nu$ that mix the light neutrino sector and $x^N_1$, $x_2^N$ mixing the RHN sector, admit a more direct physical interpretation compared to $x_1$, $x_2$ and $x_3$.

To allow the user to select between the different parameterisations, we implement the base class \texttt{ULSBase} with the additional instance attribute \texttt{which$\_$param} that can either be set to \virg{\textit{euler}} (default choice), \virg{\textit{single-imaginary}}, \virg{\textit{manual}}, respectively to adopt the parameterisation in Eq.~\eqref{eq:Euler_O}, that in Eq.~\eqref{eq:alt_O}, or to manually select the entries of the Yukawa couplings by specifying them via $Y_{\alpha j} = Y_{\alpha j}^{\rm mag}\,\text{exp}(i Y_{\alpha j}^{\rm phs})$. For each choice, the user should always input the mass of the heavy Majorana neutrinos $M_{1,2,3}$ and the lightest neutrino mass $\mnulight$, corresponding in the code to the variables named
{\ttm M1}, {\ttm M2}, {\ttm M3}, {\ttm m}. For the Euler parameterisation in Eq.~\eqref{eq:Euler_O}, the user should also input the PMNS phases $\delta$, $\alpha_{21}$, $\alpha_{31}$, the PMNS angles $\theta_{12}$, $\theta_{13}$, $\theta_{23}$ (if these are not given as input, the \texttt{NuFit-6.1} best-fit values will be used automatically) and the CI angles
$x_1$, $y_1$, $x_2$, $y_2$, $x_3$ and $y_3$, which correspond in the code to {\ttm delta}, {\ttm a21}, {\ttm a31}, {\ttm t12}, {\ttm t13}, {\ttm t23},
{\ttm x1}, {\ttm y1}, {\ttm x2}, {\ttm y2}, {\ttm x3} and {\ttm y3}, respectively.
For the alternative parameterisation of Eq.~\eqref{eq:alt_O}, the user should input the CI parameters 
$x_1^\nu$, $x_2^\nu$, $x_1^N$, $x_2^N$, $x$ and $y$ with code variables 
\texttt{x1nu}, \texttt{x2nu}, \texttt{x1N}, \texttt{x2N}, \texttt{x} and \texttt{y} 
in place of $x_1$, $y_1$, $x_2$, $y_2$, $x_3$ and $y_3$.
For the manual setting, in addition to $M_1$, $M_2$, $M_3$ and $\mnulight$, the user should input the Yukawa coupling by specifying the magnitudes $Y_{e1}^{\rm mag}$, $Y_{e2}^{\rm mag}$, $Y_{e3}^{\rm mag}$, $Y_{\mu1}^{\rm mag}$ $Y_{\mu 2}^{\rm mag}$, $Y_{\mu 3}^{\rm mag}$, $Y_{\tau 1}^{\rm mag}$, $Y_{\tau 2}^{\rm mag}$,$Y_{\tau 3}^{\rm mag}$ respectively as
{\ttm Y11\_mag}, {\ttm Y12\_mag}, {\ttm Y13\_mag}, {\ttm Y21\_mag}, {\ttm Y22\_mag}, {\ttm Y23\_mag}, {\ttm Y31\_mag}, {\ttm Y32\_mag}, {\ttm Y33\_mag},
and the phases $Y_{e1}^{\rm phs}$, $Y_{e2}^{\rm phs}$, $Y_{e3}^{\rm phs}$, $Y_{\mu1}^{\rm phs}$ $Y_{\mu 2}^{\rm phs}$, $Y_{\mu 3}^{\rm phs}$, $Y_{\tau 1}^{\rm phs}$, $Y_{\tau 2}^{\rm phs}$,$Y_{\tau 3}^{\rm phs}$ as
{\ttm Y11\_phs}, {\ttm Y12\_phs}, {\ttm Y13\_phs}, {\ttm Y21\_phs}, {\ttm Y22\_phs}, {\ttm Y23\_phs}, {\ttm Y31\_phs}, {\ttm Y32\_phs}, {\ttm Y33\_phs}. We note that in the manual case the user should separately verify that the chosen Yukawa matrix correctly reproduces the observed neutrino oscillation data, while in the other cases this is ensured by the CI parameterisation. We provide a summary of the input parameters for each case in Table \ref{tab:full_param_overview}. 

\begin{table}
 \centering
\begin{subtable}{\textwidth}
 \centering
 \begin{tabular}{c|cr}
  \hline
  \rowcolor{gray!20}
  \multicolumn{3}{c}{\textbf{Common mass parameters}} \\ \hline
  \rule{0pt}{2.5ex}
  Parameter & \multicolumn{2}{c}{Code input example} \\ \hline
  \rule{0pt}{2.5ex}
  $\log_{10}(\mnulight/\text{eV})$ & {\ttm m} & {\ttm -1.10} \\
  $\log_{10}(M_1/\text{GeV})$ & {\ttm M1} & {\ttm 12.10} \\
  $\log_{10}(M_2/\text{GeV})$ & {\ttm M2} & {\ttm 12.60} \\
  $\log_{10}(M_3/\text{GeV})$ & {\ttm M3} & {\ttm 13.00} \\
  \bottomrule
 \end{tabular}
 \label{tab:pmns_ci}
  \caption{Mass parameters common to all the CI parameterisations or the manual setting of the Yukawas.}
\end{subtable}

 \vspace{1em}

 \begin{subtable}{\textwidth}
  \centering
  \begin{tabular}{c|cr}
  \hline
  \rowcolor{gray!20}
   \multicolumn{3}{c}{\textbf{Euler parameterisation}} \\ \hline
   \rule{0pt}{2.5ex}
   Parameter & \multicolumn{2}{c}{Code input example} \\ \hline
   \rule{0pt}{2.5ex}
   $x_1~[^\circ]$ & {\ttm x1} & {\ttm 45.00} \\
   $y_1~[^\circ]$ & {\ttm y1} & {\ttm 0.00} \\
   $x_2~[^\circ]$ & {\ttm x2} & {\ttm 45.00} \\
   $y_2~[^\circ]$ & {\ttm y2} & {\ttm 0.00} \\
   $x_3~[^\circ]$ & {\ttm x3} & {\ttm 45.00} \\
   $y_3~[^\circ]$ & {\ttm y3} & {\ttm 180.00} \\
     $\delta~[^\circ]$ & {\ttm delta} & {\ttm 213.70} \\
  $\alpha_{21}~[^\circ]$ & {\ttm a21} & {\ttm 81.60} \\
  $\alpha_{31}~[^\circ]$ & {\ttm a31} & {\ttm 476.70} \\
  $\theta_{23}~[^\circ]$ & {\ttm t23} & {\ttm 48.58} \\
  $\theta_{12}~[^\circ]$ & {\ttm t12} & {\ttm 33.63} \\
  $\theta_{13}~[^\circ]$ & {\ttm t13} & {\ttm 8.52} \\
   \bottomrule
  \end{tabular}
  \hspace{5em}
  \begin{tabular}{c|cr}
  \hline
   \rowcolor{gray!20}
   \multicolumn{3}{c}{\textbf{Alternative parameterisation}} \\ \hline
   \rule{0pt}{2.5ex}
   Parameter & \multicolumn{2}{c}{Code input example} \\ \hline
   \rule{0pt}{2.5ex}
   $x_1^N~[^\circ]$ & {\ttm xN1} & {\ttm 45.00} \\
   $x_2^N~[^\circ]$ & {\ttm xN2} & {\ttm 45.00} \\
   $x_1^\nu~[^\circ]$ & {\ttm xnu1} & {\ttm 45.00} \\
   $x_2^\nu~[^\circ]$ & {\ttm xnu2} & {\ttm 45.00} \\
   $x~[^\circ]$ & {\ttm x} & {\ttm 0.00} \\
   $y~[^\circ]$ & {\ttm y} & {\ttm 180.00} \\
     $\delta~[^\circ]$ & {\ttm delta} & {\ttm 213.70} \\
  $\alpha_{21}~[^\circ]$ & {\ttm a21} & {\ttm 81.60} \\
  $\alpha_{31}~[^\circ]$ & {\ttm a31} & {\ttm 476.70} \\
  $\theta_{23}~[^\circ]$ & {\ttm t23} & {\ttm 48.58} \\
  $\theta_{12}~[^\circ]$ & {\ttm t12} & {\ttm 33.63} \\
  $\theta_{13}~[^\circ]$ & {\ttm t13} & {\ttm 8.52} \\
   \bottomrule
  \end{tabular}
  \caption{Angles specific to the Euler parameterisation of the CI matrix (left) and the alternative one (right). 
  }
  \label{tab:euler_specific}
 \end{subtable}
 
 \vspace{1em}
 
 \begin{subtable}{\textwidth}
 \centering
 \begin{tabular}{c|l r|c|l r}
 \hline
  \rowcolor{gray!20}
  \multicolumn{6}{c}{\textbf{Manual setting of the Yukawa couplings}} \\ \hline
  \rule{0pt}{2.5ex}
  Magnitude & \multicolumn{2}{c|}{Code input example} & Phase & \multicolumn{2}{c}{Code input example} \\ \hline
  \rule{0pt}{2.5ex}
  $Y_{e1}^{\rm mag}$ & {\ttm Y11\_mag} & {\ttm 0.01} & $Y_{e1}^{\rm phs}$ & {\ttm Y11\_phs} & {\ttm -1.11} \\
  $Y_{e2}^{\rm mag}$ & {\ttm Y12\_mag} & {\ttm 0.01} & $Y_{e2}^{\rm phs}$ & {\ttm Y12\_phs} & {\ttm 2.89} \\
  $Y_{e3}^{\rm mag}$ & {\ttm Y13\_mag} & {\ttm 0.01} & $Y_{e3}^{\rm phs}$ & {\ttm Y13\_phs} & {\ttm 1.32} \\
  $Y_{\mu 1}^{\rm mag}$ & {\ttm Y21\_mag} & {\ttm 0.01} & $Y_{\mu 1}^{\rm phs}$ & {\ttm Y21\_phs} & {\ttm 2.88} \\
  $Y_{\mu 2}^{\rm mag}$ & {\ttm Y22\_mag} & {\ttm 0.03} & $Y_{\mu 2}^{\rm phs}$ & {\ttm Y22\_phs} & {\ttm -0.23} \\
  $Y_{\mu 3}^{\rm mag}$ & {\ttm Y23\_mag} & {\ttm 0.05} & $Y_{\mu 3}^{\rm phs}$ & {\ttm Y23\_phs} & {\ttm -1.80} \\
  $Y_{\tau 1}^{\rm mag}$ & {\ttm Y31\_mag} & {\ttm 0.01} & $Y_{\tau 1}^{\rm phs}$ & {\ttm Y31\_phs} & {\ttm -1.72} \\
  $Y_{\tau 2}^{\rm mag}$ & {\ttm Y32\_mag} & {\ttm 0.03} & $Y_{\tau 2}^{\rm phs}$ & {\ttm Y32\_phs} & {\ttm 2.96} \\
  $Y_{\tau 3}^{\rm mag}$ & {\ttm Y33\_mag} & {\ttm 0.05} & $Y_{\tau 3}^{\rm phs}$ & {\ttm Y33\_phs} & {\ttm 1.39} \\
  \bottomrule
 \end{tabular}
 \caption{Manual input of the complex Yukawa couplings $Y_{\alpha j}$.}
 \label{tab:yukawa_manual}
\end{subtable}

 \caption{%Overview of 
 \texttt{ULYSSES}'s input parameters. Those in (a) are common to all the provided parameterisations of the Yukawa, those in (b-left) are specific to the Euler parameterisation of the CI matrix, those in (b-right) are for the alternative one, those in (c) are for the manual setting.}
 \label{tab:full_param_overview}
\end{table}

The parameterisation is selected automatically by the \texttt{ULYSSES} tools (\texttt{uls-calc}, \texttt{uls-nest}, \texttt{uls-scan}) based on the keys present in the parameter file: no additional command-line flags are required. Specifically, if the parameter file contains the Euler CI angles \texttt{x1}, \texttt{x2}, \texttt{x3}, \texttt{y1}, \texttt{y2}, \texttt{y3}, the \texttt{euler} parameterisation is used; if it contains \texttt{xnu1}, \texttt{xnu2}, \texttt{xN1}, \texttt{xN2}, \texttt{x}, \texttt{y}, the \texttt{single-imaginary} parameterisation is used; and if it contains explicit Yukawa entries \texttt{Y11\_mag}, \texttt{Y11\_phs}, etc., the \texttt{manual} setting is adopted.

\subsection{Decoupling one heavy Majorana neutrino in the Casas-Ibarra parameterisation}\label{ssec:RHNdecouple}
Within the CI parameterisation, it is possible to decouple one of the three heavy Majorana neutrinos by artificially setting the CI parameters to specific values so that the associated Yukawa couplings vanish numerically. This allows the user to keep the entire notation as in the case of three heavy neutrinos ($e.g.$, the Yukawa and CI matrices remain $3\times 3$), but effectively consider just two of them. In particular, this procedure is necessary in order to properly use the \texttt{ULYSSES} modules within a type-I seesaw scenario with two RHNs. Given how we have written the \texttt{ULYSSES} code, the modules that work with two neutrinos consider the pair $N_1$ and $N_2$. For these modules, it is necessary to decouple $N_3$. Instead, those considering only the CP-violating processes of one heavy neutrino work with $N_1$. For the latter, decoupling either $N_3$ or $N_2$ would be equivalent up to a re-labelling of the heavy neutrino indices. Then, in what follows, it suffices to show how to decouple $N_3$. 

First, since in the case of only two heavy neutrinos, $\mnulight \simeq 0$ at tree and one-loop level, the user should input a tiny entry for $\mnulight$. 
Then, $N_3$ should be taken much more massive than the other two, to ensure it is eliminated from the leptogenesis dynamics.\footnote{A heavy neutrino with exactly vanishing Yukawa couplings is surely decoupled from the dynamics regardless of its mass. However, within this approach, the Yukawa couplings can only be set to zero approximately. If the decoupled heavy neutrino mass is too small, residual (non-zero) Yukawa couplings may lead to numerical overflows. To avoid such issues, it is preferable to input the mass of the decoupled neutrino as large as possible. Nevertheless, the setup can still be used to effectively describe a scenario with two heavy Majorana neutrinos and an additional decoupled lighter one.} Lastly, using the Euler parameterisation of the CI matrix of Eq.~\eqref{eq:Euler_O}, the CI angles should be set as follows.

\begin{itemize}
 \item For NO: $x_1 = x_3 = \pi/2$, $y_1 = y_3 = 0$, keeping $x_2$ and $y_2$ free.
 \item For IO: $x_1 = x_2 = y_1 = y_2 = 0$, keeping $x_3$ and $y_3$ free.
\end{itemize}
In this way, the $R$-matrix takes the 
form
\begin{equation}
 R^{\rm (NO)} 
=
\begin{pmatrix}
0 & \cos z_2 & \sin z_2 \\
0 & -\sin z_2 & \cos z_2 \\
1 & 0 & 0
\end{pmatrix},
\quad
R^{\rm (IO)} 
=
\begin{pmatrix}
\cos z_3 & \sin z_3 & 0 \\
-\sin z_3 & \cos z_3 & 0 \\
0 & 0 & 1
\end{pmatrix},
\end{equation}
respectively in the NO and IO case, leading to $Y_{\alpha 3} = 0$ for $\alpha = e,\,\mu,\,\tau$. We note that the $2\times 2$ rotation sub-blocks contained in the $R$-matrices given above (obtained by removing the rows and columns with 0 and 1 entries) have positive unit determinant. However, the entire set of Yukawa matrices in the type-I seesaw scenario with two heavy neutrinos can be obtained only by allowing these sub-blocks to have determinant either $+1$ or $-1$. One possibility to account for all the possible $R$-matrices in this case is to multiply one of the columns of these sub-blocks by an overall factor $\varphi = \pm 1$, but this would correspond to an additional parameter to be input by the user. In \texttt{ULYSSES}, we prefer to keep the number of input parameters unchanged and simply note that both the cases of determinant equal to $\pm 1$ can be considered by extending the range of the Majorana phases from $[0, 2\pi]$ to $[0, 4\pi]$. We provide in Table \ref{tab:ParamCards_Ndecoupled} an example parameter card with the code inputs to decouple $N_3$.

We stress that the decoupling prescription described above is not equivalent to imposing a vanishing CP asymmetry $\epsilon_{j \alpha}$. One of the RHN may have the associated $\epsilon_{j\alpha}=0$ while retaining non-zero Yukawa couplings, in which case it can still contribute to the light-neutrino mass matrix and wash out an asymmetry generated by another RHN, see e.g.~\cite{Zhao:2021efc}. Such a RHN must be retained in the kinetic equations and can be treated using the multi-RHN modules and a density-matrix module should be used when flavour projections or coherences are important.
 
\begin{table}[t]
 \centering
 \begin{subtable}{0.45\textwidth}
  \centering
  \begin{tabular}{l r}
   \rowcolor{gray!20}
   \hline
   \multicolumn{2}{c}{\bf \(N_3\) decoupled, NO}\\
   \hline
   \rule{0pt}{2.5ex}
   \ttm{m}  & \ttm{-100} \\
   \ttm{M3} & \ttm{15} \\
   \ttm{x1} & \ttm{90} \\
   \ttm{x3} & \ttm{90} \\
   \ttm{y1} & \ttm{0} \\
   \ttm{y3} & \ttm{0} \\
   \ttm{x2} & Free \\
   \ttm{y2} & Free \\
   \bottomrule
  \end{tabular}
 \end{subtable}
 \hspace{1em}
 \begin{subtable}{0.45\textwidth}
  \centering
  \begin{tabular}{l r}
   \rowcolor{gray!20}
   \hline
   \multicolumn{2}{c}{\bf \(N_3\) decoupled, IO}\\
   \hline
   \rule{0pt}{2.5ex}
   \ttm{m}  & \ttm{-100} \\
   \ttm{M3} & \ttm{15} \\
   \ttm{x1} & \ttm{0} \\
   \ttm{x2} & \ttm{0} \\
   \ttm{y1} & \ttm{0} \\
   \ttm{y2} & \ttm{0} \\
   \ttm{x3} & Free \\
   \ttm{y3} & Free \\
   \bottomrule
  \end{tabular}
 \end{subtable}
 \caption{Example parameter card to achieve the decoupling of the heavy Majorana neutrino $N_3$, for both NO (left) and IO (right), using the Euler parametrisation of the CI matrix as in Eq.~\eqref{eq:Euler_O}. All unspecified parameters should be set according to Table~\ref{tab:full_param_overview}, noting that the Majorana phases can be chosen in the range $[0,4\pi]$.}
 \label{tab:ParamCards_Ndecoupled}
\end{table}

\subsection{Choosing the initial heavy neutrino abundance}

We have introduced a new instance parameter, \texttt{initial$\_$abundance}, in the \texttt{ULSBase} class that allows the user to select the initial abundance of the RHNs. Setting \texttt{initial$\_$abundance} = 0 (1) corresponds to a vanishing (thermal) initial abundance. This provides a unified and user-friendly way to control the initial conditions across all modules, avoiding potential confusion arising from different conventions in the various sets of equations. Note that the user can input any real value for \texttt{initial$\_$abundance}; the initial heavy neutrino abundance will then scale linearly with this parameter.

From the command line, the initial condition is selected via the \texttt{-{}-initial} flag, which is available in all compute tools (\texttt{uls-calc}, \texttt{uls-scan}, \texttt{uls-scan2D}, \texttt{uls-nest}):
\begin{lstlisting}[language=bash]
uls-calc -m 1BE1F --initial 0 examples/1N1F.dat  # vanishing (default)
uls-calc -m 1BE1F --initial 1 examples/1N1F.dat  # thermal
\end{lstlisting}
When using the Python API, the same parameter is passed to \texttt{selectModel}:
\begin{lstlisting}[language=bash]
L = ulysses.selectModel('1BE1F', initial_abundance=1, ...) # thermal IC
L.initial_abundance = 0.5 #intermediate values accepted, can access as property
\end{lstlisting}
The default is vanishing initial conditions. Intermediate values are also accepted and scale the equilibrium abundance linearly.

\subsection{Extending models with custom parameters}
\label{ssec:extended}
\texttt{ULYSSES} version 3 provides a procedure to pass model-specific parameters beyond the standard set listed in Table~\ref{tab:full_param_overview}, without modifying the core infrastructure. This is an extension of an example in v2 \cite{Granelli:2023vcm} can be found in the primordial black hole BEs \cite{Cheek:2021odj,Cheek:2021cfe}. This is activated via the \texttt{-{}-extended} flag: 
\begin{lstlisting}[language = bash]
uls-calc -m <ModelName> --extended examples/mycard.dat
\end{lstlisting}
In extended mode, the parameter file may contain any additional key--value pairs beyond the standard \texttt{pnames} list. All entries are collected into the dictionary \texttt{pdict} and are available inside the model's \texttt{EtaB} method. A custom module accesses them as
\begin{lstlisting}[language = bash]
def EtaB(self, *args, **kwargs):
  lam = self.pdict.get('lam', 1e-9)     # dark coupling, with default
  m_dm = self.pdict.get('m_dm', self.M1/10.) # DM mass [GeV]
  ...
\end{lstlisting}
The same mechanism is available from API by passing \texttt{extended\_mode=True} to \texttt{selectModel} and including the extra keys in the parameter dictionary. This keeps the user interface uniform: a single runcard controls both the standard leptogenesis parameters and any model-specific extensions. The dark matter module described in Sec.~\ref{sec:DM} uses this interface to accept the dark-sector coupling $\lambda$ and the dark matter mass directly from the runcard.
\subsection{Storing and plotting the ODE evolution}
\label{ssec:plots}

Beyond returning $\eta_B$, any \texttt{ULYSSES} module can store the full \emph {Ordinary Differential Equation} (ODE) trajectory for post-hoc inspection by calling \texttt{setEvolData(ys)} at the end of \texttt{EtaB} after integration is complete. The stored trajectory is available as \texttt{evolData}, an $N_z \times N_{\rm col}$ array whose first column is $z$ and whose subsequent columns are determined by the index and label methods described below. When using the Python API, \texttt{L.evolData} is accessible immediately after calling \texttt{L(params)}, enabling downstream analysis without re-running the solver.

Custom modules control which ODE columns are stored and how they are labelled by overriding four methods in the subclass (see also Tab.~\ref{tab:ulsbase_evol} in App.~\ref{app:ulsbase}):
\begin{lstlisting}[language = bash]
def flavourindices(self): return [1]  # columns for standard quantities
def flavourlabels(self): return [r"$N_{B-L}$"]
def extendedindices(self): return [2, 3] # extra columns (no effect on etaB)
def extendedlabels(self): return [r"$N_{DM}$", r"$N_{DM}^{eq}$"]
\end{lstlisting}
The indices refer to columns of the ODE solution array \texttt{ys}. Columns listed in \texttt{extendedindices()} are stored alongside the standard quantities but play no role in the baryon asymmetry calculation; they serve purely as a record of any additional quantities the module tracks. If \texttt{extendedindices()} returns an empty list (the default), only the standard quantities are stored.

Passing the \texttt{-o} flag to \texttt{uls-calc} produces an evolution plot of the stored data:
\begin{lstlisting}[language = bash]
uls-calc -m 1DME --ordering 0 examples/1N1F.dat -o evol.pdf
\end{lstlisting}
For modules where \texttt{extendedindices()} returns an empty list, the plotter produces a single panel showing the comoving abundances against $z = M_1/T$. If \texttt{extendedindices()} is non-empty, a second panel is added automatically showing the supplementary quantities --- for example, $N_{\rm DM}(z)$ alongside $N_{\rm DM}^{\rm eq}(z)$ for the module of Sec.~\ref{sec:DM}. Similarly, \texttt{uls-scan} and \texttt{uls-scan2D} produce 1D and 2D plots of $\eta_B$ as a function of the scanned parameter(s).

\subsection{Python API and Jupyter notebook}
\label{ssec:notebook}

In addition to the command-line tools, \texttt{ULYSSES} exposes a pure Python \textit{application programming interface} (API). Any model can be instantiated and called as a function:
\begin{lstlisting}[language = bash]
import ulysses

L = ulysses.selectModel('1BE1F', zmin=0.1, zmax=100,
             zsteps=1000, ordering=0, loop=False)
params = {'m': -100, 'M1': 14, 'M2': 15, 'M3': 16,
     'delta': 270, 'a21': 0, 'a31': 0,
     'x1': 180, 'y1': 1.4, 'x2': 180, 'y2': 11.2,
     'x3': 180, 'y3': 11,
     't12': 33.76, 't13': 8.62, 't23': 43.27}
eta_b = L(params)
\end{lstlisting}
After the call, the full ODE trajectory is available in \texttt{L.evolData}, enabling post-hoc inspection and the same two-panel plots described in Sec.~\ref{ssec:plots} without re-running. Parameter scans reduce to a simple loop over a list of dictionaries.

A self-contained Jupyter notebook (\texttt{ULYSSES\_intro.ipynb}, included in the repository root) demonstrates four representative scenarios end-to-end:
\begin{enumerate}
 \item \texttt{1BE1F} - unflavoured single-RHN leptogenesis using \texttt{euler} CI parametrisation.
 \item \texttt{1DME} - fully flavoured leptogenesis via density matrix equations, illustrating the evolution of individual lepton-flavour asymmetries, using \texttt{euler} CI parametrisation.
 \item \texttt{BEARS\_3RHN} - low-scale leptogenesis with three nearly degenerate RHNs, using the \\ \texttt{single\_imaginary} CI parametrisation.
 \item \texttt{1BE1F\_DM\_FreezeIn} - simultaneous leptogenesis and freeze-in dark matter production using the \texttt{extended\_mode} interface.
\end{enumerate}

\section{Unified treatment of leptogenesis via oscillations and resonant leptogenesis}
\label{sec:lowlep}
Low-scale leptogenesis mechanisms such as leptogenesis via oscillations~\cite{Akhmedov:1998qx, Asaka:2005pn} and resonant leptogenesis~\cite{Pilaftsis:1997jf, Pilaftsis:2003gt} make leptogenesis a testable baryogenesis scenario.
The two mechanisms were originally formulated in opposite kinematic regimes, leptogenesis via oscillations in the relativistic limit and described from the outset using DMEs,
whereas resonant leptogenesis was formulated in terms of CP-violating decays of non-relativistic RHNs.
First-principles studies~\cite{Garny:2011hg,Garbrecht:2011aw,BhupalDev:2014pfm} of resonant leptogenesis subsequently converged towards a DME description,
ultimately showing that both mechanisms can consistently be described by a common set of DMEs~\cite{Klaric:2020phc, Klaric:2021cpi}.

In \texttt{ULYSSES} version 2 \cite{Granelli:2023vcm}, we implemented the momentum-averaged DMEs relevant to leptogenesis via oscillations in the case of two heavy neutrinos. While these equations are comprehensive, accounting for lepton number conserving (LNC) and lepton number violating (LNV) processes in the relativistic regime, as well as including non-linear terms involving heavy neutrino abundances and lepton chemical potentials, they become inaccurate for masses of the heavy neutrinos above 100 GeV. Above this mass threshold, non-relativistic effects become significant and one must improve the computation of the production and destruction rates of heavy neutrinos accordingly, see e.g.~\cite{Klaric:2020phc, Klaric:2021cpi}. In this third release of \texttt{ULYSSES}, we add such improvements extending the validity of the already implemented DMEs well above the scale of $100 \,\text{GeV}$, while also accounting for three heavy neutrinos.
The framework presented here is therefore also applicable in the resonant leptogenesis regime. We accompany these improvements with additional refinements aimed at making the baryon computation more accurate and faster.

\subsection{Density matrix equations with three heavy neutrinos}
\label{sec:DMEs}
%%%%%%%%%%%%%%%%%%%%%%%%%%%%
%

The momentum-averaged DMEs relevant to the scenario of low-scale leptogenesis and implemented in \texttt{ULYSSES} are given by (see \cite{Drewes:2016gmt,Hernandez:2016kel,Ghiglieri:2017gjz, Eijima:2017anv} for derivations)
\allowdisplaybreaks
\begin{subequations}
\begin{align}
\label{eq:DME_N}
\frac{d\delta r_N}{dx} = &-\frac{i}{Hx}\left[\langle \mathcal{H}\rangle,\delta r_N\right]-\frac{1}{N_{N}^{\rm eq}|_{x=0}}\frac{dN_{N}^{\rm eq}}{dx}- \frac{\langle\gamma_{\LNC}^{(0)}\rangle}{2Hx}\left\{Y^\dagger Y,\delta r_N\right\} +\nonumber\\&+ \frac{N_N^{\rm eq}}{N_N^{\rm eq}|_{x=0}}{}\frac{\langle\tilde{\gamma}_{\LNC}^{(1)}\rangle}{Hx} Y^\dagger \mu Y -\frac{\langle\gamma_{\LNC}^{(2)}\rangle}{2Hx}\left\{Y^\dagger\mu Y,\delta r_N\right\} - \frac{\langle S_{\LNV}^{(0)}\rangle}{2T^2Hx}\left\{ MY^{\rm T}Y^*M,\delta r_N\right\}+\nonumber\\
& - \frac{N_N^{\rm eq}}{N_N^{\rm eq}|_{x=0}}\frac{\langle \tilde{S}_{\LNV}^{(1)}\rangle}{T^2Hx}\,M Y^{\rm T}\mu Y^* M + \frac{\langle S_{\LNV}^{(2)}\rangle}{2T^2Hx}\left\{MY^{\rm T}\mu Y^* M,\delta r_N\right\},\\
\label{eq:DME_Nbar}
\frac{d\delta r_{\bar{N}}}{dx} = &-\frac{i}{Hx}\left[\langle \mathcal{H^*}\rangle,\delta r_{\bar{N}}\right]-\frac{1}{N_{N}^{\rm eq}|_{x=0}}\frac{dN_{N}^{\rm eq}}{dx}- \frac{\langle\gamma_{\LNC}^{(0)}\rangle}{2Hx}\left\{Y^{\rm T} Y^*,\delta r_{\bar{N}}\right\} +\nonumber\\&- \frac{N_N^{\rm eq}}{N_N^{\rm eq}|_{x=0}}\frac{\langle\tilde{\gamma}_{\LNC}^{(1)}\rangle}{Hx} Y^T \mu Y^* +\frac{\langle\gamma_{\LNC}^{(2)}\rangle}{2Hx}\left\{Y^{\rm T}\mu Y^*,\delta r_{\bar{N}}\right\} - \frac{\langle S_{\LNV}^{(0)}\rangle}{2T^2Hx}\left\{ MY^{\dagger}YM,\delta r_{\bar{N}}\right\}+\nonumber\\
& + \frac{N_N^{\rm eq}}{N_N^{\rm eq}|_{x=0}}\frac{\langle \tilde{S}_{\LNV}^{(1)}\rangle}{T^2Hx}\,M Y^{\dagger}\mu Y M - \frac{\langle S_{\LNV}^{(2)}\rangle}{2T^2Hx}\left\{MY^{\dagger}\mu Y M,\delta r_{\bar{N}}\right\},\\
%%%%
\kappa \frac{d\mu_{\Delta_\alpha}}{dx} = &-\frac{\langle\gamma_{\LNC}^{(0)}\rangle}{2Hx}(Y\delta r_NY^\dagger-Y^* \delta r_{\bar{N}}Y^{\rm T})_{\alpha\alpha} + \frac{N_N^{\rm eq}}{N_N^{\rm eq}|_{x=0}}\frac{\langle\tilde{\gamma}_{\LNC}^{(1)}\rangle}{Hx}(YY^\dagger)_{\alpha\alpha}\mu_\alpha+\nonumber\\
&-\frac{\langle\gamma_{\LNC}^{(2)}\rangle}{2Hx}[Y\delta r_NY^\dagger+Y^*\delta r_{\bar{N}}Y^{\rm T}]_{\alpha\alpha}\mu_\alpha+\nonumber\\
&{}+ \frac{\langle S_{\LNV}^{(0)}\rangle}{2T^2Hx}(Y^*M\delta r_NMY^{\rm T}-YM\delta r_{\bar{N}}MY^\dagger)_{\alpha\alpha}+ \frac{N_N^{\rm eq}}{N_N^{\rm eq}|_{x=0}}\frac{\langle \tilde{S}_{\LNV}^{(1)}\rangle}{T^2Hx}(YM^2Y^\dagger)_{\alpha\alpha}\mu_\alpha+ \nonumber\\
&{}-\frac{\langle S_{\LNV}^{(2)}\rangle}{2T^2Hx}[YM\delta r_{\bar{N}}MY^\dagger+Y^*M\delta r_NMY^{\rm T}]_{\alpha\alpha}\mu_\alpha\,,
\label{eq:DME_mu}
 \end{align}
 \end{subequations}
 where $\alpha = e,\,\mu,\,\tau$. The evolution variable $x$ is defined as $x\equiv T_{\sph}/T$, with $T_{\sph}\simeq 131.7\,\text{GeV}$ being the temperature of sphaleron freeze-out \cite{DOnofrio:2014rug}; $H \simeq 1.66 \sqrt{g_{*}}\,T^2/M_{\rm P}$ is the Hubble rate for the Universe's expansion; $g_* = 106.75$ is the effective number of relativistic degrees of freedom; $M_{\rm P}\simeq 1.22 \times 10^{19}\,\text{GeV}$ the Planck mass. The DMEs written above govern the evolution of the asymmetry in the $\Delta_\alpha \equiv B/3-L_\alpha$ number, that is, of the associated chemical potential (normalised to temperature) $\mu_{\Delta_\alpha}$ and of the comoving number density matrix of the heavy Majorana neutrinos with positive helicity $\rho_N$, having defined  $\delta r_N\equiv (\rho_N-N_N^{\rm eq}\, \mathds{1})/N_{N}^\text{eq}|_{x=0}$, with $N_{N}^\text{eq}$ the number of heavy Majorana neutrinos, per generation, in a comoving volume when in thermal equilibrium, $N_{N}^\text{eq}|_{x=0}$ is that at infinite temperature.\footnote{In the second version of the code (Ref.~\cite{Granelli:2023vcm}) we adopted a different convention, writing the equations in terms of $R_N\equiv \rho_N/N_N^{\rm eq}$ and $R_{\bar N}\equiv \rho_{\bar N}/N_N^{\rm eq}$. With the convention used here we avoid potential overflows in the non-relativistic limit when one neutrino is decoupled.} The equations for the density matrix of the heavy Majorana neutrinos with opposite helicity $\rho_{\bar{N}}$ in Eq.~\eqref{eq:DME_Nbar}, are obtained by substituting $\delta r_N\to \delta r_{\bar{N}}$, $\mu \to-\mu$ and $Y\to Y^*$ in Eq.~\eqref{eq:DME_N}, with $\delta r_{\bar{N}}=(\rho_{\bar{N}}-N_N^{\rm eq}\,\mathds{1})/N_{N}^\text{eq}|_{x=0}$.  The matrices $M=\text{diag}(M_1,M_2, M_3)$ and $\mu=\text{diag}(\mu_e,\, \mu_\mu,\,\mu_\tau)$ are respectively the matrix of the heavy Majorana neutrinos in the mass basis and that of the lepton chemical potentials $\mu_\alpha = -2\sum_{\beta =e,\,\mu,\,\tau}\chi_{\alpha\beta}\mu_{\Delta_\beta}$, where the susceptibility matrix $\chi$ accounts for the effects of spectator processes and reads \cite{Eijima:2017anv, Ghiglieri:2017gjz}
 \begin{equation}
 \chi= \frac{1}{711}\begin{pmatrix} 257&20&20\\
 20&257&20\\
 20&20&257
 \end{pmatrix}\,.
\end{equation}
The thermally averaged Hamiltonian $\left\langle\mathcal{H}\right\rangle$ is given by \cite{Ghiglieri:2017gjz, Antusch:2017pkq} (see also \cite{Hernandez:2016kel, Abada:2018oly})
\begin{equation}
\frac{\langle \mathcal{H}\rangle}{T} \simeq \left\langle\frac{1}{y_0}\right\rangle \frac{\Delta M_\text{diag}^2}{2T^2} + (Y^\dagger Y)\langle h^{\LNC}\rangle + (Y^T Y^*)\langle h^{\LNV}\rangle\,, 
\label{eq:thermH}
\end{equation}
where we have included both LNC and LNV thermal contributions, respectively encoded in $\langle h^{\LNC}\rangle$ and $\langle h^{\LNV}\rangle$, full energy-momentum relations and Fermi-Dirac statistics in the thermal averages; $y_0=E_N/T$, $E_N$ being the heavy neutrinos' energy; $\Delta M_\text{diag}^2 \equiv \text{diag}(0,M_2^2-M_1^2, M_3^2-M_1^2)$. We have subtracted an overall $M_1^2$ factor from the zero-temperature contribution without affecting the dynamics.

The quantities $\gamma_{\LNC}^{(0)}, \tilde{\gamma}_{\LNC}^{(1)}, \gamma_{\LNC}^{(2)}$ and $S_{\LNV}^{(0)}, \tilde{S}_{\LNV}^{(1)}, S_{\LNV}^{(2)}$, are the rates respectively for the LNC and LNV processes involving the heavy Majorana neutrinos. The dependence on the momentum and temperature of the LNV and LNC rates was computed numerically in the relativistic limit in \cite{Ghiglieri:2017csp} (see also \cite{Ghiglieri:2017gjz}) and tabulated in \href{http://www.laine.itp.unibe.ch/leptogenesis/}{\ttfamily{this site}}. To extend the validity of the DMEs above the $100\,\text{GeV}$ mass scale mass and include the non-relativistic corrections to the LNV and LNC rates into account, we follow the procedure presented in \cite{Klaric:2021cpi} (see also \cite{Klaric:2020phc, Drewes:2021nqr}).
We then perform the thermal average, denoted by $\left\langle \cdot \right \rangle$, including the full energy-momentum relation and Fermi-Dirac statistics. We give more details on $\langle\gamma_{\LNC}^{(0)}\rangle,\, \langle\tilde\gamma_{\LNC}^{(1)}\rangle,\,\langle\gamma_{\LNC}^{(2)}\rangle,\,\langle S_{\LNV}^{(0)}\rangle,\, \langle \tilde{S}_{\LNV}^{(1)}\rangle$ and $\langle S_{\LNV}^{(2)}\rangle$, as well as on $\langle 1/y_0\rangle$, $\langle h^{\LNC}\rangle$, $\langle h^{\LNC}\rangle$, in Appendix \ref{app:Rates}. We also include indirect contributions to $\langle h^{\LNC}\rangle$ and $\langle h^{\LNV} \rangle$ in the broken phase. 
%\AG{Remember to cancel the flag also in the table at the end of the paper or keep it always true by default and we don't mention it} \DP{It is always true by default, do we completely remove the option of setting it by hand? Maybe even for ourselves or as an easter egg we can keep it.} \AG{I would keep it in the code without mentioning it anywhere, it might be useful for ourselves.} 
The thermally averaged rates and indirect corrections to the Hamiltonian are precomputed on temperature and mass grids and stored in interpolation tables (in the folder \texttt{data/ARS\_rates}) that are used by default in the 3RHN module when solving the DMEs. To facilitate modifications and extensions, we provide the code \texttt{ProdRatesCalc.py} used to generate these tables, together with the code \texttt{ProdRates.py} containing all the necessary functions used for the evaluation. The necessary numerical integrations are accelerated using a bespoke \texttt{numba}-compatible wrapper around \texttt{cquadpack}. {All the thermally averaged rates and corrections to the Hamiltonian are computed in the code at the same mass $\Mav \equiv M_1$. The corrections to the rates induced by non-zero mass splittings is thus not captured by this approach, which nevertheless remains valid as long as the mass splittings are sufficiently small, or the heavy neutrinos are sufficiently relativistic. Thus, \emph{the user should not solve this set of equations for arbitrarily large heavy neutrino mass splittings.}} 

We adopt a normalisation of the comoving volume such that it contains a single photon when $z\ll 1$, with $z\equiv \Mav x/T_{\sph}$
and $N_N^{\rm eq}|_{x=0} = N^\text{eq}_{N}(z\ll 1) = 3/8$. We also consider the analytical approximation $ N^\text{eq}_{N}(z) \simeq (3/16) z^2 K_2(z)$, where $K_n(z)$, $n = 1\,,2\,,\,...$, is the modified $n^\text{th}$ Bessel function of the second kind.\footnote{We note that $z^2K_2(z)\to 2$ for $z\to +\infty$ so that the normalisation at $z\ll1$ of $N_{N}^\text{eq}(z)$ is restored.} Consequently, the overall constant factor $\kappa$ appearing in Eq.~\eqref{eq:DME_mu} reads $\kappa \simeq 2\pi^2/[9\zeta(3)]$,
 while
 \begin{equation}
  \frac{1}{N_{N}^\text{eq}|_{x=0}}\frac{dN_{N}^\text{eq}}{dx} \simeq -\frac{\Mav}{2T_{\sph}}z^2K_1(z)\,.
 \end{equation}
We implement the DMEs with three quasi-degenerate heavy Majorana neutrinos discussed in this section in a new module called \texttt{etabARS$\_$3RHN.py}.
The code, after solving numerically the DMEs, computes the present baryonic yield as
%%%%%%%%%%%%%%%%%%%%%%%%%%%%%%%%%
 \begin{equation}
  Y_B = c_s\,\frac{15}{2\pi^2 g_{*,s}}
 \sum_{\alpha = e,\,\mu,\,\tau} \mu_{\Delta_\alpha}^{\sph}\,,
 \label{eq:YBl}
 \end{equation}
%%%%%%%%%%%%%%%%%%%%%%%%%%%%%%%%%%%
%
where $\mu_{\Delta_{e,\mu,\tau}}^{\sph}$ are the chemical potentials $\mu_{\Delta_{e,\mu,\tau}}$ evaluated at $T_{\sph}$ ($i.e.$~at $x=1$), $c_s = 28/79$ is the sphaleron conversion coefficient and $g_{*,s} = 106.75$ are the entropic effective degrees of freedom during leptogenesis. Also, the code converts the asymmetry into the baryon-to-photon ratio $\eta_{B}$ and the baryon density parameter $\Omega_Bh^2$ according to
\begin{equation}
 \eta_B = Y_{B} \cdot \frac{\pi^4 g_{*,s}^{(0)}}{45\zeta(3)},\quad \Omega_Bh^2 = \eta_B\cdot \frac{m_p n_\gamma^{(0)}}{\rho_c h^{-2}},
\end{equation}
where $g_{*,s}^{(0)} = 43/11$ are the entropic effective degrees of freedom today, $n_\gamma^{(0)}\simeq 410.7\,\text{cm}^{-3}$ is the present photon number density, $m_p = 1.672621898\times 10^{-24}\,\text{g}$ is the proton mass and $\rho_c = 1.87840 \times 10^{-29}h^2\,\text{g}\,\text{cm}^{-3}$ is the critical density of the Universe, $h$ is the Hubble expansion rate normalised to $100\,\text{km}\,\text{s}^{-1}\,\text{Mpc}^{-1}$ \cite{ParticleDataGroup:2024cfk}. 

To numerically solve the DMEs we implement the function \texttt{solve$\_$ivp} of the \texttt{SciPy} Python package \cite{Virtanen:2019joe} with the \textit{Backward Differentiation Formula} (BDF) method, setting the absolute tolerance $\texttt{atol} = 10^{-13}$ and relative tolerance $\texttt{rtol} = 10^{-6}$ and the default range of $x$ to $[10^{-6},1]$.

With the improvements in the rates and Hamiltonian included in this module, we extend the validity of the DMEs much above the scale of $100\,\text{GeV}$. However, for sufficiently large masses and splittings, a more comprehensive treatment of leptogenesis involving decoherence effects for the flavour asymmetries might become necessary. To get an idea of when these effects start playing a role, one can compare the temperature at which oscillations set on, 
\begin{equation}
     T_{\rm osc}\simeq \left(\frac{|\Delta M|\bar{M} M_P}{9.96 \sqrt{g_*}}\right)^{1/3} \simeq 10^{9}\,\text{GeV}\,\left(\frac{|\Delta M|}{\bar{M}}\right)^{1/3}\left(\frac{\bar{M}}{100\,{\rm TeV}}\right)^{2/3},
     \label{eq:Tosc}
\end{equation} $\Delta M$ being the largest mass splittings between the non-decoupled heavy neutrinos, with the temperature $T_{\text{2-flav}}\approx 10^9\,\text{GeV}$ above which the $\mu$-Yukawa interaction are out-of-equilibrium, i.e.~when leptogenesis proceeds in the two-flavour regime \cite{Abada:2006ea} (see also, e.g., \cite{Moffat:2018smo, Granelli:2021fyc}). 

\begin{figure}[t!]
\centering
 \includegraphics[width=0.85\textwidth]{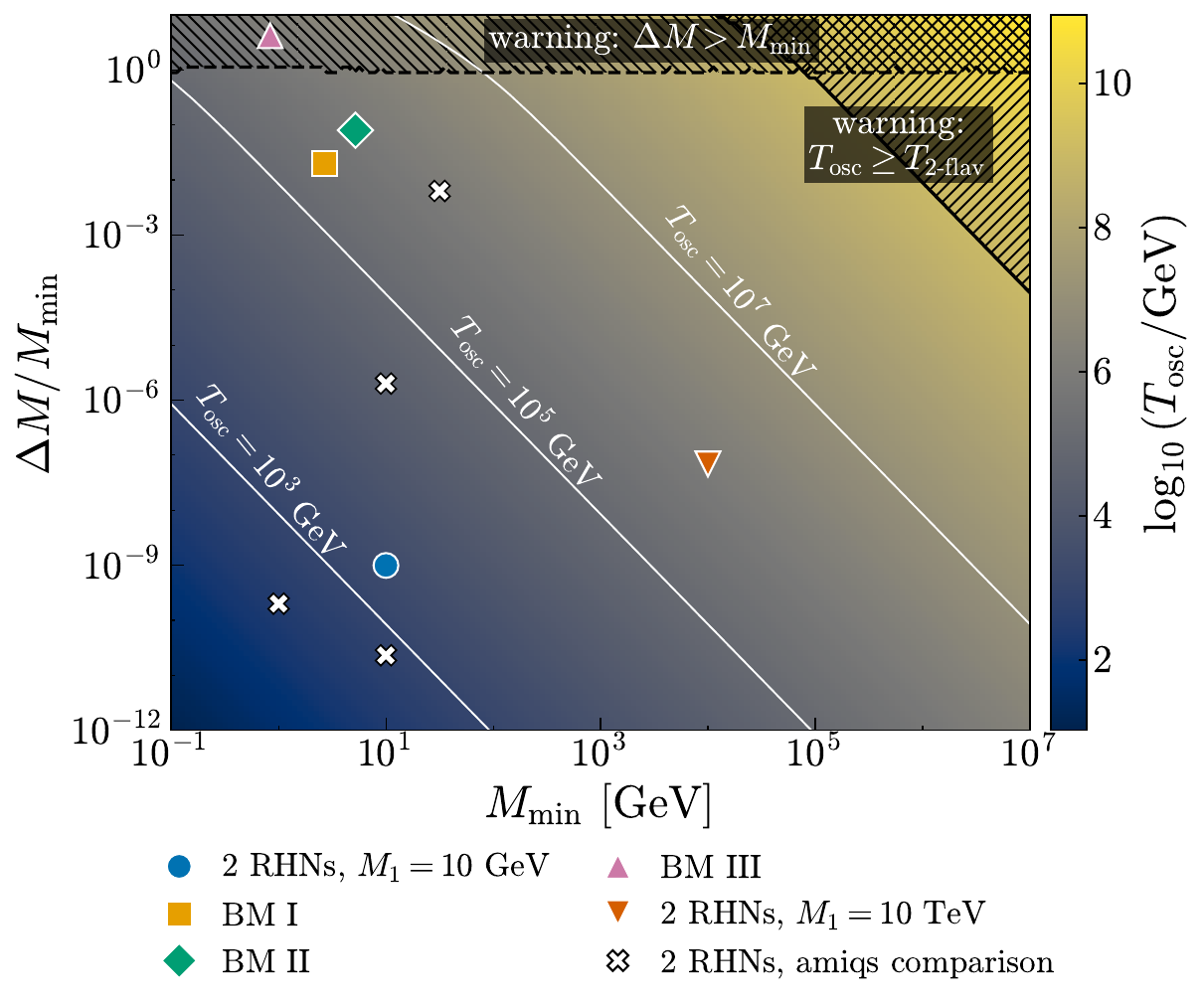}
  \caption{Oscillation temperature $T_{\rm osc}$ of Eq.~\eqref{eq:Tosc}  in the plane of the lightest non-decoupled heavy-neutrino mass $M_{\rm min}$ and the largest relative mass splitting $\Delta M/M_{\rm min}$. The hatched regions are those in which \texttt{etabARS$\_$3RHN.py} issues the warning for $\Delta M>M_{\rm min}$ ($T_{\rm osc}\geq T_{\text{2-flav}}=10^{9}\,\text{GeV}$). Markers show the benchmarks of Figs.~\ref{fig:Check_ULS}--\ref{fig:BMCs} and the two-RHN benchmarks of Ref.~\cite{Hernandez:2022ivz} used for the comparison with \texttt{amiqs}.}
 \label{fig:validity_heatmap}
\end{figure}

The module \texttt{etabARS\_3RHN.py} issues a warning when (i) the largest mass splitting $\Delta M$ amongst the non-decoupled heavy neutrinos exceeds the mass of the lightest mode, as all rates are evaluated at common mass $M_1$ and (ii) when $T_{\text{osc}}>T_{\text{2-flav}}$. A heatmap representing the regime of validity and the points at which the benchmarks presented in the paper lie is presented in Fig.~\ref{fig:validity_heatmap}. The benchmark III of Ref.~\cite{Abada:2018oly}, for which the heavy neutrinos have masses between $0.8$ and $4.3\,\text{GeV}$, lies in the region where the first warning is issued. In this case the heavy neutrinos remain relativistic during the whole evolution, $\Mav\ll T_{\sph}$, so that the rates are essentially independent of the heavy-neutrino masses and the result agrees with the independent code \texttt{amiqs} (see Sec.~\ref{sec:checks}). The warnings should therefore be understood as a prompt to check the result for the parameter point at hand rather than as a verdict.

\subsection{Speeding-up the evaluation of the density matrix equations}\label{app:numerical}
The DMEs are stiff and their numerical resolution tends to be computationally demanding and time-consuming. This effectively limits the ability to perform valid scans of the leptogenesis parameter space over reasonable times, unless specific treatments and approximations are utilised. In this section, we outline the specific procedure we have implemented to speed-up the numerical calculation and make it more accurate, improving on the \virg{stitching} option introduced in the second version of the code \cite{Granelli:2023vcm}. We start by noting that the matrices $\delta r_N$ and $\delta r_{\bar{N}}$ are hermitian. As such, we can decompose them according to $\delta r_N = \sum_{j=1}^9 c_N^{(j)}\lambda_j$ and $\delta r_{\bar{N}} = \sum_{j=1}^9 c_{\bar{N}}^{(j)}\lambda_j$, with the coefficients $c_N^{(j)}$ and $c_{\bar{N}}^{(j)}$ real and $\lambda_j$ the following hermitian matrices:
\begin{eqnarray}
\lambda_1 &=& \begin{pmatrix} 1 & 0 & 0 \\ 0 & 0 & 0 \\ 0 & 0 & 0 \end{pmatrix},
\quad~~
\lambda_2 = \begin{pmatrix} 0 & 0 & 0 \\ 0 & 1 & 0 \\ 0 & 0 & 0 \end{pmatrix},\quad~~
\lambda_3 = \begin{pmatrix} 0 & 0 & 0 \\ 0 & 0 & 0 \\ 0 & 0 & 1 \end{pmatrix}\\
\lambda_4 &=& \begin{pmatrix} 0 & 1 & 0 \\ 1 & 0 & 0 \\ 0 & 0 & 0 \end{pmatrix},
\quad~~\lambda_5 = \begin{pmatrix} 0 & -i & 0 \\ i & 0 & 0 \\ 0 & 0 & 0 \end{pmatrix},
\quad
\lambda_6 = \begin{pmatrix} 0 & 0 & 1 \\ 0 & 0 & 0 \\ 1 & 0 & 0 \end{pmatrix}\\
\quad
\lambda_7 &=& \begin{pmatrix} 0 & 0 & -i \\ 0 & 0 & 0 \\ i & 0 & 0 \end{pmatrix},
\quad
\lambda_8 = \begin{pmatrix} 0 & 0 & 0 \\ 0 & 0 & 1 \\ 0 & 1 & 0 \end{pmatrix},
\quad~~
\lambda_9 = \begin{pmatrix} 0 & 0 & 0 \\ 0 & 0 & -i \\ 0 & i & 0 \end{pmatrix}.
\end{eqnarray}
differing from the standard Gell-Mann matrices only by a non-standard choice of $\lambda_1$, $\lambda_2$ and $\lambda_3$. This decomposition is analogous to the one performed in terms of Pauli matrices in \cite{Klaric:2021cpi} for the case with two quasi-degenerate in mass heavy Majorana neutrinos. The coefficients $c_N^{(j)}$ are real and can be obtained from the matrices $\delta r_N$ according to
\begin{eqnarray}
c_N^{(1)} &=& 
(\delta r_N)_{11}, \quad c_N^{(2)} = (\delta r_N)_{22},\quad c_N^{(3)} = (\delta r_N)_{33},\label{eqs:c0c1c2}  \\
c_N^{(4)} &=& \frac{1}{2} \left[ (\delta r_N)_{12} + (\delta r_N)_{21} \right],\quad
c_N^{(5)} = \frac{i}{2} \left[ (\delta r_N)_{12} - (\delta r_N)_{21} \right], \quad \label{eqs:c3c4}\\
c_N^{(6)} &=& \frac{1}{2} \left[ (\delta r_N)_{13} + (\delta r_N)_{31} \right], \quad \label{eqs:c5c6}
c_N^{(7)} = \frac{i}{2} \left[ (\delta r_N)_{13} - (\delta r_N)_{31} \right], \\
c_N^{(8)} &=& \frac{1}{2} \left[ (\delta r_N)_{23} + (\delta r_N)_{32} \right], \quad \label{eqs:c7c8}
c_N^{(9)} = \frac{i}{2} \left[ (\delta r_N)_{23} - (\delta r_N)_{32} \right], 
\end{eqnarray}
and similarly for the coefficients $c_{\bar{N}}^{(j)}$ in terms of the entries of $\delta r_{\bar{N}}$.
By defining the $21$-dimensional vector $q \equiv (c_N^{(1)}, c_N^{(2)},...,c_N^{(9)}, c_{\bar{N}}^{(1)}, c_{\bar{N}}^{(2)}, ..., c_{\bar{N}}^{(9)}, \mu_{\Delta_e}, \mu_{\Delta_\mu}, \mu_{\Delta_\tau})$, the DMEs can be written more compactly as 
\begin{equation}\label{eq:DMEs_linear}
\frac{d q_\kappa}{dx} = \sum_{\rho=1}^{21} A_{\kappa \rho} q_\rho + \frac{1}{2} \sum_{\rho = 1}^{21}\sum_{\sigma=1}^{21}B_{\kappa \rho \sigma} q_\rho q_{\sigma} + C_\kappa\,,
\end{equation}
$\kappa = 1,...,21$, with all the terms being real so that the total number of degrees of freedom contained in the density matrices of the heavy Majorana neutrinos is halved. The coefficients $A_{\kappa \rho}$, $B_{\kappa \rho \sigma}$ and $C_\kappa$ can be obtained from the DMEs in Eq.~\eqref{eq:DME_N} by using the decomposition of $\delta r_N$ (and $\delta r_{\bar{N}}$) and applying the relations in Eqs.~(\ref{eqs:c0c1c2}-\ref{eqs:c7c8}). 

\begin{figure}[t!]
\centering
\includegraphics[width=\textwidth]{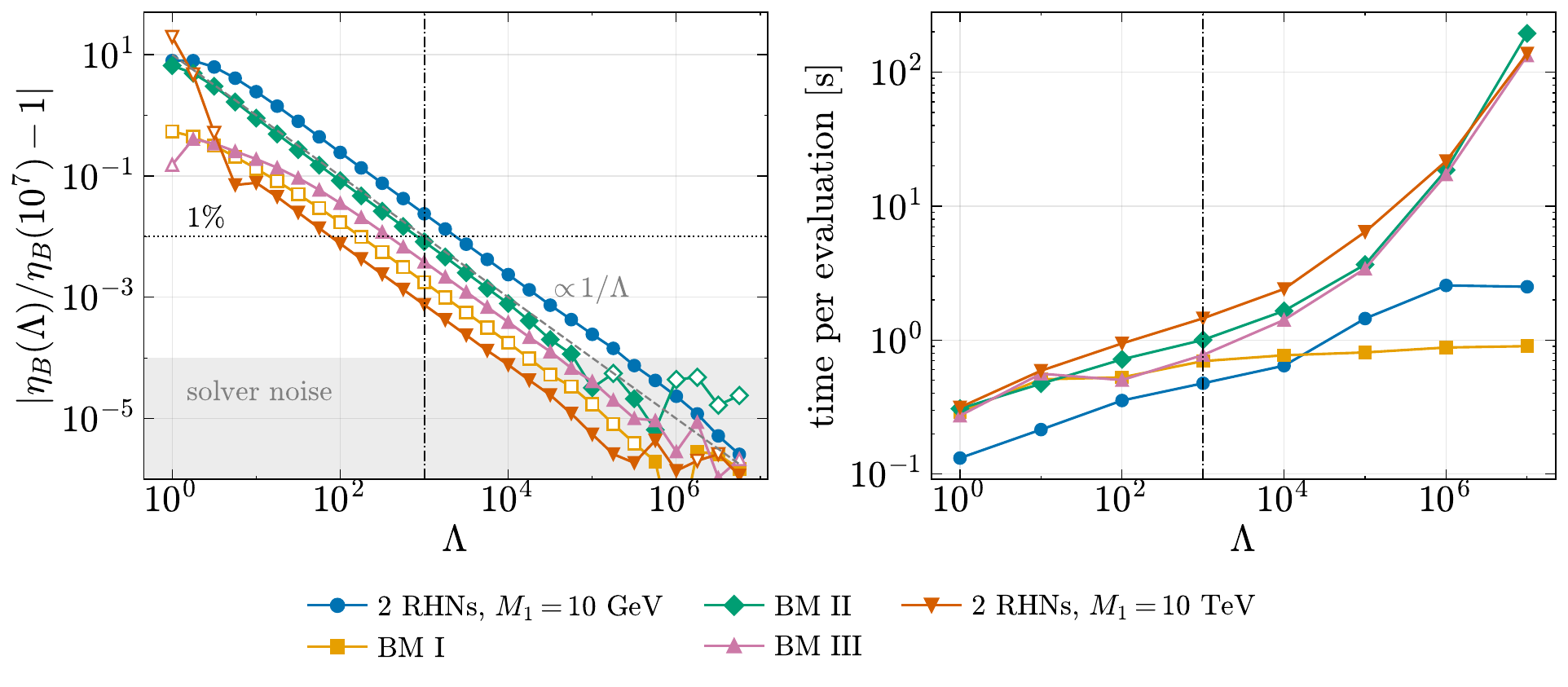}
\caption{Left: relative deviation of $\eta_B(\Lambda)$ from $\eta_B(\Lambda=10^7)$ for the benchmark of Figs.~\ref{fig:Check_ULS}--\ref{fig:BMCs}; filled (open) markers denote positive (negative) deviations. The grey dashed line shows the $1/\Lambda$ scaling, the dotted line marks $1\%$ and the shaded band the numerical noise of the solver, estimated by repeating the runs with a different number of threads. Right: wall-clock time per evaluation of \texttt{etabARS$\_$3RHN.py} on a single core. The absolute times are indicative and depend on the machine load as well as CPU performance. The vertical dash-dotted line marks the default $\Lambda=10^3$.}
\label{fig:Lambda}
\end{figure}

Fast modes in the linear terms can be averaged out via the following prescription \cite{Klaric:2021cpi}:
\begin{equation}
 A \to A \left[\mathds{1}-A x/\Lambda\right]^{-1},\quad 
 C \to \left[\mathds{1}-A x/\Lambda\right]^{-1} C\,,
\end{equation}
with $A$ being the matrix with entries $A_{\kappa\rho}$ and $\Lambda>0$ an arbitrary regulator that determines the scale at which the fast modes in $A$ are forced to enter in the quasi-static regime. Its value can be set by the user through the instance parameter \texttt{Lambda}. For large enough values of $\Lambda$, the regulator is ineffective and the prescription reduces to the unmodified linear terms, so that the fast modes are essentially retained. Conversely, for smaller values of $\Lambda$, the fast modes are effectively averaged out and their effects to the dynamics suppressed.We warn the user that  \emph{the choice of 
$\Lambda$ should not be made blindly}: in practice, one should verify that for representative points of the parameter space, the choice of $\Lambda$ should be sufficiently large that increasing it by an order of magnitude does not substantially modify the asymmetries.
This provides a check that the prescription is under control and does not artificially distort the dynamics.
For the specific examples that we are going to discuss, we found $\Lambda = 10^3$ as an optimal trade-off between computational speed and numerical accuracy. This is shown in Fig.~\ref{fig:Lambda}, where we display, for the five benchmarks of Figs.~\ref{fig:Check_ULS}--\ref{fig:BMCs}, the relative deviation of $\eta_B$ from its value at $\Lambda=10^7$ (left panel) and the corresponding execution time (right panel). In all cases the deviation decreases as $1/\Lambda$, as expected from $A\left[\mathds{1}-Ax/\Lambda\right]^{-1}= A + A^2x/\Lambda+\mathcal{O}(\Lambda^{-2})$ once the regulator no longer affects the relevant modes. For $\Lambda=10^3$ the deviations range between $0.08\%$ and $2.4\%$, and they fall below $1\%$ for all benchmarks when $\Lambda\gtrsim3\times10^{3}$. The execution time grows mildly up to $\Lambda\sim10^{4}$ (by up to a factor of about three from $\Lambda=10^3$ to $10^4$, depending on the benchmark and on the machine load) and steeply for $\Lambda\gtrsim10^{5}$, when the fast modes are no longer averaged and must be resolved by the solver. Users requiring sub-percent precision can therefore increase $\Lambda$ to $10^{4}$ at a modest cost. Relative deviations below $\sim10^{-4}$ are not resolved.

To further speed-up the evaluation, we neglect by default the non-linear terms, $i.e.$~those proportional to $\langle\gamma_{\LNC}^{(2)}\rangle$ and $\langle S_{\LNV}^{(2)}\rangle$. The contributions from non-linear terms are in general suppressed by the smallness of the entries of $\delta r_N$ and $\mu_{\Delta_{\alpha}}$ 
(this is true for the benchmarks that we have considered in the next subsection).
Moreover, the linear form of the equations, $i.e.$~Eq.~\eqref{eq:DMEs_linear} with $B_{\kappa \rho \sigma} = 0$, allows us to input an analytical Jacobian matrix in the \texttt{solve$\_$ivp} function (the matrix $A\left[\mathds{1}-A x/\Lambda\right]^{-1}$), thus significantly improving both the computational efficiency and numerical stability of the differential equation solver. {Also,  to our knowledge, an interpolation of the non-linear rates to the non-relativistic regime is currently missing in the literature.} However, we highlight that we have included all the infrastructure in \texttt{ULYSSES} to account for the non-linear terms, 
{facilitating further developments along these lines, but we leave their full implementation to future releases.}

\subsection{Internal compatibility and cross-checks with benchmark points}
\label{sec:checks}
\begin{figure}[t!]
\centering
 \includegraphics[width=0.48\textwidth]{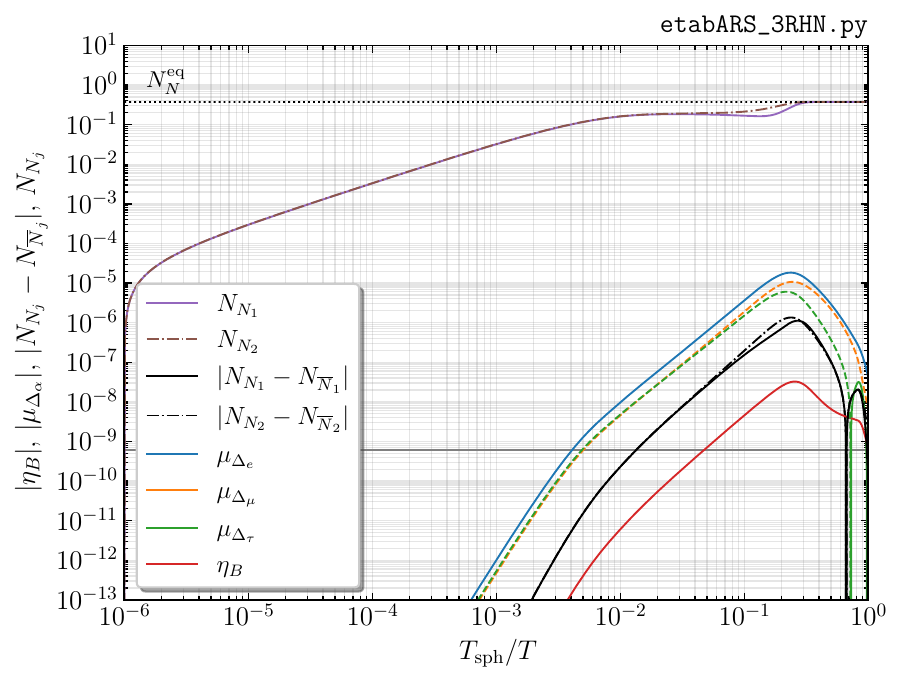}
  \includegraphics[width=0.48\textwidth]{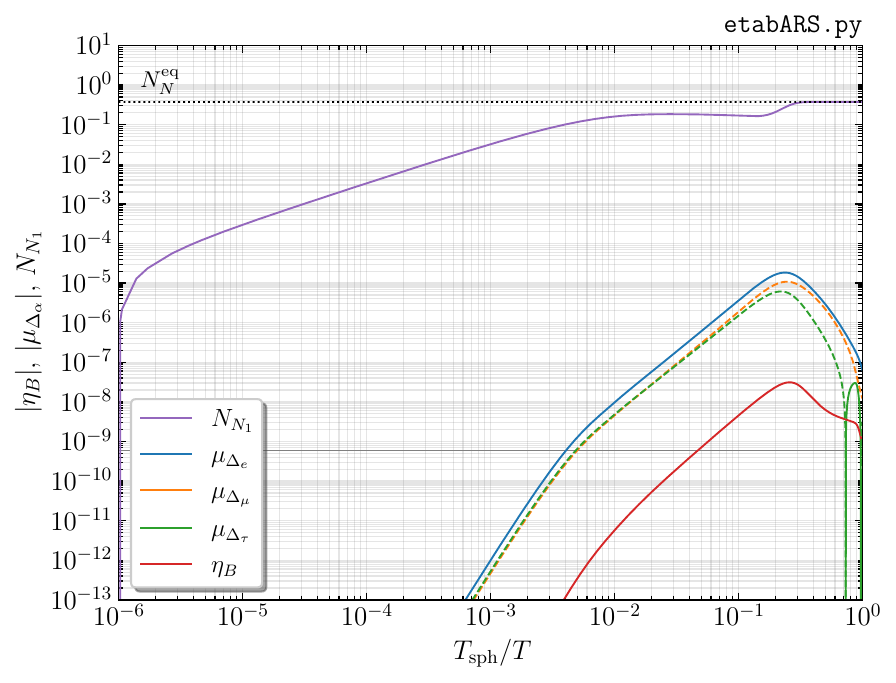}
 \caption{Compatibility check of the new low-scale leptogenesis module \texttt{etabARS$\_$3RHN.py} (left panel) and \texttt{etabARS.py} released with the previous version of the code (right panel). Shown are the lightest heavy neutrino abundance $N_{N_1}(z) \equiv (3/8)(\delta r_{N})_{11} + N_N^\text{eq}(z)$ (purple), the chemical potentials $\mu_{\Delta_e}$ (blue), $\mu_{\Delta_\mu}$ (orange), $\mu_{\Delta_\tau}$ (green) and the final BAU $\eta_B$ (red) against $x = T_{\text{sph}}/T$. Solid (dashed) curves represent positive (negative) contributions. Also shown in black are the asymmetries in the heavy neutrinos, that is $|N_{N_1}-N_{\bar{N}_1}|$ (solid) and $|N_{N_2}-N_{\bar{N}_2}|$ (dot-dashed). The thicker grid line marks the observed BAU, while the black horizontal dotted line corresponds to $N_N^\text{eq}(z) = (3/16)z^2K_2(z)$. The parameters chosen for this benchmark are $M_1 = 10\,\text{GeV}$, $(M_2-M_1)/M_1 = 10^{-9}$, $\delta = 212^\circ$,  $\alpha_{21} = 180^\circ$,  $\alpha_{31} = 0$, $x= 0$ and $y = -2.975$, assuming normal ordering with $\theta_{23} = 43.3^\circ$, $\theta_{12} = 33.68^\circ$ and $\theta_{13} = 8.56^\circ$, $N_3$ is decoupled and we have included loop corrections according to Eq.~\eqref{eq:Mloop}. The runcard for this benchmark is provided in \texttt{examples/ARS\_2RHN\_vs\_3RHN.dat}. On a single core, this evaluation of \texttt{etabARS\_3RHN.py} is on average 25x faster than the \texttt{etabARS.py} module.}
 \label{fig:Check_ULS}
\end{figure}

To validate the results of \texttt{ULYSSES} we have performed a number of cross-checks.
Firstly, we checked the compatibility of the code with itself by solving the set of DMEs in the scenario with two RHNs using the module \texttt{etabARS.py} released with \texttt{ULYSSESv2} and that of this new version, \texttt{etabARS$\_$3RHN.py}. We always find good agreement as long as $\Mav \lesssim 100\,\text{GeV}$, which is the operative limit of the \texttt{etabARS.py} \cite{Granelli:2023vcm}. We show such validation for a specific benchmark point in which $N_3$ is decoupled in Fig.~\ref{fig:Check_ULS}, by depicting the evolution of the final asymmetry, the chemical potentials $\mu_{\Delta_\alpha}$, $\alpha = e,\,\mu,\,\tau$ and the lightest heavy neutrino abundance against $x=T_{\rm sph}/T$. The parameters of the considered benchmark point are described in the caption of the figure. The final asymmetry obtained with the \texttt{etabARS$\_$3RHN.py} module reads $\eta_B \simeq 6.11\times 10^{-10}$, while with \texttt{etabARS.py} we get $\eta_B \simeq 8.35\times 10^{-10}$. The $\sim 1.37$ factor of difference in the final result originates from the updated versions of the rates and the Hamiltonian included in \texttt{etabARS$\_$3RHN.py}. The approximation of neglecting terms non-linear in $\delta r_N$ and $\mu$ in Eqs.~(\ref{eq:DME_N}–\ref{eq:DME_mu}), which is not adopted in \texttt{etabARS.py}, does not affect the result in this benchmark at an appreciable level, while it allows for a faster evaluation. Owing to the improvements implemented in the the 3 RHN scenario, the module \texttt{etabARS$\_$3RHN.py} is significantly more efficient than the previous one \texttt{etabARS.py} with 2 RHNs. We therefore recommend using the former as the default choice also for the scenario with 2 RHNs (provided $N_3$ is decoupled, see Sec.~\ref{ssec:RHNdecouple}), although the two modules remain fully compatible.

\begin{figure}[t!]
    \centering
    \includegraphics[width = .49 \textwidth]{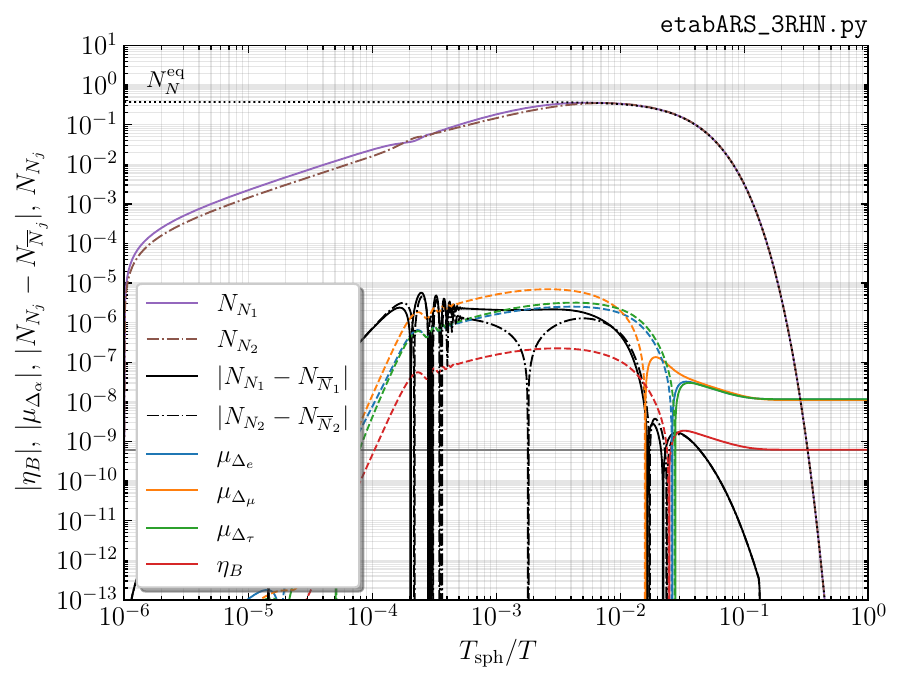}
    \includegraphics[width = .49 \textwidth]{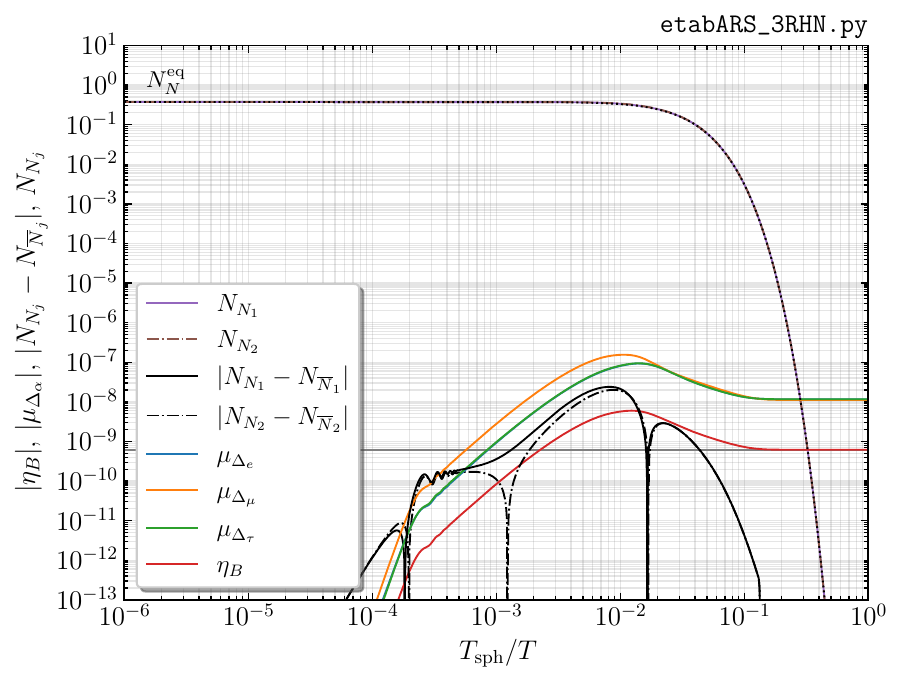}
    \includegraphics[width = .49 \textwidth]{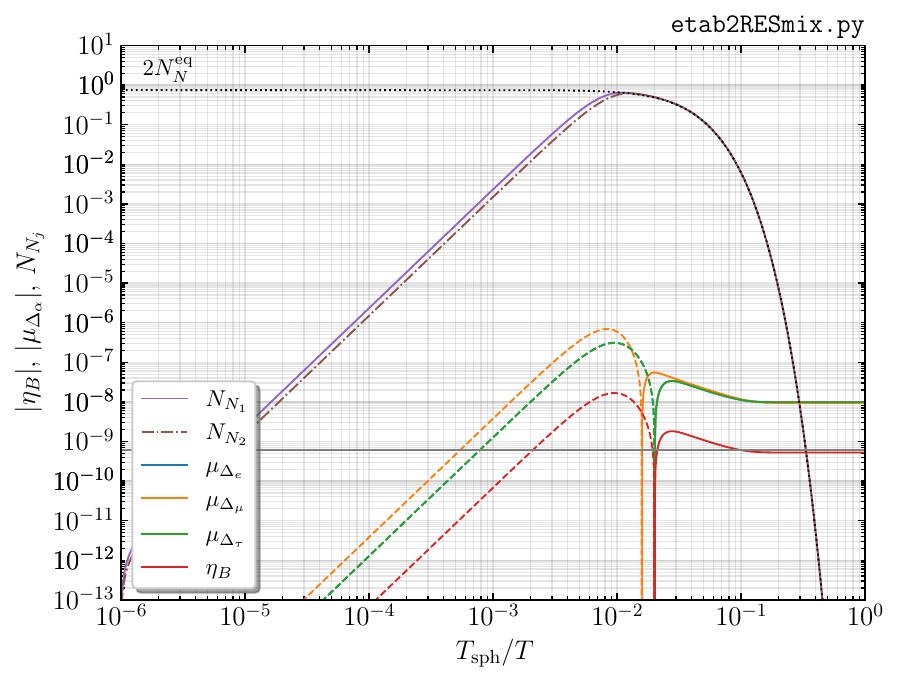}
    \includegraphics[width = .49 \textwidth]{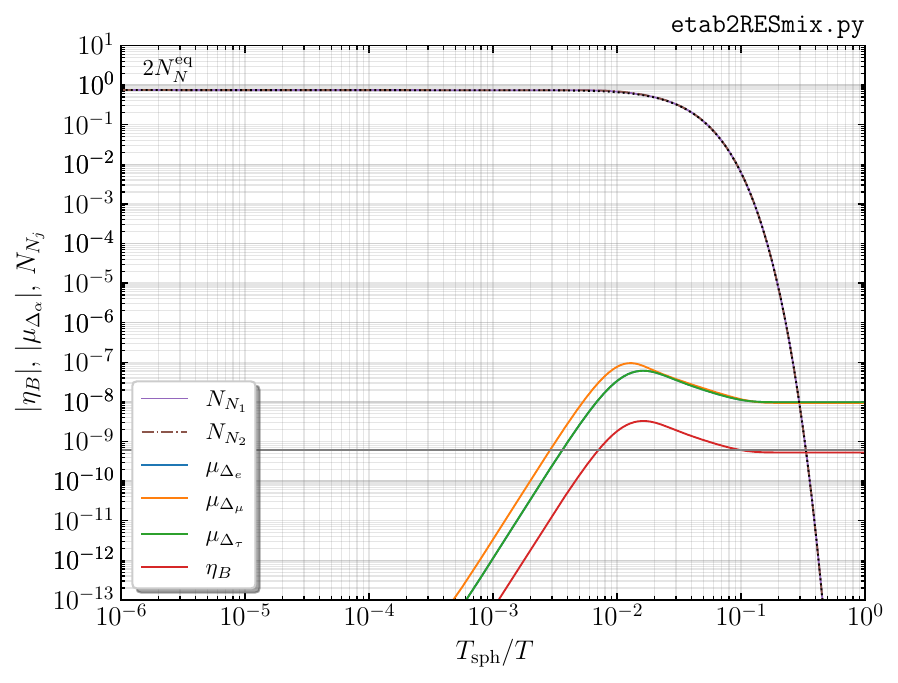}
    \caption{ Compatibility check of the new low-scale leptogenesis module \texttt{etabARS$\_$3RHN.py} (top panels) against \texttt{etab2RESmix.py} for resonant leptogenesis released with the first version of the code (bottom panels). The parameters chosen for this benchmark are $M_1 = 10\,\text{TeV}$,  $(M_2-M_1)/M_1 = 6.9\times 10^{-8}$, $Y_{e1}^{\rm mag}=Y_{\tau2}^{\rm mag}=10^{-6}$, $Y_{e2}^{\rm mag}=Y_{\mu2}^{\rm mag}=Y_{\tau1}^{\rm mag}=2\times 10^{-6}$, $Y_{\mu1}^{\rm mag}=3\times 10^{-6}$, $Y_{e1}^{\rm phs}=1$, $Y_{e2}^{\rm phs}=Y_{\mu 1}^{\rm phs}=2$, $Y_{\mu 2}^{\rm phs}=Y_{\tau 1}^{\rm phs}=3$,  $Y_{\tau 2}^{\rm phs}=4$ and $Y_{e3}^{\rm mag}=Y_{\mu3}^{\rm mag}=Y_{\tau3}^{\rm mag}=Y_{e3}^{\rm phs}=Y_{\mu3}^{\rm phs}=Y_{\tau3}^{\rm phs}=0$; $N_3$ is decoupled with $M_3=10^{10}\,\text{GeV}$ and we have included loop corrections. The runcard for this benchmark is provided in
\texttt{examples/ARS\_vs\_RES.dat}. The left (right) panels are for vanishing (thermal) initial abundance for the heavy neutrinos. For this benchmarks, the evaluation of the DMEs using \texttt{etabARS\_3RHN.py} is on average 3x slower than the Boltzmann equation evolver of \texttt{etab2RESmix.py} module. However, it should be noted that in intermediate regimes the DME evolution is more reliable than the BE evolution and hence one should make choices based on the region of interest}.
    \label{fig:ARS_vs_RES}
\end{figure}

Secondly, we have verified the compatibility of \texttt{etabARS\_3RHN.py} with the resonant leptogenesis module \texttt{etab2RESmix.py}
that shipped with previous versions of \texttt{ULYSSES}.
The aforementioned resonant leptogenesis module relies on semi-classical BEs from~\cite{DeSimone:2007edo},
combined with \virg{mixing} contribution to the CP-asymmetries from~\cite{BhupalDev:2014oar}.
\footnote{The previous versions of \texttt{ULYSSES} also includes a module treating \virg{mixing and oscillation} asymmetries as separate phenomena from~\cite{BhupalDev:2014oar}.
%Since there is no consensus whether these should be treated as separate sources of CP violation (see e.g.), we 
There is ongoing debate whether including these terms separately constitutes as double-counting (see e.g.~\cite{Dev:2017wwc,Jukkala:2021sku} and references therein). In the comparison we therefore only include one term, and focus on the regime of large mass splittings where the discrepancies between the CP-asymmetries available in the literature do not play a significant role.
}

\begin{figure}[t!]
 \centering
 \includegraphics[width=0.48\textwidth]{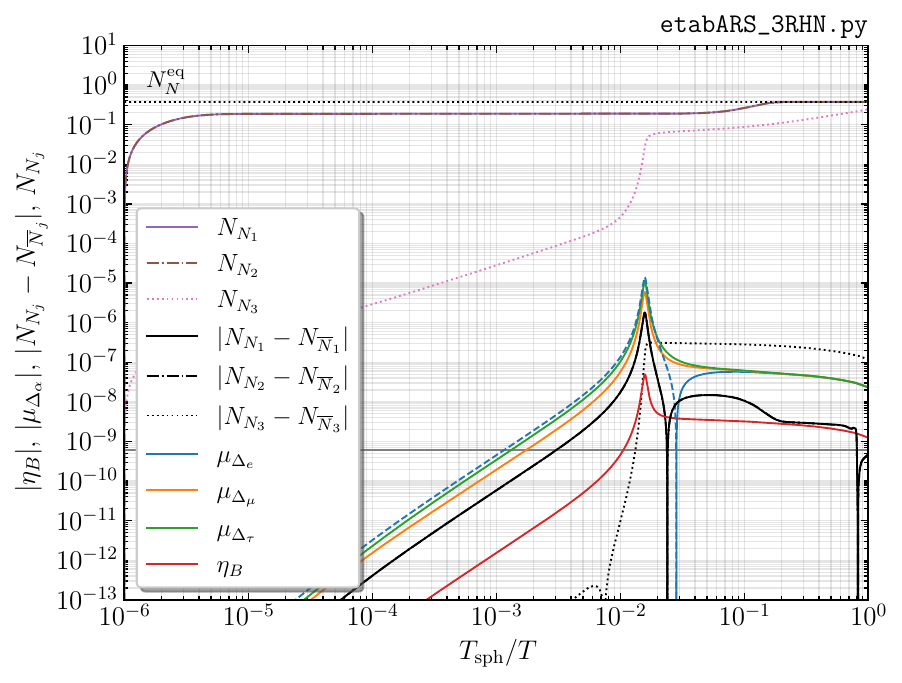}\\
  \includegraphics[width=0.48\textwidth]{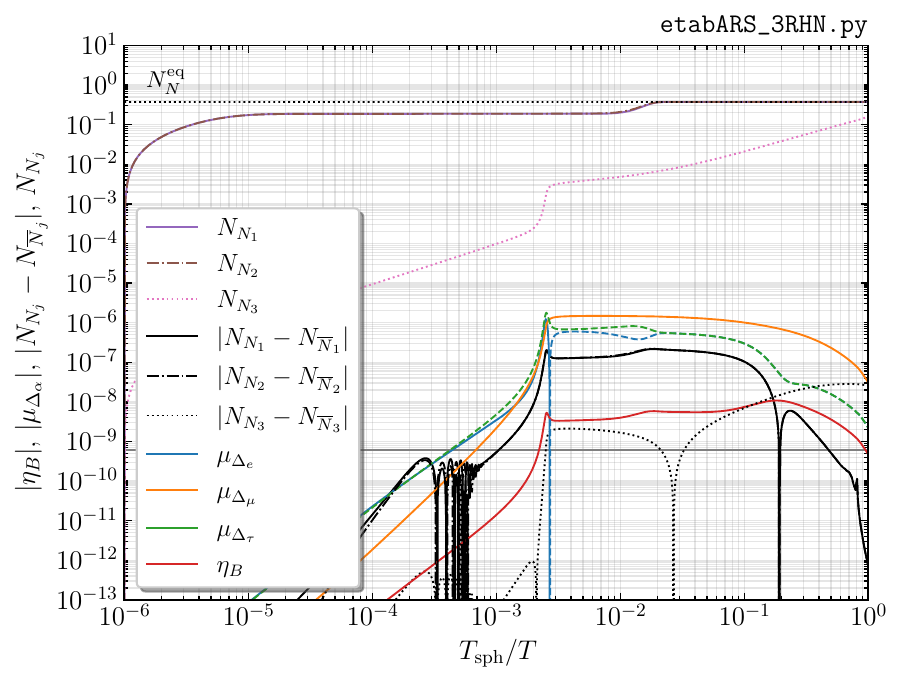}
  \includegraphics[width=0.48\textwidth]{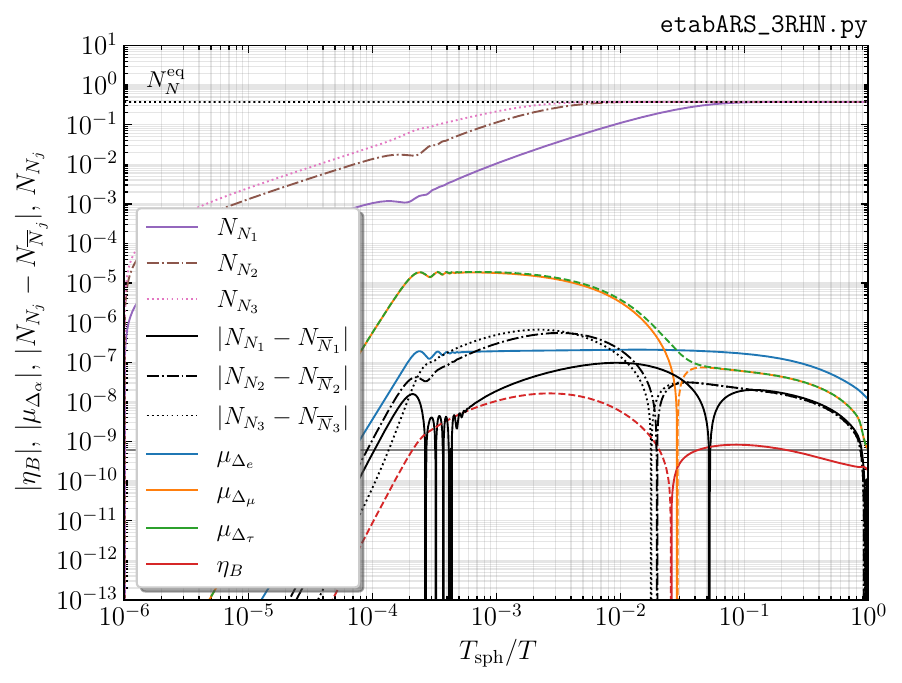}
 \caption{The results of \texttt{ULYSSES} for benchmark point I (top panel), II (bottom-left panel) and III (bottom-right panel) of \cite{Abada:2018oly}. Shown are the heavy neutrino abundances $N_{N_{j}}(z) \equiv (3/8)(\delta r_{N})_{jj}+ N_N^\text{eq}(z)$, $j = 1,\,2,\,3$ (purple, brown and pink), the chemical potentials $\mu_{\Delta_e}$ (blue), $\mu_{\Delta_\mu}$ (orange), $\mu_{\Delta_\tau}$ (green) and the final BAU $\eta_B$ (red) against $x = T_{\text{sph}}/T$. Solid (dashed) curves represent positive (negative) contributions. Also shown in black are the asymmetries in the heavy neutrinos, that is $|N_{N_1}-N_{\bar{N}_1}|$ (solid), $|N_{N_2}-N_{\bar{N}_2}|$ (dot-dashed) and $|N_{N_3}-N_{\bar{N}_3}|$ (dotted). The thicker grid line marks the observed BAU, while the black horizontal dotted line corresponds to $N_N^\text{eq}(z) = (3/16)z^2K_2(z)$. Loop corrections are included.}
 \label{fig:BMCs}
\end{figure}

\begin{table}[h!]
 \centering
 \begin{tabular}{l r|l r|l r}
 \hline
  \rowcolor{gray!20}
  \multicolumn{6}{c}{\textbf{Benchmark I, Inverted Ordering}} \\
  \hline
  {\ttm Y11\_mag} & {\ttm 8.149e-05} & {\ttm Y11\_phs} & {\ttm -1.819} & & \\
  {\ttm Y12\_mag} & {\ttm 8.149e-05} & {\ttm Y12\_phs} & {\ttm -0.248} & & \\
  {\ttm Y13\_mag} & {\ttm 9.669e-08} & {\ttm Y13\_phs} & {\ttm -1.384} & & \\
  {\ttm Y21\_mag} & {\ttm 2.997e-05} & {\ttm Y21\_phs} & {\ttm -0.449} & {\ttm m} & {\ttm -15.159} \\
  {\ttm Y22\_mag} & {\ttm 2.997e-05} & {\ttm Y22\_phs} & {\ttm 1.122} & {\ttm M1} & {\ttm 0.4313637639} \\
  {\ttm Y23\_mag} & {\ttm 5.385e-08} & {\ttm Y23\_phs} & {\ttm -0.547} & {\ttm M2} & {\ttm 0.4313637644} \\
  {\ttm Y31\_mag} & {\ttm 2.927e-05} & {\ttm Y31\_phs} & {\ttm -3.005} & {\ttm M3} & {\ttm 0.4399639359} \\
  {\ttm Y32\_mag} & {\ttm 2.927e-05} & {\ttm Y32\_phs} & {\ttm -1.434} & & \\
  {\ttm Y33\_mag} & {\ttm 4.123e-08} & {\ttm Y33\_phs} & {\ttm 2.897} & & \\
  \hline
  \rowcolor{gray!20}
  \multicolumn{6}{c}{\textbf{Benchmark II, Normal Ordering}} \\
  \hline
  {\ttm Y11\_mag} & {\ttm 2.828e-05} & {\ttm Y11\_phs} & {\ttm -0.142} & & \\
  {\ttm Y12\_mag} & {\ttm 2.828e-05} & {\ttm Y12\_phs} & {\ttm 1.429} & & \\
  {\ttm Y13\_mag} & {\ttm 1.360e-08} & {\ttm Y13\_phs} & {\ttm -1.272} & & \\
  {\ttm Y21\_mag} & {\ttm 3.105e-07} & {\ttm Y21\_phs} & {\ttm -2.881} & {\ttm m} & {\ttm -17.879} \\
  {\ttm Y22\_mag} & {\ttm 3.105e-07} & {\ttm Y22\_phs} & {\ttm -1.310} & {\ttm M1} & {\ttm 0.7159765903} \\
  {\ttm Y23\_mag} & {\ttm 6.907e-08} & {\ttm Y23\_phs} & {\ttm 2.171} & {\ttm M2} & {\ttm 0.7160300954} \\
  {\ttm Y31\_mag} & {\ttm 4.916e-05} & {\ttm Y31\_phs} & {\ttm 3.060} & {\ttm M3} & {\ttm 0.7494270991} \\
  {\ttm Y32\_mag} & {\ttm 4.916e-05} & {\ttm Y32\_phs} & {\ttm -1.652} & & \\
  {\ttm Y33\_mag} & {\ttm 6.046e-08} & {\ttm Y33\_phs} & {\ttm 2.168} & & \\
  \hline
  \rowcolor{gray!20}
  \multicolumn{6}{c}{\textbf{Benchmark III, Normal Ordering}} \\
  \hline
  {\ttm Y11\_mag} & {\ttm 5.522e-08} & {\ttm Y11\_phs} & {\ttm -2.189} & & \\
  {\ttm Y12\_mag} & {\ttm 2.025e-07} & {\ttm Y12\_phs} & {\ttm -2.145} & & \\
  {\ttm Y13\_mag} & {\ttm 2.884e-07} & {\ttm Y13\_phs} & {\ttm 2.554} & & \\
  {\ttm Y21\_mag} & {\ttm 6.140e-07} & {\ttm Y21\_phs} & {\ttm -1.290} & {\ttm m} & {\ttm -11.996} \\
  {\ttm Y22\_mag} & {\ttm 2.184e-06} & {\ttm Y22\_phs} & {\ttm -1.292} & {\ttm M1} & {\ttm -0.0786517297} \\
  {\ttm Y23\_mag} & {\ttm 3.008e-06} & {\ttm Y23\_phs} & {\ttm -2.872} & {\ttm M2} & {\ttm 0.4572231462} \\
  {\ttm Y31\_mag} & {\ttm 5.325e-07} & {\ttm Y31\_phs} & {\ttm -0.598} & {\ttm M3} & {\ttm 0.6363875858} \\
  {\ttm Y32\_mag} & {\ttm 1.860e-06} & {\ttm Y32\_phs} & {\ttm -0.633} & & \\
  {\ttm Y33\_mag} & {\ttm 2.581e-06} & {\ttm Y33\_phs} & {\ttm -2.191} & & \\
  \bottomrule
 \end{tabular}
 \caption{Parameter cards for the benchmark points I (top), II (middle) and III (bottom) used to get Fig.~\ref{fig:BMCs} (loop corrections included), respectively also in the runcards \texttt{ARS\_BMCI.dat}, \texttt{ARS\_BMCII.dat}, and \texttt{ARS\_BMCIII.dat} in the \texttt{examples} folder.}
 \label{tab:BMCs_params}
\end{table}

The DMEs for low-scale leptogenesis can reproduce the BEs employed in the resonant scenario in the non-relativistic regime and for sufficiently large mass splittings (see e.g. \cite{Klaric:2021cpi}).
Here, we provide a benchmark point at $M_1 = 10\,\text{TeV}$ to explicitly show this matching.
The parameters are fixed as in runcard~\texttt{examples/ARS\_vs\_RES.dat}, adopting the \texttt{manual} parametrisation for the Yukawa couplings.
We have set the Yukawa couplings associated with $N_3$ exactly to zero, ensuring it is properly decoupled.
The evolution plot for this benchmark point as obtained with \texttt{etabARS\_3RHN.py} (\texttt{etab2RESmix.py}) is shown in the top (bottom) panels of Fig.~\ref{fig:ARS_vs_RES}, on the left (right) for vanishing (thermal) initial abundance.
The evolution at early times is clearly much different because the DMEs for low-scale leptogenesis capture a number  of effects (helicity, finite duration of oscillations) that are typically missed by the BE treatment.
Nevertheless, the evolution in the two cases starts getting similar from $T \gtrsim 100 \,T_{\sph}$.
The final asymmetry obtained with the \texttt{etabARS\_3RHN.py} module reads $\eta_B\simeq  6.12 \times 10^{-10}$ while with \texttt{etab2RESmix.py} we get $\eta_B\simeq  5.15 \times 10^{-10}$, for either vanishing or thermal initial abundance for the heavy neutrinos. Remarkably, the modules agree up to an overall factor of $\sim 1.2$.
Such an overall agreement is expected for the chosen benchmark, which exhibits a strong washout in all flavours, thereby erasing any asymmetries generated prior to the non-relativistic RHN decays.
In general, the DMEs in the \texttt{etabARS\_3RHN.py} will provide more accurate results, as they include a number of relativistic and coherence effects that are absent from the BEs in \texttt{etab2RESmix.py}.

Additionally, we have validated \texttt{ULYSSES} against results in literature. We solved the equations with three RHNs for the specific benchmarks provided in \cite{Abada:2018oly}. We give the parameters used for this validation in Table~\ref{tab:BMCs_params} (runcards provided in the \texttt{example} folder). We show in Fig.~\ref{fig:BMCs} the behaviour of the final asymmetry $\eta_B$, the chemical potentials $\mu_{\Delta_e}$, $\mu_{\Delta_\mu}$ and $\mu_{\Delta_\tau}$ and the heavy neutrino abundances against $x=T_{\sph}/T$ for these benchmarks, as obtained with \texttt{etabARS$\_$3RHN.py}. We find an overall good agreement by comparing with the results of \cite{Abada:2018oly}, albeit with $\mathcal{O}(1)$ differences. We have checked that these differences arise due to the more sophisticated rates and Hamiltonian contributions implemented in this version of \texttt{ULYSSES} and that the agreement is improved when such corrections are not included. We emphasize that, unlike \texttt{etabARS$\_$3RHN.py}, the \texttt{etabARS.py} module and the machinery of \cite{Abada:2018oly} both include non-linear terms but do not implement LNV corrections to the Hamiltonian. Moreover, all rates and averaged quantities are evaluated without (fully) considering non-relativistic corrections. Taken together, these differences lead to minor deviations in the results, while the overall agreement remains satisfactory. We also confirm the agreement of \texttt{ULYSSES} with the \texttt{amiqs} code available \href{https://github.com/stefanmarinus/amiqs}{here}. Using the three RHN code implemented in this version after decoupling $N_3$, we checked the benchmarks in the two RHN scenario provided in \cite{Hernandez:2022ivz} in Table 7 (see also Fig.~7 there). We also verified the compatibility with \texttt{amiqs} for the same benchmarks of \cite{Abada:2018oly} in the 3 RHN scenario discussed earlier. After noting that in \cite{Hernandez:2022ivz} -- also in \cite{Sandner:2023tcg} and in the \texttt{amiqs} code -- a different convention for the chemical potential $\mu_{\Delta_\alpha}$ is used, leading to $\mu_{\Delta_\alpha}^\texttt{amiqs}= 2\mu_{\Delta_\alpha}^\texttt{ULYSSES}$, where $\mu_{\Delta_\alpha}^\texttt{amiqs}$ is the definition in \cite{Hernandez:2022ivz, Sandner:2023tcg} and $\mu_{\Delta_\alpha}^\texttt{ULYSSES}$ ours, we find a perfect matching between \texttt{ULYSSES} and their full numerical results.\footnote{We match the results of \cite{Hernandez:2022ivz} and \texttt{amiqs} despite we have neglected non-linear terms proportional to $\langle\gamma_{\LNC}^{(2)}\rangle$ and $\langle S_{\LNV}^{(2)}\rangle$ in \texttt{etabARS$\_$3RHN.py}, meaning that these are indeed negligible for the benchmarks considered.} We provide the runcards for the benchmarks of \cite{Hernandez:2022ivz} in \texttt{examples/ARS\_2RHN\_BMC\#.dat}, with \texttt{\#} = \texttt{I}, \texttt{II}, \texttt{III}, \texttt{IV}. Finally, we have verified the compatibility with the private code (not publicly available) used in Refs.~\cite{Klaric:2021cpi,Drewes:2021nqr}.

\section{$\Delta L = 1$ scatterings in full phase-space vanilla leptogenesis}\label{sec:scat}

\begin{figure}
\centering
\begin{minipage}{0.32\textwidth}
\centering
\begin{tikzpicture}[baseline=(current bounding box.center),]
\begin{feynman}
\vertex (V1);
\vertex (Ni) [above left=0.6cm and 0.9cm of V1] {\(N_1\)};
\vertex (l) [below left=0.6cm and 0.9cm of V1] {\(\ell, \bar{\ell}\)};
\vertex (V2) [right=1.2cm of V1];
\vertex (t) [above right=0.6cm and 0.9cm of V2] {\(\bar{t}, t\)};
\vertex (q) [below right=0.6cm and 0.9cm of V2] {\(q_3,\bar{q_3}\)};
\diagram*{
  (Ni) -- (V1),
  (l) -- (V1),
  (V1) -- [scalar, edge label=\(\Phi\)] (V2),
  (V2) -- (t),
  (V2) -- (q)
};
\end{feynman}
\end{tikzpicture}
\\[0.3em]
(a) \(s\)-channel
\end{minipage}
\begin{minipage}{0.34\textwidth}
\centering
\begin{tikzpicture}[baseline=(current bounding box.center)]
\begin{feynman}
\vertex (V1);
\vertex (Ni) [left=1.1cm of V1] {\(N_1\)};
\vertex (l) [right=1.1cm of V1] {\(\ell\)};
\vertex (V2) [below=1.4cm of V1];
\vertex (Q) [left=1.1cm of V2] {\(q_3, \bar{t}\)};
\vertex (t) [right=1.1cm of V2] {\(t, \bar{q_3}\)};
\diagram*{
  (Ni) -- (V1) -- (l),
  (Q) -- (V2) -- (t),
  (V1) -- [scalar, edge label'=\(\Phi\)] (V2)
};
\end{feynman}
\end{tikzpicture}
\centering
\begin{tikzpicture}[baseline=(current bounding box.center)]
\begin{feynman}
\vertex (V1);
\vertex (Ni) [left=1.1cm of V1] {\(N_1\)};
\vertex (l) [right=1.1cm of V1] {\(\bar{\ell}\)};
\vertex (V2) [below=1.4cm of V1];
\vertex (Q) [left=1.1cm of V2] {\(t, \bar{q_3}\)};
\vertex (t) [right=1.1cm of V2] {\(q_3, \bar{t}\)};
\diagram*{
  (Ni) -- (V1) -- (l),
  (Q) -- (V2) -- (t),
  (V1) -- [scalar, edge label'=\(\Phi\)] (V2)
};
\end{feynman}
\end{tikzpicture}
\\[0.3em]
(b) \(t\)-channel
\end{minipage}
\caption{$\Delta L = 1$ scattering diagrams for the processes $N_{1}\ell (\bar{\ell}) \rightarrow \bar{t} q_3 (t \bar{q_3})$, $N_{1} q_3 \rightarrow \ell t $, $N_{1} \bar{t} \rightarrow \ell \bar{q_3}$,  $N_{1}\bar{q_3} \rightarrow \bar{\ell} \bar{t}$ and $N_{1}t \rightarrow \bar{\ell} q_3$ mediated by the Higgs doublet: (a) $s$-channel contribution, (b) $t$-channel contributions.}
\label{fig:DL1}
\end{figure}
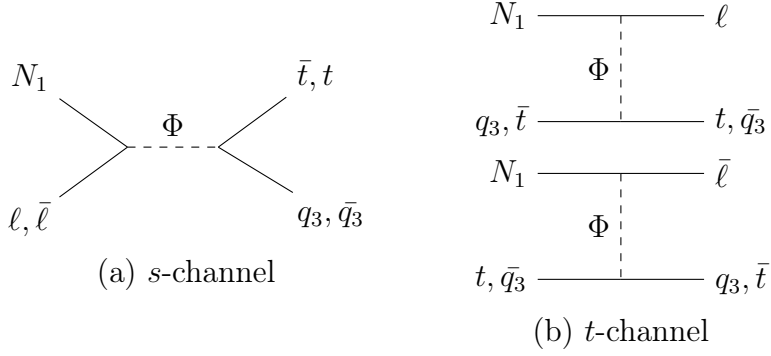

%%%%%%%%%%
In this third version of \texttt{ULYSSES}, we incorporate kinetic equations that account for the $\Delta L = 1$ tree-level scattering process involving the top quark and mediated by the Higgs doublet $\Phi$ in the scenario of unflavoured high-scale leptogenesis with one decaying heavy neutrino $N_1$. Specifically, we consider scatterings of the type $N_{1}\ell (\bar{\ell}) \rightarrow \bar{t} q_3 (t \bar{q_3})$, $N_{1} q_3 \rightarrow \ell t $, $N_{1} \bar{t} \rightarrow \ell \bar{q_3}$,  $N_{1}\bar{q_3} \rightarrow \bar{\ell} \bar{t}$ and $N_{1}t \rightarrow \bar{\ell} q_3$ and inverse processes, where $\ell$ ($\bar{\ell}$) is a generic (anti)lepton ${\rm SU}(2)$ doublet -- we work in the single-flavour regime --, $q_3$ ($\bar{q_3}$) is the third-generation (anti)quark ${\rm SU}(2)$ doublet and $t$ ($\bar{t}$) is the right-handed top (anti)quark ${\rm SU}(2)$ singlet. We sketch the corresponding Feynman diagrams in Fig.~\ref{fig:DL1} (see also \cite{Giudice:2003jh}). In this work, we do not include processes involving gauge bosons nor assume CP-violation for $\Delta L=1$ processes, as the corresponding contributions are
negligible in the strong washout regime. However, in the weak washout regime, CP-violation tends to suppress asymmetry production; a full treatment of this effect has been explored in \cite{Abada:2006ea} and we leave its implementation to a future version of the code.

The module \texttt{etaB1BE1F\_CaseS2.py} enables the evolution of the full phase-space distributions of the heavy neutrino $N_1$ and of the $B-L$ asymmetry. The kinetic equations read:\\
\begin{eqnarray}\label{eq:RHN distribution function}
   H   z \frac{\partial f_N}{\partial z}&=&C_D\left[f_N\right]+2 C_{S, s}\left[f_N\right]+4 C_{S, t}\left[f_N\right],\\
   \label{eq:BL distribution function}
 H z \frac{\partial f_{B-L}}{\partial z}&=&C_D\left[f_{B-L}\right]+2 C_{S, s}\left[f_{B-L}\right]+4 C_{S, t}\left[f_{B-L}\right],
\end{eqnarray}
and are implemented \emph{without assuming kinetic equilibrium nor Maxwell-Boltzmann statistics}. The factors of two in front of the scattering terms come from summing over the processes with particles and antiparticles, 
and the terms for the $t$-channel proportional to $C_{S,t}$ have an additional factor of two because of the corresponding $u$-channel diagrams. Here, $f_N$ ($f_{B-L}$) is the phase-space distribution associated to the heavy neutrino $N_1$ ($B-L$ asymmetry) as a function of the momentum and $z={M_1}/{T}$.
 The decay functional $C_D[f_N]$ in Eq.~\eqref{eq:RHN distribution function} is the following integral \cite{Hahn-Woernle:2009jyb} (already implemented in \texttt{ULYSSES} version 2):
\begin{equation}
\label{eq:RHN decay}
 \begin{aligned}
  C_D\left[f_N\right]= & \frac{1}{2 E_N} \int \frac{d^3 \vec{p}_\ell}{2 E_\ell(2 \pi)^3} \frac{d^3\vec{p}_{\Phi}}{2 E_{\Phi}(2 \pi)^3}(2 \pi)^4 \delta^4\left(p_N-p_\ell-p_{\Phi}\right) \\
  & \times\left[f_{\Phi} f_\ell\left(1-f_N\right)\left(\left|\mathcal{M}_{\Phi \ell\rightarrow N_1}\right|^2+\left|\mathcal{M}_{ \Phi^* \bar{\ell}\rightarrow N_1}\right|^2\right)\right. \\
  & \left.-f_N\left(1-f_\ell\right)\left(1+f_{\Phi}\right)\left(\left|\mathcal{M}_{N_1 \rightarrow \ell\Phi }\right|^2+\left|\mathcal{M}_{N_1 \rightarrow \bar{\ell}\Phi^* }\right|^2\right)\right]\,,
 \end{aligned}
\end{equation}
\noindent  where $p_N=(E_N,\vec{p}_N)$ is the four-momentum (energy) of the heavy neutrino $N_1$; $p_\ell=(E_\ell,\vec{p}_\ell)$ and $p_\Phi=(E_\Phi,\vec{p}_\Phi)$ denote that of the lepton doublet and Higgs doublet, respectively, while $f_\ell$ and $f_\Phi$ are the corresponding phase-space distribution functions; the quantities $\mathcal{M}$ denote the Feynman amplitudes for the decay and inverse-decay processes $N \leftrightarrow \ell\Phi$ and $N \leftrightarrow \bar{\ell}\Phi^*$.
The scattering functionals $C_{S,s}[f_N]$ and $C_{S,t}[f_N]$ are respectively the 
$s$-channel (Fig.~\ref{fig:DL1}a) and $t$-channel contributions (Fig.~\ref{fig:DL1}b) from the $\Delta L=1$ scattering, and are given by: 
\begin{equation}
\label{eq:s-channel scattering}
 \begin{aligned}
C_{S, s}\left[f_N\right] = &\frac{1}{2 E_N} \int \prod_{i=\ell, q, t} \frac{d^3 \vec{p_i}}{(2 \pi)^3 2 E_i}(2 \pi)^4 \delta^4\left(p_N+p_\ell-p_t-p_q\right)\left|\mathcal{M}_s\right|^2 \\
& \times\left[\left(1-f_N\right)\left(1-f_\ell\right) f_t f_q-f_N f_\ell\left(1-f_t\right)\left(1-f_q\right)\right]\,.
\end{aligned}
\end{equation}
\begin{equation}
\label{eq:t-channel scattering}
 \begin{aligned}
C_{S, t}\left[f_N\right] = &\frac{1}{2 E_N} \int \prod_{i=\ell, q, t} \frac{d^3 \vec{p}_i}{(2 \pi)^3 2 E_i}(2 \pi)^4 \delta^4\left(p_N+p_t-p_\ell-p_q\right)\left|\mathcal{M}_t\right|^2 \\
& \times\left[\left(1-f_N\right)\left(1-f_t\right) f_\ell f_q-f_N f_t\left(1-f_\ell\right)\left(1-f_q\right)\right]\,.
\end{aligned}
\end{equation}
Here, $p_q=(E_q,\vec p_q)$ and $p_t=(E_t,\vec p_t)$ denote the four-momenta of the quark doublet $q_3$ and right-handed top quark $t$, respectively, while $f_q$ and $f_t$ are their corresponding phase-space distribution functions, $\mathcal{M}_s$ and $\mathcal{M}_t$ are the amplitudes for the $s$- and $t$-channel $\Delta L=1$ scattering processes.

Analogously, we can write the expressions for the decay and scattering functionals applied to $f_{B-L}$:
\begin{equation}
\label{eq: decay B-L}
 \begin{aligned}
  C_D\left[f_{B-L}\right]= & \frac{1}{2 E_N} \int \frac{d^3 \vec{p}_N}{2 E_N(2 \pi)^3} \frac{d^3 \vec{p}_{\Phi}}{2 E_{\Phi}(2 \pi)^3}(2 \pi)^4 \delta^4\left(p_N-p_\ell-p_{\Phi}\right) \\
  & \times\left[f_{B-L}(f_N+f_{\Phi})+2\epsilon(f_\ell^{eq}\left(f_{\Phi}+f_N\right)-f_N(1+f_\Phi)\right)],
 \end{aligned}
\end{equation}
\begin{equation}
\label{eq:s-channel scattering B-L}
 \begin{aligned}
C_{S, s}\left[f_{B-L}\right] = &\frac{1}{2 E_N} \int \prod_{i=N, q, t} \frac{d^3 \vec{p}_i}{(2 \pi)^3 2 E_i}(2 \pi)^4 \delta^4\left(p_N+p_\ell-p_t-p_q\right)\left|\mathcal{M}_s\right|^2 \\
& \times\left[f_{B-L}\left(f_N\left(f_t+f_q-1\right)-f_tf_q\right)\right],
\end{aligned}
\end{equation}
\begin{equation}
\label{eq:t-channel scattering B-L}
 \begin{aligned}
C_{S, t}\left[f_{B-L}\right] = &\frac{1}{2 E_N} \int \prod_{i=N, q, t} \frac{d^3 \vec{p}_i}{(2 \pi)^3 2 E_i}(2 \pi)^4 \delta^4\left(p_N+p_t-p_\ell-p_q\right)\left|\mathcal{M}_t\right|^2 \\
& \times\left[f_{B-L}\left(f_q\left(f_t+f_N-1\right)-f_tf_N\right)\right],
\end{aligned}
\end{equation}
 where $\epsilon$ is the usual unflavoured CP-asymmetry parameter, while $f_\ell^{\rm eq}$ is the equilibrium phase-space distribution of the lepton doublet. We employ momentum binning to evaluate the distribution function $f_N$ in Eq.~\eqref{eq:RHN distribution function}, using Bose-Einstein and Fermi-Dirac statistics for the particles involved. 
Then, to get the equation for the $B-L$ asymmetry $\eta_{B-L}$, we integrate the equation for $f_{B-L}$ over $d^3\vec{p}_\ell$ and normalise appropriately. 

The numerical implementation proceeds in two decoupled stages. The first stage solves the RHN phase-space ODE (Eq.~\eqref{eq:RHN distribution function}), while the second uses the resulting distributions as stored input to solve the $B-L$ asymmetry ODE (Eq.~\eqref{eq:BL distribution function}). The main computational challenge is that, after the analytic reduction, the $s$- and $t$-channel scattering collision integrals $C_{S,s}[f_N]$ and $C_{S,t}[f_N]$ reduce to ten sets of two-dimensional integrals per momentum mode and per $z$-step, each of which must be evaluated numerically. Performing these integrations during every ODE evaluation would be prohibitively expensive.

To address this, we developed a bespoke \texttt{numba}-compatible wrapper around \texttt{cquadpack}, enabling fast numerical quadrature within \texttt{numba}-compiled routines and substantially reducing the computational cost. The full phase space modules \texttt{1BE1F\_Case2},\texttt{1BE1F\_Case3}, and \texttt{1BE1F\_Case4} also use this wrapper now and are an order of magnitude faster than in version 2 for the benchmark \texttt{examples/1N1F.dat} shipped with \texttt{ULYSSES}.

We also exploit the fact that the full RHS of the $f_N$ equation is \emph{linear} in $f_N$. The scattering contributions can therefore be written as
\begin{equation}
  Hz\frac{\partial f_N}{\partial z}\bigg|_{\rm scat} = K\bigl[A(z,y_N)\,f_N + B(z,y_N)\bigr],
\end{equation}
where $A$ and $B$ are scalar functions that depend only on the kinematics, not on $f_N$ itself, and $y_N = |\vec{p}_N|/T$ and $K$ is the usual decay parameter \cite{Buchmuller2005305}. They are extracted by evaluating the full RHS at $f_N = 0$ to obtain $B(z,y_N)$, and at $f_N = 1$ to obtain $A(z,y_N) + B(z,y_N)$, with $K$ factored out. Both functions are precomputed offline over a dense grid and stored. At runtime these are loaded once and evaluated via bilinear interpolation.

\begin{figure}
  \centering
   \includegraphics[width=1\textwidth]{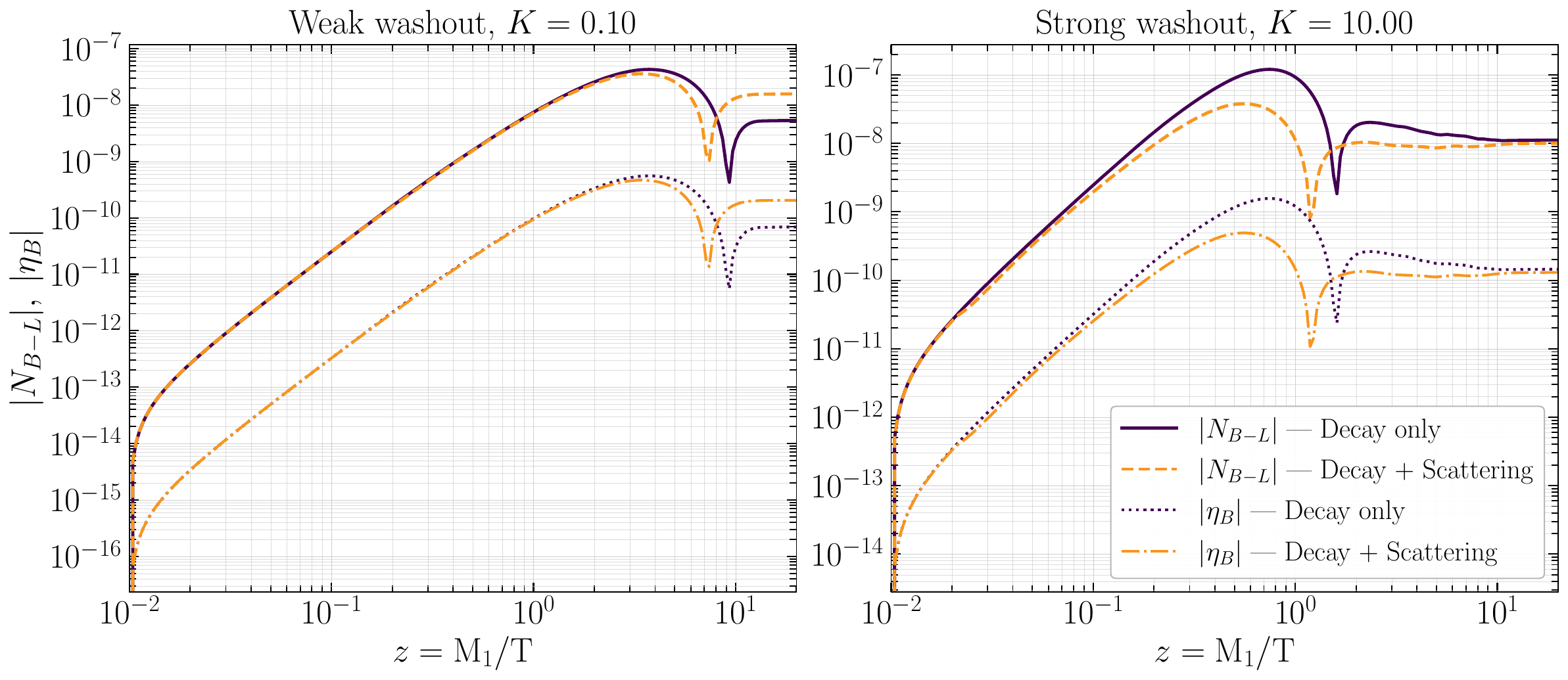}
  \caption{ Time evolution of $|N_{B-L}|$ and $|\eta_B|$ against $z = M_1/T$ respectively in the regime of weak washout (left panel) with $K = 0.1$ and strong washout (right panel) with $K=10$ \cite{Buchmuller2005305}.} 
  \label{fig:N_B-L_CaseS2}
 \end{figure}

Working in the rescaled variable $g(z,y_N) \equiv f_N / f_N^{\rm eq}$, where $f_N^{\rm eq} = (e^{E_N/T}+1)^{-1}$, improves numerical stability near equilibrium. The all-mode RHN ODE is solved as a single vectorised system. In the strong washout regime ($K \gtrsim 3$), the washout term $K \cdot A(z, y_N)$ is large relative to the Hubble rate, creating a wide separation of timescales that renders the ODE stiff. Explicit integrators would then require prohibitively small time steps to maintain stability, so we use the Radau implicit solver instead, supplying the analytic diagonal Jacobian to maximise its efficiency. In the weak washout regime no such stiffness arises and the explicit RK45 integrator is used.
With the $f_N(z, y_N)$ grid stored, the $B-L$ asymmetry ODE is solved by evaluating the full momentum integral in Eq.~\eqref{eq:BL distribution function} at each $z$-step. The $t$-channel infrared divergence at $t = 0$ is regulated by the cutoff $a_h = m_\Phi/M = 10^{-5}$ \cite{Hahn-Woernle:2009jyb}.

A comparison between the decay-only scenario and the case including $\Delta L = 1$ scattering is shown in Fig.~\ref{fig:N_B-L_CaseS2}, for weak (strong) washout on the left (right) panel. For the weak washout regime, the sign flip in the asymmetry occurs earlier due to the enhanced production of RHNs via $\Delta L = 1$ scattering, resulting in a larger final asymmetry. In contrast, in the strong washout regime, the scattering processes contribute to additional washout, and hence the two competing terms lead to approximately the same asymmetry as the term without $\Delta L=1$ processes.

\section{A module for simultaneous leptogenesis and freeze-in dark matter}
\label{sec:DM}

As a concrete demonstration of the \texttt{-{}-extended} parameter interface introduced in Sec.~\ref{ssec:extended}, we present a toy module \texttt{etab1BE1F\_DM\_FreezeIn.py}. Here we simultaneously solve the standard vanilla leptogenesis equations with an additional equation for the abundance of a feebly interacting particle produced through the out-of-equilibrium decays of $N_1$. We identify this particle as dark matter (DM). The module is intended as a proof of concept illustrating how the \texttt{-{}-extended} interface enables coupling additional Boltzmann equations to the standard leptogenesis solver without modifying the core infrastructure.

In particular, we focus on a DM toy model
in which DM is minimally coupled to the lightest heavy neutrino $N_1$ and is produced from the out-of-equilibrium decays $N_1$ via a \textit{freeze-in} mechanism.
This module 
solves three coupled Boltzmann equations simultaneously, extending the vanilla 
leptogenesis equations (see $e.g$ Ref.~\cite{Granelli:2020pim}, module \texttt{1BE1F.py}) with a third equation for the DM abundance. The system of evolution equations reads
\begin{eqnarray}
\label{eq:DM_BE1}
 \frac{dN_1}{dz} &=& -(D + D_{\rm dark})\left(N_1 - N_1^{\rm eq}\right)\,, \\
 \label{eq:DM_BE2}
 \frac{dN_{B-L}}{dz} &=& \epsilon\, D\left(N_1 - N_1^{\rm eq}\right) - W\, N_{B-L}\,, \\
 \label{eq:DM_BE3}
 \frac{dN_{\rm DM}}{dz} &=& D_{\rm dark}\, N_1\,,
\end{eqnarray}
where $z = M_1/T$, $\epsilon$ is the unflavoured CP-asymmetry, $W$ is the %standard 
inverse-decay washout term and $D$ is the standard decay parameter while $D_{\rm dark}$ is that associated to DM. The decay parameters are decomposed as
\begin{equation}
 D = (1 - {\rm Br}_{\rm dark})\,D_{\rm total}\,, \qquad D_{\rm dark} = {\rm Br}_{\rm dark}\,D_{\rm total}\,,
\end{equation}
where $D_{\rm total} \equiv D + D_{\rm dark}$ is proportional to the total width of $N_1$. The branching ratio into the dark sector is
\begin{equation}
\label{eq:Brdark}
 {\rm Br}_{\rm dark} = \frac{\Gamma_{\rm dark}}{\Gamma_{\rm SM} + \Gamma_{\rm dark}}\,, \qquad \Gamma_{\rm dark} = \frac{\lambda^2 M_1}{8\pi}\,,
\end{equation}
where $\lambda$ is a dark-sector coupling controlling the partial width of $N_1$ into the dark sector and $\Gamma_{\rm dark}$ is the decay width of $N_1$ into the dark sector. In the FIMP (Feebly Interacting Massive Particle) limit $\lambda \ll$ Yukawa couplings, the DM never reaches thermal equilibrium and no washout term appears in Eq.~\eqref{eq:DM_BE3}. The leptogenesis sector is therefore negligibly perturbed by the presence of the dark sector, and the standard vanilla leptogenesis
is recovered in the limit $\lambda \to 0$.

Once the system is solved up to $z_{\rm final}$ (code variable \texttt{zmax}), the DM comoving number density $N_{\rm DM}(z_{\rm final})$ is converted to the DM yield $Y_{\rm DM} = N_{\rm DM}/s$ and the relic abundance
\begin{equation}
 \Omega_{\rm DM} h^2 = \frac{m_{\rm DM}\, s_0\, Y_{\rm DM}}{\rho_c/h^2}\,,
\end{equation}
using the present-day entropy density $s_0 = 2891.2\,\mathrm{cm}^{-3}$. 

\subsection{Implementation in \texttt{ULYSSES}}

The DM module is invoked by selecting the model \texttt{1BE1F\_DM\_FreezeIn} with the \texttt{-{}-extended} flag described in Sec.~\ref{ssec:extended}. The parameter file contains all standard vanilla leptogenesis inputs (see Sec.~\ref{sec:setup}) plus two model-specific keys read from \texttt{pdict}:
\begin{figure}[t!]
  \centering
   \includegraphics[width=.48\textwidth]{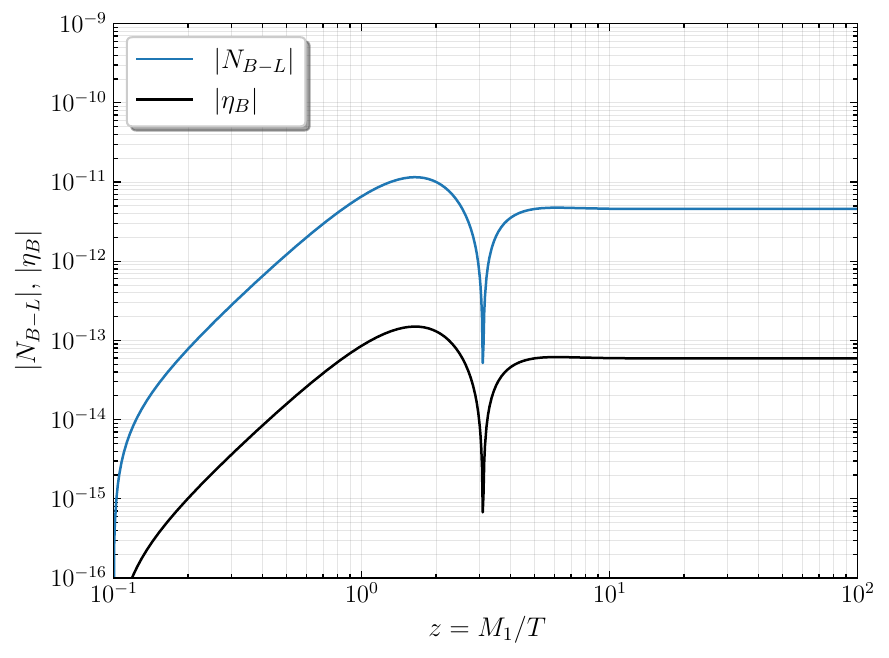}
  \includegraphics[width=.48\textwidth]{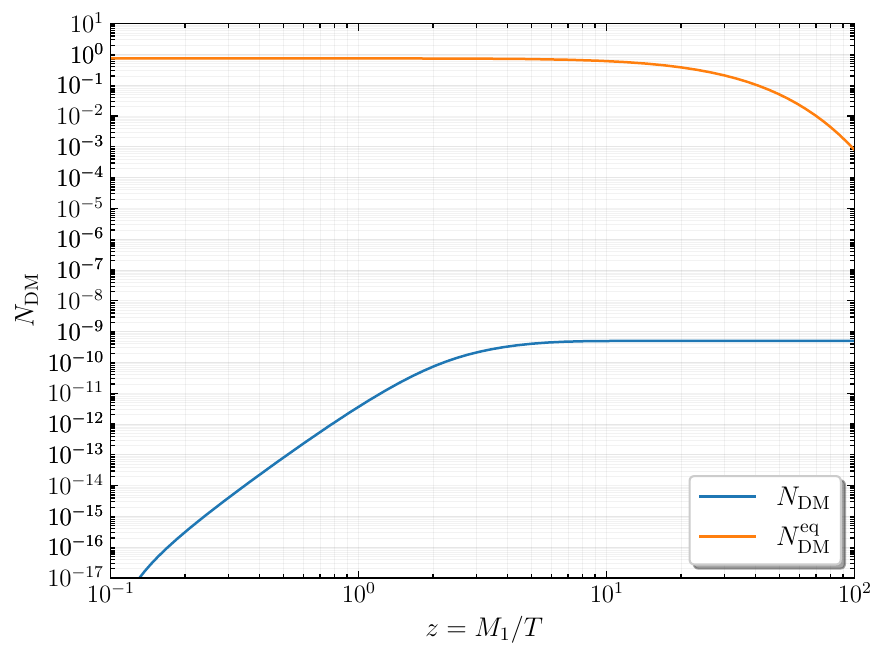}
  \caption{Evolution plot for the freeze-in dark matter plus vanilla leptogenesis scenario. The left panel shows the $B-L$ asymmetry, $|N_{B-L}|$ in blue and $|\eta_{B-L}|$ in black, versus $z$; the right panel shows the DM abundance $N_{\rm DM}(z)$ obtained from the evolution equation (blue) and the equilibrium abundance $N_{\rm DM}^{\rm eq}(z)$ (orange). The parameters are as in the example runcard given in the text. We also describe how to obtain such plot in the Jupyter notebook.}
     \label{fig:DM_freezein}
 \end{figure}
\begin{itemize}
 \item \texttt{lam}: the dark-sector coupling $\lambda$ (dimensionless), controlling the DM production rate via Eq.~\eqref{eq:Brdark}. Typical freeze-in values are $\lambda \sim 10^{-7}$--$10^{-9}$.
 \item \texttt{m\_dm}: the DM mass in GeV, used to compute $\Omega_{\rm DM} h^2$. Defaults to $M_1/10$ if omitted.
\end{itemize}
After the run, \texttt{ULYSSES} prints $\eta_B$ alongside $Y_{\rm DM}$ and $\Omega_{\rm DM} h^2$. The DM abundance $N_{\rm DM}(z)$ is stored via \texttt{setEvolData} as a supplementary column (see Sec.~\ref{ssec:plots}) and has no effect on the calculation of the returned baryon asymmetry.
An example parameter file extending the standard vanilla leptogenesis card is given below (\texttt{examples/1N1F\_dm.dat}):
\lstset{label={lst:exampleDM}}
\begin{lstlisting}[language = bash]
m   -100   # log10(m_lightest / eV)
M1  12    # log10(M1 / GeV)
M2  15    # log10(M2 / GeV)
M3  16    # log10(M3 / GeV)
x1  180    # Casas-Ibarra angle [deg]
y1  1.4
x2  180
y2  11.2
x3  180
y3  11
delta 270    # Dirac CP phase [deg]
a21  0     # Majorana phase [deg]
a31  0
t23  49.7   # PMNS angles [deg]
t12  33.82
t13  8.61
lam  1e-7   # dark-sector coupling
m_dm 1e11    # DM mass [GeV]
\end{lstlisting}
The run is executed as:
\begin{bash}
uls-calc -m 1BE1F_DM_FreezeIn --extended examples/1N1F_dm.dat
\end{bash}
To additionally generate an evolution plot saved to a PDF:
\begin{bash}
uls-calc -m 1BE1F_DM_FreezeIn --extended -o evol.pdf examples/1N1F_dm.dat
\end{bash}
 The evolution plot in Fig.~\ref{fig:DM_freezein} displays two panels: the left shows the standard leptogenesis evolution ($N_1$, $N_{B-L}$, and $\eta_B$), while the right shows the DM abundance $N_{\rm DM}(z)$ alongside its equilibrium value $N_{\rm DM}^{\rm eq}(z)$, confirming the out-of-equilibrium nature of the freeze-in production.

\section{Summary}\label{sec:conclusions}
We have released the third version of the Universal LeptogeneSiS Equation Solver
\texttt{ULYSSES} \href{https://github.com/earlyuniverse/ulysses}{\faGithub}. In this manual, we have described the several significant new features and
infrastructure improvements of the new version.
We introduced an alternative parameterisation
of the CI matrix in terms of a single complex angle and four real angles
\cite{Drewes:2021nqr},  instances to select the initial abundance of the heavy neutrinos, an \texttt{-{}-extended} parameter interface enabling users to include model-specific inputs beyond the standard leptogenesis runcard, without modifying the core infrastructure, and adopts the best-fit flavour neutrino oscillation mixing angles and light neutrino squared mass differences from the
\texttt{NuFit-6.1} global analysis~\cite{Esteban:2024eli} as defaults.

On the theoretical side, the central new addition is a state-of-the-art
implementation of the density matrix equations for low-scale leptogenesis via
oscillations with three RHNs. Compared to the two-neutrino
implementation in version~2~\cite{Granelli:2023vcm}, the new equations incorporate
non-relativistic corrections to the Hamiltonian and to the heavy neutrino production
rates from both lepton-number-conserving and lepton-number-violating processes,
significantly extending the validity of the code to heavy neutrino masses beyond the $100\,\text{GeV}$ scale. We have provided tables for the thermally averaged rates that enter the density matrix equations, together with the scripts used to compute them, in order to make the implementation more transparent and accessible, also allowing users to adapt the code and calculate their own rates if needed. We have accompanied
this implementation with numerical speed-up routines for the density matrix equations. Beyond these code’s new features, this release presents the only comparative analysis of results produced by different research groups on low-scale leptogenesis. In particular, we compared our results with the benchmarks provided in \cite{Abada:2018oly, Hernandez:2022ivz}, as well as with the public \texttt{C++} code \href{https://github.com/stefanmarinus/amiqs}{\texttt{amiqs}} and the private one used, e.g., for \cite{Drewes:2021nqr}, finding satisfactory agreement -- the residual differences that arose when comparing with \cite{Abada:2018oly}
can be attributed to the more updated rates implemented in \texttt{ULYSSES}. We have also validated the results of the low-scale density matrix equation module beyond the $100\, \text{GeV}$ scale against the two-RHN resonant leptogenesis Boltzmann equation module shipped with previous versions. In the regime above $100\,\text{GeV}$ heavy neutrino mass scale, large mass splittings and strong wash-out where the two descriptions of leptogenesis are expected to coincide, we find a remarkable $\mathcal{O}(1)$ agreement despite the significant differences in the evolution equations. Overall, this benchmarking effort not only  demonstrates 
the  robustness of the \texttt{ULYSSES} infrastructure, but also provides a vital reference point for future code developments in the GeV-to-TeV mass regime as well as offering a solid structural basis for the development of the final leptogenesis equations.

For the high-scale vanilla leptogenesis scenario, we have implemented $\Delta L = 1$ scattering processes involving the top quark and mediated by the Higgs doublet to the full phase-space kinetic equation, improving the evolution of both the RHN and lepton asymmetry without assuming kinetic equilibrium nor Maxwell-Boltzmann statistics. Moreover, we have showcased the potentiality of the \texttt{-{}-extended} parameter interface that allows users to pass model-specific parameters beyond the standard leptogenesis runcard, enabling custom model implementations without modifying the core infrastructure; as a proof of concept we have included a toy freeze-in dark matter module (\texttt{1BE1F\_DM\_FreezeIn}) illustrating the flexibility of the package as a general-purpose Boltzmann equation solver.

Future extensions of \texttt{ULYSSES} could include a more comprehensive implementation of scattering processes and resonance effects to the scenarios of high-scale leptogenesis (including scatterings involving gauge bosons, as well as $\Delta L = 2$ processes; scattering and resonance effects in the flavoured density matrix equations), non-instantaneous sphaleron freeze-out via a temperature-dependent susceptibility matrix and extensions to more complex dark matter scenarios, such as co-annihilation and semi-annihilation.

\section*{Code Availability}

\texttt{ULYSSES} is distributed under the MIT licence and is \href{https://github.com/earlyuniverse/ulysses}{publicly available}.
The release accompanying this paper is tagged \texttt{v3.0.0}, and all results, figures and benchmark tables presented above were produced with this tag. The runcards used for every figure are shipped in the repository under \texttt{examples/}, with file names matching the labels used in the text (\texttt{ARS\_2RHN\_BMC[I,II,III].dat}, \texttt{ARS\_vs\_RES.dat}, \texttt{ARS\_BMCI.dat}, etc.). The same release also contains the scripts that generate the thermally-averaged rate tables of Appendix~\ref{app:Rates} and the self-contained Jupyter notebook \texttt{ULYSSES\_intro.ipynb} reproducing the basic-usage examples.

\texttt{ULYSSES~v3.0} requires Python in the range $3.9\,\le\,\text{Python}\,<\,3.15$ and depends on \texttt{numpy}~$\ge 1.22$, \texttt{scipy}~$\ge 1.9$, \texttt{matplotlib}~$\ge 3.5$, \texttt{pandas}~$\ge 1.5$, \texttt{numba}~$\ge 0.56$, \texttt{mpmath}~$\ge 1.2$, \texttt{mpltern}~$\ge 0.4$, \texttt{multiprocess}~$\ge 0.70.12$, \texttt{tqdm}~$\ge 4.60$, \texttt{progressbar}~$\ge 2.5$, \texttt{python-dateutil}~$\ge 2.8$ and \texttt{termcolor}~$\ge 1.1$. Nested-sampling parameter scans optionally rely on \texttt{pymultinest}~$\ge 2.11$. Pre-built wheels are distributed via PyPI, so a standard \texttt{pip install ulysses} requires no additional system dependencies; a C compiler is only needed when building from source, for the bundled \texttt{NumbaQuadpack} extension module used both by the $\Delta L = 1$ phase-space integrator of Section~\ref{sec:scat} and by \texttt{ProdRates.py} (Section~\ref{sec:lowlep} and Appendix~\ref{app:Rates}) when users wish to recompute or modify the thermally-averaged rate tables. The results shown in this paper were cross-checked on a variety of operating systems and Python versions within the supported range, producing identical $\eta_B$ values to within the solver tolerances.

\section*{Acknowledgements}
We thank Stefan Sandner for useful comparisons with the \texttt{amiqs} code. We are grateful to Marco Drewes for helpful initial discussions. We also thank Arturo de Giorgi, Zara Graham-Jones, Joanne Roper, Federico Silvetti, and Joseph Tudor for testing the installation during the beta phase of this work, and Maximilian Berbig for inspiring suggestions on the title of this manuscript.
A.G.~is supported by the Spanish grant PID2023-148162NB-C21 (MCIN/AEI/10.13039/501100011033) co-funded by the European Union (FEDER), and in part by the  
European Union's Horizon Europe programme under the Marie Skłodowska-Curie Actions – Staff Exchanges (SE) grant agreement No.~101086085-ASYMMETRY. 

\appendix

\section{Thermally averaged terms in the density matrix equations}\label{app:Rates}
In this section, we provide further details on the implementation of the functions $\langle 1/y_0\rangle$, $\langle h^{\LNC}\rangle$, $\langle h^{\LNV}\rangle$, as well as on the rates $\langle\gamma_{\LNC}^{(a)}\rangle$ and $\langle S_{\LNV}^{(a)}\rangle$ appearing in the DMEs for low-scale leptogenesis via oscillations given in Eqs.~(\ref{eq:DME_N}-\ref{eq:DME_mu}). The results reported in this section are cast in the file \texttt{ProdRates.py}. We start by specifying that the average $\left\langle \cdot \right\rangle$ refers to the following integration:
\begin{equation}\label{eq:therm_average}
 \left\langle \cdot \right \rangle = \frac{1}{n_N^{\rm eq}}\int \frac{d^3\vec{k}}{(2\pi)^3} \,(\cdot) f_F(y_0) = \frac{1}{ 2\pi^2 n_N^{\rm eq}}\int_0^{+\infty} dy\, (\cdot) y^2 f_F(y_0),
\end{equation}
where $(\cdot)$ should be replaced with the function to average; $ f_F(y_0) = 1\,/(e^{y_0}+1)$ is the Fermi-Dirac phase-space distribution function; $k = (E_N, \vec{k})$ is the heavy neutrinos' four-momentum with $E_N \simeq (\Mav^2 + |\vec{k}|^2)^{1/2}$ (we are assuming that the three heavy neutrinos share equal energy and 3-momentum, and neglect mass splittings so that $\Mav \equiv M_1 \simeq M_2 \simeq M_3$); $y \equiv |\vec{k}|/T$ and $y_0 \equiv E_N/T = (\Mav^2/T^2 + y^2)^{1/2}$; $n_N^{\rm eq}$ is defined as
\begin{equation}
n_N^{\rm eq} \equiv \frac{1}{(2\pi)^3}\int \, d^3\vec{k}\,f_F(y_0) = \frac{1}{2\pi^2}\int_0^{+\infty} dy\, y^2 f_F(y_0).
\end{equation}

\subsection{Thermally averaged terms in the Hamiltonian}
\begin{figure}
 \centering
\includegraphics[width=0.48\textwidth]{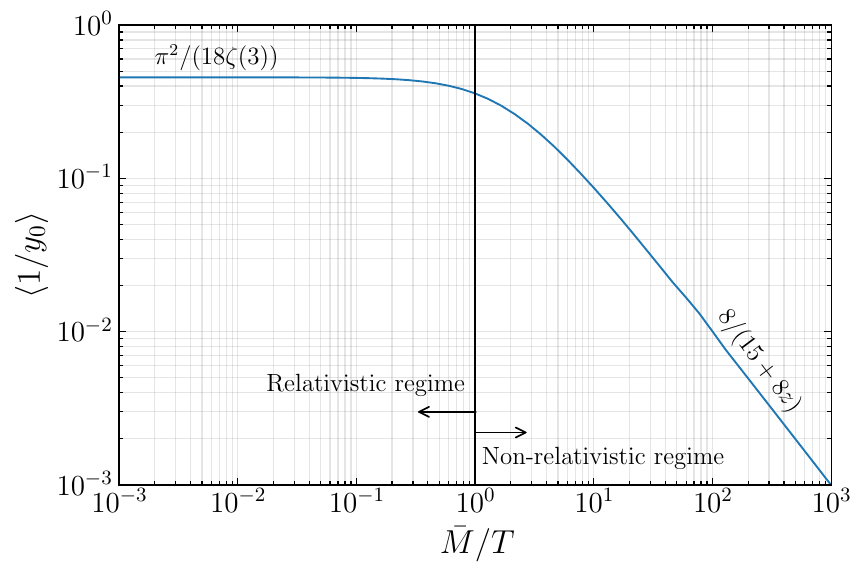}
\includegraphics[width=0.48\textwidth]{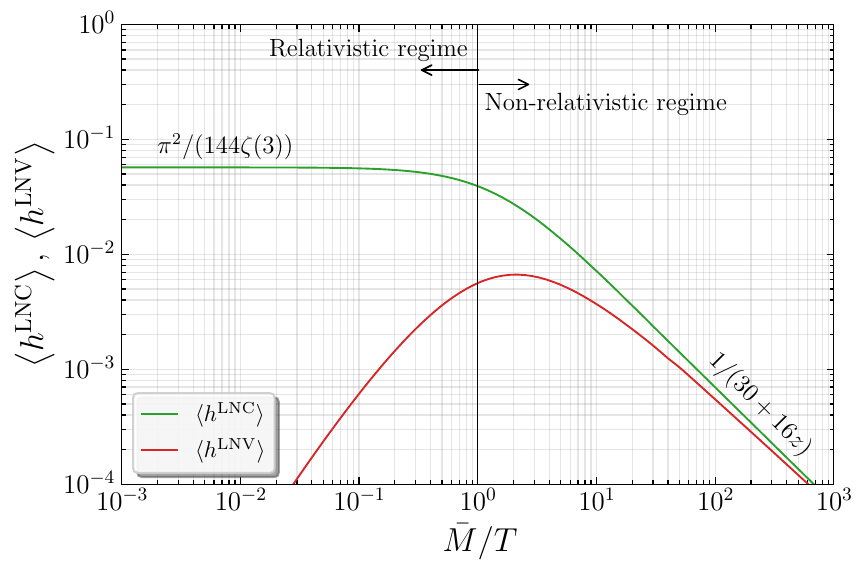}
 \caption{The quantities $\left\langle 1/y_0 \right\rangle$ (left, blue), $\left\langle h^{\rm LNC}\right\rangle$ (right, green) and $\left\langle h^{\rm LNV}\right\rangle$ (right, red) entering the averaged Hamiltonian $\left\langle \mathcal{H}\right\rangle$, see Eq.~\eqref{eq:thermH}, as evaluated numerically inside \texttt{ULYSSES}, here shown as functions of $z=\Mav/T$.}
 \label{fig:1/y0}
\end{figure}

Following from the above definitions, the function $\left\langle 1/y_0 \right \rangle$ can be computed as a ratio of two integrals 
\begin{equation}
\left\langle \frac{1}{y_0} \right \rangle \equiv \frac{I_1}{I_2},
\end{equation}
where
\begin{eqnarray}
 I_1 &\equiv& \int_z^{+\infty} dy\, \frac{y^2}{y_0} f_F(y_0) =z^2 \int_1^{+\infty} d\gamma \sqrt{\gamma^2 - 1} f_F(\gamma z),\\
 I_2 &\equiv& \int_z^{+\infty} dy\, y^2 f_F(y_0)= z^3 \int_1^{+\infty} d\gamma \gamma \sqrt{\gamma^2 - 1} f_F(\gamma z),
\end{eqnarray}
having defined $\gamma \equiv E_N/M_1$ and $z \equiv M_1/T$. We compute both integrals numerically. We note that in the non-relativistic regime the Fermi-Dirac statistics resembles the Boltzmann one so that $f_F(\gamma z) \approx e^{-\gamma z}$. Using this approximation, the integrals above can be written in terms of Bessel functions of the second kind as $I_{1} = z K_{1}(z)$ and $I_2 = z^2 K_2(z)$, so that $\left\langle1/y_0\right\rangle = K_1(z)/(z K_2(z))\simeq 8/(15+8z)$ \cite{Buchmuller2005305}. We adopt this approximation in the code whenever $z$ gets too large, avoiding numerical overflows. Conversely, in the ultra-relativistic limit such that $y\simeq y_0$ the integrals can be evaluated analytically, giving $I_1 = \pi^2/12$ and $I_2 = 3\zeta(3)/2$, with $\left\langle 1/y_0 \right \rangle \simeq \pi^2/(18\zeta(3))$, matching the value used in the previous version of \texttt{ULYSSES} \cite{Granelli:2023vcm}.
We evaluate the functions $\left\langle h^{\rm LNC} \right \rangle$ and $\left \langle h^{\rm LNV}\right \rangle$ numerically according to the following integrals:\footnote{We derived the expressions for $\left\langle h^{\rm LNC} \right \rangle$ and $\left \langle h^{\rm LNV}\right \rangle$ from those given in \cite{Antusch:2017pkq}, but multiplied the latter by $1/2$ to match with the relativistic LNC term used in the second version of \texttt{ULYSSES} \cite{Granelli:2023vcm} and other works (e.g., \cite{Abada:2018oly, Hernandez:2022ivz}).} 
\begin{eqnarray}
 \left\langle h^{\rm LNC} \right\rangle &=& \frac{1}{16 I_2} \int_0^{+\infty} dy\, \frac{y^2}{y_0} \left[1 + \frac{y_0}{y}-\frac{z^2}{2y^2} \log\left(\frac{1+y/y_0}{1-y/y_0}\right) \right] f_F(y_0)\,, \\
 \left\langle h^{\rm LNV} \right\rangle &=& \frac{1}{16 I_2} \int_0^{+\infty} dy\, \frac{y^2}{y_0} \left[ 1 - \frac{y_0}{y} + \frac{z^2}{2y^2}\log\left(\frac{1+y/y_0}{1-y/y_0}\right) \right] f_F(y_0). 
\end{eqnarray}
In the relativistic limit, $z\to 0$, we find
\begin{equation}\label{eq:hLNC_rel}
 \langle h^{\LNC} \rangle \simeq \frac{1}{8}\left\langle \frac{1}{y_0}\right\rangle = \frac{\pi^2}{144\zeta(3)},
\end{equation} while for $\langle h^{\LNV} \rangle$ the following approximated relation holds \cite{Antusch:2017pkq}
\begin{equation}\label{eq:hLNV_rel}
 \langle h^{\LNV}\rangle \simeq\, 2.5\times 10^{-3} z^2\left[3.50-0.47\log(z^2)+3.47\log^2(z)\right].
\end{equation}
We use the approximations in Eqs.~\eqref{eq:hLNC_rel} and \eqref{eq:hLNV_rel} when $z\leq 10^{-3}$ to make the numerical evaluation faster.
In the non-relativistic limit, $z\gg 1$, it can be shown that
\begin{equation}\label{eq:h_nonrel}
 \langle h^{\LNC}\rangle \simeq \langle h^{\LNV}\rangle \simeq \frac{1}{16}\left\langle \frac{1}{y_0}\right\rangle \simeq \frac{K_1(z)}{16z K_2(z)} \simeq \frac{1}{30+16 z}.
\end{equation}
We adopt the approximation in Eq.~\eqref{eq:h_nonrel} in the limit of large $z$ to avoid numerical instabilities.
In Fig.~\ref{fig:1/y0} we plot the results for $\left\langle 1/y_0 \right \rangle$ (left panel), as well as $\left\langle h^{\rm LNC} \right \rangle$ and $\left\langle h^{\rm LNV} \right \rangle$ (right panel), as functions of $z$.

We also include indirect contributions to $\langle h^{\LNC}\rangle$ and $\langle h^{\LNV}\rangle$ in the broken
phase, which adds up to the contributions given here. We discuss these indirect contributions in the next subsection.

\subsection{Thermally averaged heavy neutrino production rates}
Our procedure to obtain the functions $ \gamma_{\LNC}^{(0)}$ and $ S_{\LNV}^{(0)}$ follows that outlined in \cite{Klaric:2021cpi}. The contributions differ between the symmetric and broken phases, depending on whether the neutral component of the Higgs doublet has acquired a non-zero vacuum expectation value (broken phase) or not (symmetric phase). The symmetric and broken phases are defined according to the temperature $T_{\rm ew} \simeq 160\,\text{GeV}$: the system is in the symmetric phase for 
$T\gtrsim T_{\rm ew}$ and in the broken phase for $T\lesssim T_{\rm ew}$. The rates $ \gamma_{\LNC}^{(0)}$ and $ S_{\LNV}^{(0)}$ are calculated as
\begin{eqnarray}\label{eq:G0}
 \gamma_{\LNC}^{(0)} &=& \left(1 + \frac{y}{y_0}\right)\left(\Sigma_\mathcal{A}^0 - \hat{k}^i\Sigma_\mathcal{A}^i \right) + \gamma^{\rm ind}_+, \\
  z^2S_{\LNV}^{(0)} &=& \left(1 - \frac{y}{y_0}\right)\left(\Sigma_\mathcal{A}^0 + \hat{k}^i\Sigma_\mathcal{A}^i\right) + \gamma^{\rm ind}_-,
  \label{eq:S0}
\end{eqnarray}
where $\Sigma_\mathcal{A} = (\Sigma_\mathcal{A}^0, \Sigma_\mathcal{A}^i)$ is the 4-vector describing the heavy neutrino anti-hermitian self-energy while $\gamma^{\rm ind}_\pm$ is the contribution due to indirect processes that produce neutrinos through active neutrino mixing, which is non-zero only in the broken phase. As in \cite{Klaric:2021cpi}, we decompose the heavy neutrino self-energy as
\begin{equation}\label{eq:selfenergytot}
 \Sigma_\mathcal{A}(\Mav, |\vec{k}|, T) = \Sigma_\mathcal{A}^{\rm rel}(|\vec{k}|, T) + \Theta\left(\Mav - m_\Phi(T)\right) \Sigma_\mathcal{A}^{\rm 1\leftrightarrow 2}(\Mav, |\vec{k}|, T),
\end{equation}
where $m_\Phi(T) = (T/4)(g_1^2 + 3g_2^2 + 4h_t^2 + 8\lambda_H)^{1/2}$ is the Higgs's thermal mass, with $g_1 = 0.35$, $g_2 = 0.65$, $h_t = 0.993$ and $\lambda_H = 0.129$ being the ${\rm U}(1)_Y$, ${\rm SU}(2)_L$, top Yukawa and Higgs self-couplings, respectively; $\Sigma_\mathcal{A}^{\rm rel}(|\vec{k}|, T)$ is the mass-independent relativistic contribution to the self-energy and $\Sigma_\mathcal{A}^{\rm 1\leftrightarrow 2}(\Mav, |\vec{k}|, T)$ is the mass-dependent contribution from $1\leftrightarrow 2$ processes. 
The latter, in general, receive contributions from heavy neutrino and Higgs doublet decays in the symmetric phase, $\Sigma_\mathcal{A}^{1\leftrightarrow 2}(T\gtrsim T_{\rm ew})= \Sigma_N + \Sigma_\Phi$  and heavy neutrino decays into Higgs and electroweak bosons in the broken phase, $\Sigma_{\mathcal{A}}^{1\leftrightarrow 2} (T\lesssim T_{\rm ew})= \Sigma_H + 2\Sigma_W + \Sigma_Z$. However, the Higgs decay contribution in the symmetric phase is only active in the relativistic regime for $z \leq m_\Phi(T)/T\simeq 0.63$. Its contribution is going to be included in $\Sigma_\mathcal{A}^{\rm rel}(|\vec{k}|, T)$ and thus, effectively, we only consider $\Sigma_N$ in $\Sigma_{\mathcal{A}}^{1\leftrightarrow 2}$ as it follows from the insertion of a Heaviside function in Eq.~\eqref{eq:selfenergytot}. We give more details on each of these contributions, as well as on $\Sigma_\mathcal{A}^{\rm rel}$ and $\gamma_\pm^{\rm ind}$ in what follows. We neglect the (bare and thermal) masses of the leptons in our treatment, except for the relativistic and indirect contributions.

\paragraph{Symmetric phase -- heavy neutrino decay.} The contribution from the heavy neutrino decay into the lepton and Higgs doublets to $\Sigma_\mathcal{A}^{1\leftrightarrow 2}$ is non-zero only when $\Mav \geq m_\Phi(T)$ and reads \cite{Frossard:2012pc} (see also \cite{Klaric:2021cpi})
\begin{eqnarray}
 \Sigma_N^0(y, z) &=& \frac{T}{8\pi y}\,\mathcal{I}^N_1(y, z)\,,\\
 \hat{k}^i\Sigma_N^i(y, z) &=& \frac{T}{16\pi y^2}\,\left[2y_0 \mathcal{I}^N_1(y, z) - z^2 \left(1 -x_\Phi(z)\right) \mathcal{I}^N_0(y, z)\right],
\end{eqnarray}
where $x_\Phi(z) \equiv m_H^2(T)/\Mav^2\propto z^{-2}$ and $\mathcal{I}^N_n(y, z) = \mathcal{I}_n(y_0; x^N_-(y, z), x^N_+(y,z))$, with
\begin{equation}
 \mathcal{I}_n(a; x_-, x_+) \equiv \int_{x_{-}}^{x_+} dx\, x^n [1-f_F(x)+f_B(a-x)];
\end{equation}
$x^N_{\pm}(y,z) = (1/2)[y_0(1-x_\Phi(z)) \pm y \lambda^{1/2}(1, 0, x_\Phi(z))]$, $\lambda$ the Källén function defined as $\lambda(a, b, c) = a^2 + b^2 + c^2 - 2ab - 2bc - 2ac$ and $f_B(x) = 1/(e^{x}-1)$ the Bose-Einstein phase-space distribution function.
We implement the following analytical expressions for $\mathcal{I}_0(a; x_-,x_+)$ and $\mathcal{I}_1(a; x_-,x_+)$ in the code and use it to evaluate $\mathcal{I}^N_0(y,z)$ an $\mathcal{I}^N_1(y,z)$:
\begin{eqnarray}\label{eq:mathcalI0}
 \mathcal{I}_0(a; x_-, x_+) &=& \log\left(\frac{e^{x_-}-e^{a}}{ e^{x_+}-e^{a}}\right)+ \log\left(\frac{1 + e^{x_+}}{1 + e^{x_-}}\right),\\
 \mathcal{I}_1(a; x_-, x_+) &=& x_+ \log\left(\frac{1 + e^{x_+}}{1 - e^{x_+ - a}}\right)- x_- \log\left(\frac{1 + e^{x_-}}{1 - e^{x_- -a}}\right) + \nonumber \\
 &&+\, \text{Li}_2\left(e^{x_- - a}\right) - \text{Li}_2\left(e^{x_+ - a}\right) + \text{Li}_2\left(-e^{x_+}\right) - \text{Li}_2\left(-e^{x_-}\right),\label{eq:mathcalI1}
\end{eqnarray}
where $\text{Li}_2(x)$ is the dilogarithm (polylogarithm of order 2).

\paragraph{Broken phase -- heavy neutrino decays.} In the broken phase, the heavy neutrinos can decay into a lepton accompanied by a real Higgs boson, a $W$-boson, or a $Z$-boson. Each decay channel is included only when it is kinematically accessible, that is, when the condition $\Mav \geq m_H(T),\,m_W(T),\,m_Z(T)$, where $m_H(T) = \sqrt{2\lambda_H} v(T)$, $m_W(T) = g_2 v(T)/\sqrt{2}$ and $m_Z(T) = \sqrt{g_1^2 + g_2^2} v(T)/\sqrt{2}$, with $v(T)$ being the temperature-dependent Higgs' vacuum expectation value. We consider the following temperature dependence for $v(T)$ as  in \cite{Hambye:2016sby}:
\begin{equation}
 v(T) = \sqrt{1 - \frac{T^2}{T_{\rm ew}^2}} \,\Theta(T_{\rm ew}-T)\, v,
\end{equation}
with $v(0) = v = 174\,\text{GeV}$. We compute the corresponding contributions to $\Sigma_\mathcal{A}^{1 \leftrightarrow 2}$ as
\begin{eqnarray}
 \Sigma_B^0(y, \Mav, T) &=& \frac{T}{32\pi y}\,\mathcal{I}^B_1(y, \Mav, T)\,,\\
 \hat{k}^i\Sigma_B^i(y, \Mav, T) &=& \frac{T}{64\pi y^2}\,\left[2y_0 \mathcal{I}^B_1(y, \Mav, T) - z^2 \left(1 -x_B(\Mav, T)\right) \mathcal{I}^B_0(y, \Mav, T)\right],
\end{eqnarray}
 where $x_B( \Mav, T) \equiv m_B^2(T)/\Mav^2$, $\mathcal{I}^B_n(y, \Mav, T) = \mathcal{I}_n(y_0; x^B_-(y, \Mav, T), x^B_+(y, \Mav, T))$ and 
$x^B_{\pm}(y, \Mav, T) = (1/2)[y_0(1-x_B( \Mav, T)) \pm y \lambda^{1/2}(1, 0, x_B( \Mav, T))]$, with $B = H, W, Z$. Compared to the expressions in the symmetric phase, these contributions are smaller by a factor of $1/4$ that takes into account that only one
isospin runs in the loop of the self-energy diagrams, as well as the $1/\sqrt{2}$ suppression that appears in the coupling to the real Higgs compared to the doublet.

\paragraph{Broken phase -- indirect heavy neutrino production.} 
In the broken phase, the heavy neutrinos mix with the active neutrino states, opening up additional production channels. Here we use the expression from~\cite{Klaric:2021cpi}, which is generalised from \cite{Eijima:2017anv} (see also \cite{Ghiglieri:2016xye}), so that it is applicable in the non-relativistic regime as well. This rate significantly differs for the two heavy neutrino helicities and is given by 
%\AG{I had to add an additional factor $.5(1\pm y/y_0)$ to match with Juraj's result, please check if it is correct}:
\begin{equation}\label{eq:gammaindpm}
  \gamma_\pm^\mathrm{ind} = \frac{v^2}{4T^2} \left(1\pm\frac{y}{y_0}\right) \frac{(y_0 \pm y) \Gamma_k^\nu + \Gamma_u^\nu}{[b_\nu/T+(1+a_\nu) (y_0 \pm y)]^2 + [\Gamma_u^\nu + \Gamma_k^\nu (y_0 \pm y)]^2/(4T^2)}\,,
 \end{equation}
where the scattering rates $\Gamma^\nu_{u,k}$ 
and thermal mass
coefficients $a_\nu$ and $b_\nu$ should be evaluated on the heavy neutrino mass shell $y_0^2 - y^2 = z^2$.

Since we are working in the broken phase of the SM, we have to include the $W$ and $Z$ boson masses, hence we use the generalised expressions from~\cite{Quimbay:1995jn}, with:
\begin{eqnarray}
 a_\nu(E_N,k)&=&\frac14(g_1^2 + g_2^2) A(0,m_Z)
 + \frac14 g_2^2 A(0,m_W)
 ,\\
 b_\nu(E_N,k)&=&\frac{T}{4}(g_1^2 + g_2^2) B(0,m_Z)
 + \frac{T}{4} g_2^2 B(0,m_W)\,,
\end{eqnarray}
with $m_W = m_W(T)$ and $m_Z = m_Z(T)$ as the gauge boson masses that directly depend on the expectation-value of the Higgs field.
The functions $A$ and $B$ are given by the integrals:
 \begin{align}
 A(z_1, z_2)=\frac{1}{y^2}\int^\infty_0\frac{d\tilde y}{8\pi^2}
 &\left(\left[-\frac{(y_0^2+y^2+\Delta)}{2y}\frac{\tilde y}
 {y_2}L_2^{+}(\tilde y)-\frac{y_0 \tilde y}{y}L_2^{-}(\tilde y)+\frac{4{\tilde y}^2}
 {y_2}\right]f_B(y_2)\right.\nonumber\\
 &+\left.\left[\frac{(y_0^2-y^2-\Delta)}{2y}\frac{\tilde y}
 {y_1}L_1^{+}(\tilde y)-\frac{y_0 \tilde y}{y}L_1^{-}(\tilde y)+\frac{4{\tilde y}^2}
 {y_1}\right]f_F(y_1)\right),\\
 B(z_1,z_2)=\frac{1}{y^2}\int^\infty_0\frac{d\tilde y}{8\pi^2}
 &\left(\left[\frac{(y_0^2-y^2)}{y}{\tilde y}L_2^{-}(\tilde y)+
 \frac{(y_0^2-y^2+\Delta)}{2}\frac{y_0}{y}\frac{\tilde y}
 {y_2}L_2^{+}(\tilde y)-\frac{4y_0 {\tilde y}^2}{y_2}\right]
 f_B(y_2)\right.\nonumber\\
 &+\left.\left[\frac{(y_0^2-y^2)}{y}{\tilde y}L_1^{-}(\tilde y)-
 \frac{(y_0^2-y^2-\Delta)}{2}\frac{y_0}{y}\frac{\tilde y}
 {y_1}L_1^{+}(\tilde y)-\frac{4y_0 {\tilde y}^2}
 {y_1}\right]f_F(y_1)\right),
 \end{align} 
where $\Delta = z_2^2 - z_1^2$, $y_{1,2} = \sqrt{{\tilde y}^2 + z_{1,2}^2}$,
and the functions $L_{1,2}$ are given by
 \begin{eqnarray}
   L_1^{\pm}(\tilde y) &=&
   \log\left[
   \frac{y_0^2-y^2- \Delta-2y_1y_0-2y\tilde y}
 {y_0^2-y^2-\Delta-2y_1y_0+2y\tilde y}\right]\pm
 \log\left[
   \frac{y_0^2-y^2-\Delta+2y_1y_0-2y\tilde y}
 {y_0^2-y^2-\Delta+2y_1y_0+2y\tilde y}\right],\\
 L_2^{\pm}(\tilde y) &=&\log\left[\frac{y_0^2-y^2+
 \Delta+2y_2y_0+2y\tilde y}
 {y_0^2-y^2+\Delta+2y_2y_0-2y\tilde y}\right]\pm
 \log\left[\frac{y_0^2-y^2+\Delta-2y_2y_0+2y\tilde y}
 {y_0^2-y^2+\Delta-2y_2y_0-2y\tilde y}
   \right].
 \end{eqnarray}

The rates $\Gamma^\nu_{u,k}$ include the $2\leftrightarrow2$ scatterings, as well as the $1\leftrightarrow2$ processes.
The $2\leftrightarrow2$ scattering rate is given by Eqs. (5.23-5.24)
from~\cite{Ghiglieri:2018wbs}
 \begin{align}
  \Gamma^\mathrm{HTL}_{\! u} & \approx 
  \frac{T}{16\pi} 
  \biggl\{
  2 g_2^2 \log\biggl( \frac{1 + 4 y^2T^2 / m_W^2 }{ 1 + 4 y^2T^2 / \tilde{m}_W^2 } \biggr)
  \label{Gamma_HTL_u} \\ 
  & + \,
  (g_1^2 + g_2^2) 
  \biggl[
  \cos^2(\theta - \tilde{\theta}) 
   \log \biggl( \frac{1 + 4 y^2T^2 / m_Z^2}{1 + 4 y^2T^2 / \tilde{m}_Z^2 } \biggr)
  + 
  \sin^2(\theta - \tilde{\theta}) 
   \log \biggl( \frac{1 + 4 y^2T^2 / m_Z^2}{1 + 4 y^2T^2 / \tilde{m}_Q^2 } \biggr)
  \biggr]
  \biggr\} 
  \;, 
  \nonumber 
 %%%%%%%%%%%%%%%%%%%%%%%
 \\
  \Gamma^\mathrm{HTL}_{\! u} + 2 y \Gamma^\mathrm{HTL}_{\!k}
  & \approx 
  \frac{T}{8\pi} 
  \biggl\{
  2 g_2^2 \log\biggl( \frac{ 1 + 4 y^2T^2 / m_W^2 }{1 + 4 y^2T^2 / \bar{m}_W^2 } \biggr)\\
  & + \,
  (g_1^2 + g_2^2) 
  \biggl[
  \cos^2(\theta - \bar{\theta}) 
   \log \biggl( \frac{1 + 4 y^2T^2 / m_Z^2 }{ 1 + 4 y^2T^2 / \bar{m}_Z^2 } \biggr)
  + 
  \sin^2(\theta - \bar{\theta}) 
   \log \biggl( \frac{1 + 4 y^2T^2 / m_Z^2 }{ 1 + 4 y^2T^2 / \bar{m}_Q^2 } \biggr)
  \biggr]
  \biggr\} 
  \;, \nonumber
 \end{align}
where the $\tilde{m}$ and $\bar{m}$ masses correspond to the different gauge boson polarisations, and depend on the Debye masses
\begin{align}
m^2_{E1} &= \left(\frac{n_S}{6} + \frac{5n_G}{9}\right)g_1^2 T^2, \qquad
m^2_{E2} = \left(\frac{2}{3} + \frac{n_S}{6} + \frac{n_G}{3}\right)g_2^2 T^2\,.
\end{align}
where $n_S = 1$ number of scalar Higgs doublets and
$n_G=3$ the number of fermion generations. This sets the \emph{tilde} quantities as:
\begin{align}
\sin(2\tilde\theta) &= \frac{\sin(2\theta)\, m_Z^2}{\sqrt{\sin^2(2\theta)\, m_Z^4 
  + [\cos(2\theta)\, m_Z^2 + m_{E2}^2 - m_{E1}^2]^2}} \\
\tilde{m}_\pm^2 &= \frac{1}{2}\left\{m_Z^2 + m_{E1}^2 + m_{E2}^2 
  \pm \sqrt{\sin^2(2\theta)\, m_Z^4 
  + [\cos(2\theta)\, m_Z^2 + m_{E2}^2 - m_{E1}^2]^2}\right\} \\
\tilde{m}_W^2 &= m_W^2 + m_{E2}^2, \qquad
\tilde{m}_Z^2 = \tilde{m}_+^2, \qquad
\tilde{m}_Q^2 = \tilde{m}_-^2
\end{align}
and the \emph{barred} quantities as:
\begin{align}
\sin(2\bar\theta) &= \frac{\sin(2\theta)\, m_Z^2}{\sqrt{\sin^2(2\theta)\, m_Z^4 
  + [\cos(2\theta)\, m_Z^2 + (m_{E2}^2 - m_{E1}^2)/2]^2}} \\
\bar{m}_\pm^2 &= \frac{1}{2}\left\{m_Z^2 + \frac{m_{E1}^2+m_{E2}^2}{2} 
  \pm \sqrt{\sin^2(2\theta)\, m_Z^4 
  + \left[\cos(2\theta)\, m_Z^2 
  + \frac{m_{E2}^2-m_{E1}^2}{2}\right]^2}\right\} \\
\bar{m}_W^2 &= m_W^2 + \frac{m_{E2}^2}{2}, \qquad
\bar{m}_Z^2 = \bar{m}_+^2, \qquad
\bar{m}_Q^2 = \bar{m}_-^2
\end{align}
where the vacuum mixing angle is given by the standard expression
$\sin(2\theta) = 2g_1 g_2/(g_1^2+g_2^2)$.

The $1\leftrightarrow2$ Born rate is given by the expressions (5.27-5.29),
 \begin{equation}
 \Gamma^\mathrm{Born}_{\!u,k}
   =
   (g_1^2 + g_2^2)\,
   \tilde{\Gamma}_{\!u,k}^\mathrm{Born}
   (m_Z/T, \mu_{a}, 0)
   + 
   2 g_2^2\, \tilde{\Gamma}_{\!u,k}^\mathrm{Born}
   (m_W/T, \mu_\alpha - \mu_Q,\mu_Q)
   \;,
 \end{equation}
where
 \begin{align}
 \tilde{\Gamma}^\mathrm{Born}_{\!u}(z,\mu^{ }_1,\mu^{ }_2) 
 &=
 \frac{z^2 T}{32\pi y^2}
 \biggl[
   l_{1f} \biggl( \frac{z^2}{4 y} + \mu^{ }_1 \biggr) 
 - 
 l_{1b} \biggl( y + \frac{z^2}{4y} - \mu^{ }_2 \biggr) 
 \biggr]\,,\\
 \bigl( \tilde{\Gamma}^\mathrm{Born}_{\!u}
 + 2 y\, \tilde{\Gamma}^\mathrm{Born}_{\!k}
 \bigr) (z,\mu^{ }_1,\mu^{ }_2)
 &=
 \frac{T}{8\pi y}
 \biggl[
   l_{2b} \biggl(y+ \frac{z^2}{4 y} - \mu^{ }_2 \biggr) 
 - 
 l_{2f} \biggl( \frac{z^2}{4y} + \mu^{ }_1 \biggr) 
 \biggr]\,,
 \end{align}
with the functions $l_{1,2;b,f}$ given in terms of logarithms and dilogarithms
\begin{align}
  l_{1b,f}(y) \; \equiv \; \log \Bigl( 1 \mp e^{-y} \Bigr)
 \;, \quad
 l_{2b,f}(y) \; \equiv \; \mathrm{Li}^{ }_2 \Bigl(\pm e^{-y}\Bigr)\,.
 \end{align}
 Here $\mu_\alpha$ denotes the chemical potential normalised to temperature associated with the lepton flavour $\alpha=e,\mu,\tau$ (i.e.\ the corresponding lepton doublet), while $\mu_Q$ is that of the left-handed quark doublet $Q$.
We shall focus on the rates at leading order in the chemical potentials, and therefore set $\mu_\alpha=\mu_Q=0$ in the expressions above. 

Equivalently, indirect contributions also modify the heavy neutrino thermal masses
%\AG{I have included a factor of $.5(1\pm y/y_0)$ here, please check if it is correct.}
 \begin{equation}
  h^\mathrm{ind}_\pm = \frac{v^2}{4T^2}\left(1\pm\frac{y}{y_0}\right) 
  \frac{(y_0 \pm y) (1+a_\nu) + b_\nu/T}{[b_\nu/T+(1+a_\nu) (y_0 \pm y)]^2 + [\Gamma_u^\nu + \Gamma_k^\nu (y_0 \pm y)]^2/(4T^2)}\,.
 \end{equation}
The required modifications to the thermally averaged Hamiltonian are given by:
\begin{eqnarray}
  \langle h^\mathrm{LNC} \rangle & \rightarrow& \langle h^\mathrm{LNC} \rangle + \langle h^\mathrm{ind}_+ \rangle \,,\\
  \langle h^\mathrm{LNV} \rangle & \rightarrow& \langle h^\mathrm{LNV} \rangle + \langle h^\mathrm{ind}_- \rangle\,,
\end{eqnarray}
with the average defined by the same momentum-dependent integral~(\ref{eq:therm_average}).

\paragraph{Relativistic contributions.} We include the relativistic contributions to the LNV and LNC rates as computed in \cite{Ghiglieri:2017gjz, Ghiglieri:2017csp} (see also \cite{Ghiglieri:2018wbs}), where the complete set of leading-order processes was taken into account: $1 \leftrightarrow 2$ decays and inverse decays, $2 \leftrightarrow 2$ scatterings and collinearly enhanced $1 \leftrightarrow 2$ processes with multiple soft gauge boson scatterings, resummed to incorporate the Landau-Pomeranchuk-Migdal (LPM) effect. The authors of \cite{Ghiglieri:2017gjz, Ghiglieri:2017csp, Ghiglieri:2018wbs} have evaluated such rates as functions of $|\vec{k}|/T$ and temperature and tabulated their numerical results in \href{http://www.laine.itp.unibe.ch/leptogenesis/}{\ttfamily{this site}}. In the table provided in the latter site, the LNC and LNV linear contributions are respectively referred to as $Q_+$ and $Q_-$ and are mass-independent. In our notation, these correspond to
\begin{equation}
 \Sigma_{\mathcal{A}}^{\rm rel, 0} - \hat{k}^i\Sigma_{\mathcal{A}}^{\rm rel, i} = \frac{Q_+}{2},\quad \Sigma_{\mathcal{A}}^{\rm rel, 0} + \hat{k}^i\Sigma_{\mathcal{A}}^{\rm rel, i} = 2y^2 Q_-.
\end{equation}
which we plug into Eqs.~\eqref{eq:G0} and \eqref{eq:S0}.

 In the symmetric phase we include both relativistic contributions originating from $Q_+$ and $Q_-$. However, in the broken phase, the results of \cite{Ghiglieri:2017gjz, Ghiglieri:2017csp, Ghiglieri:2018wbs} incorporate both direct and indirect contributions to the rates. 
 \begin{figure}[t!]
 \centering
 \includegraphics[width=0.32\textwidth]{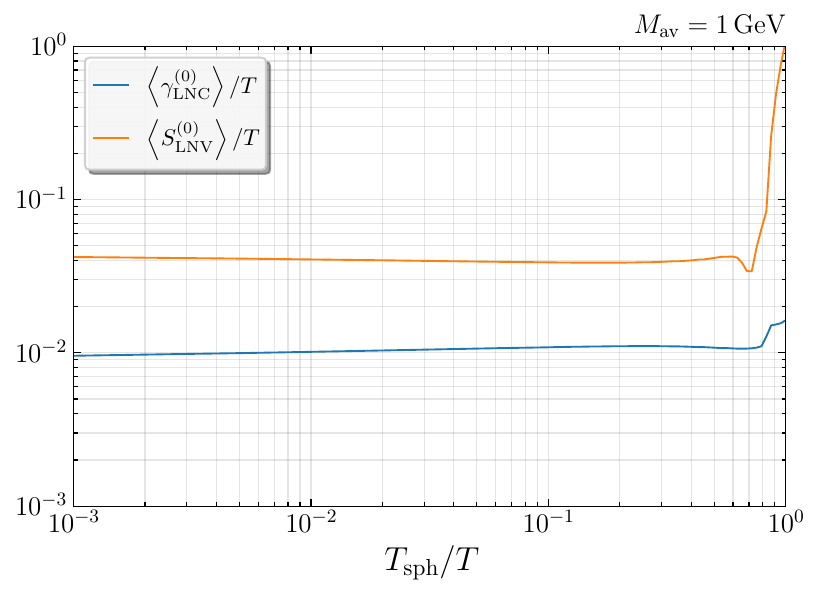}
  \includegraphics[width=0.32\textwidth]{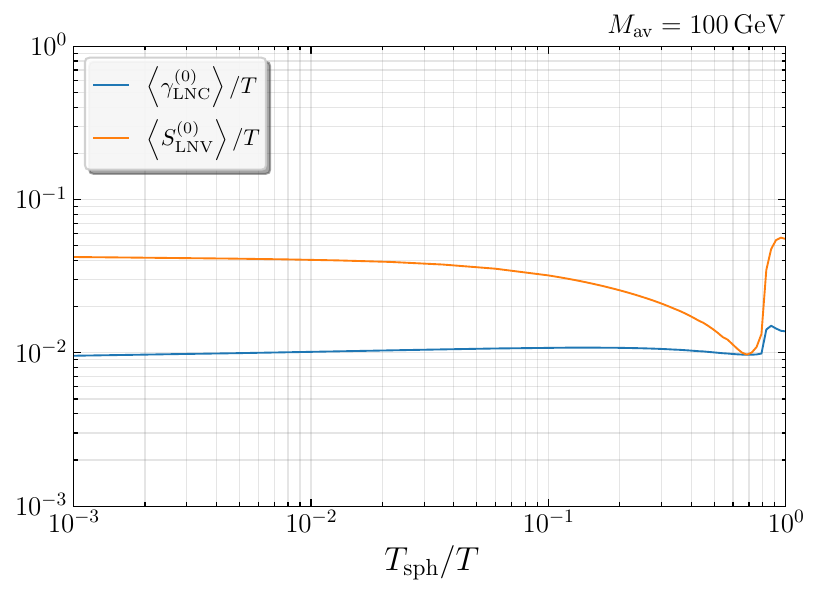}
  \includegraphics[width=0.32\textwidth]{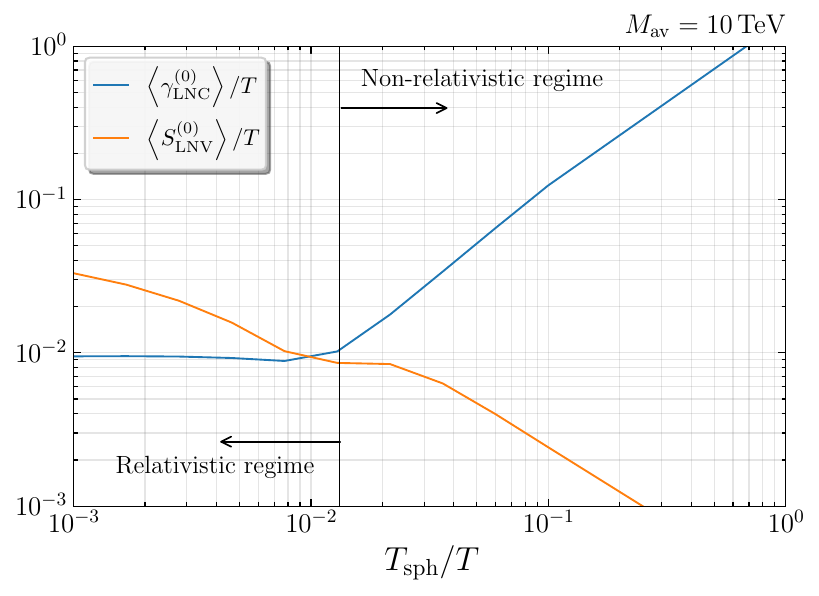}
 \caption{The rates $\langle \gamma_{\LNC}^{(0)}\rangle/T$ (blue) and $\langle S_{\LNV}^{(0)}\rangle/T$ (orange) as functions of $x = T_{\sph}/T$ for three benchmark masses $\Mav = 1\,\text{GeV}$ (left panel), $100\,\text{GeV}$ (middle panel) and $10\,\text{TeV}$ (right panel).}
 \label{fig:G0_and_S0}
\end{figure}
\begin{figure}[h!]
 \centering
 \includegraphics[width=0.32\textwidth]{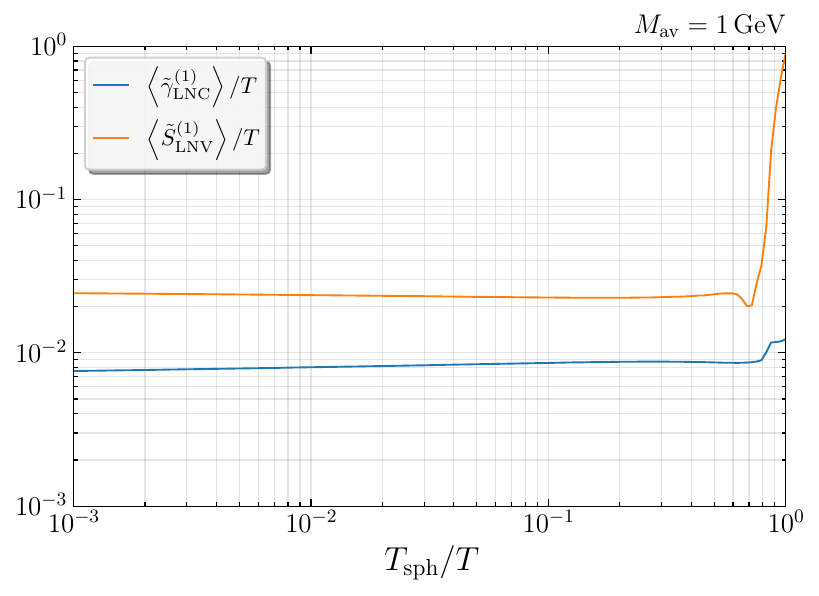}
  \includegraphics[width=0.32\textwidth]{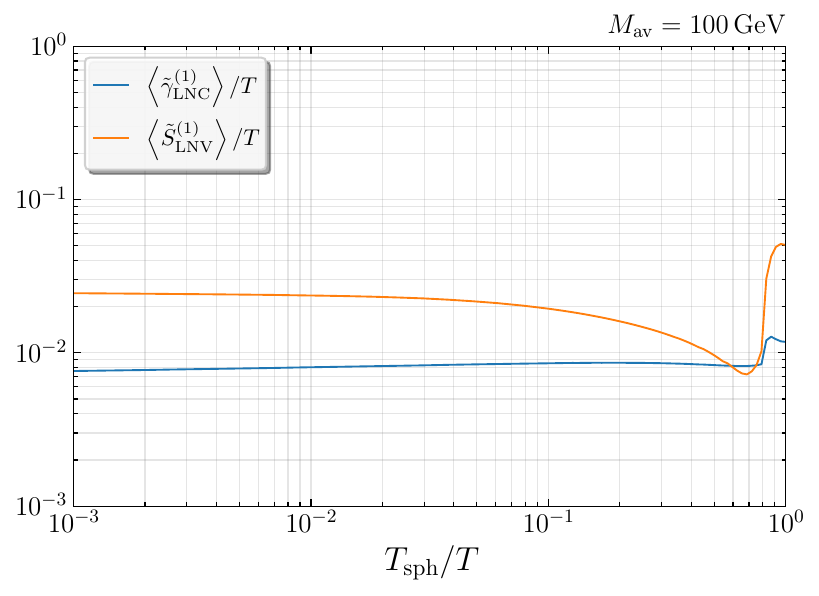}
  \includegraphics[width=0.32\textwidth]{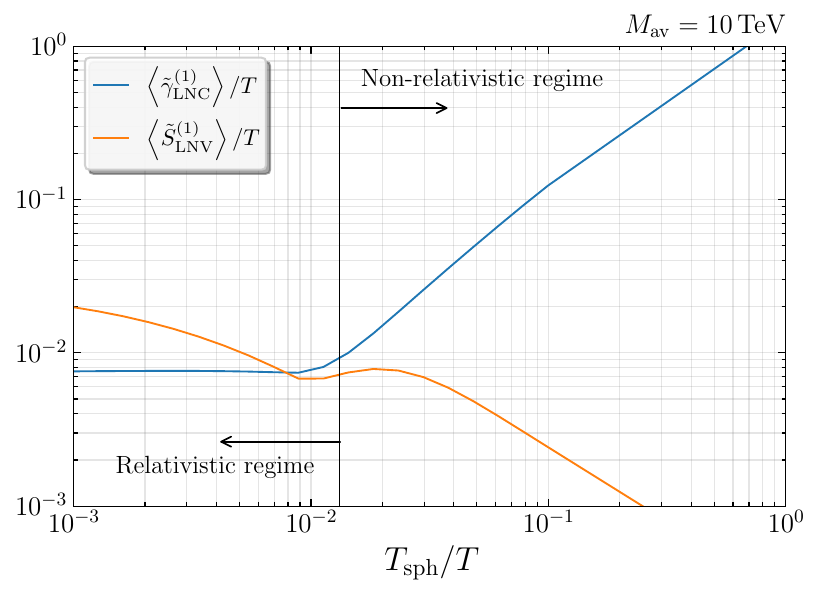}
 \caption{The terms $\langle \tilde{\gamma}_{\LNC}^{(1)}\rangle/T$ (blue) and $\langle \tilde{S}_{\LNV}^{(1)}\rangle/T$ (orange) as functions of $x = T_{\sph}/T$ for three benchmark masses $\Mav = 1\,\text{GeV}$ (left panel), $100\,\text{GeV}$ (middle panel) and $10\,\text{TeV}$ (right panel). We remind that $\tilde{\gamma}_{\LNC}^{(1)} = -\gamma_{\LNC}^{(0)}f'_F(y_0)/f_F(y_0)$ and $\tilde{S}_{\LNV}^{(1)} = -S_{\LNV}^{(0)}f'_F(y_0)/f_F(y_0)$.}
 \label{fig:G1_and_S1}
\end{figure}
\begin{figure}[h!]
 \centering
 \includegraphics[width=0.32\textwidth]{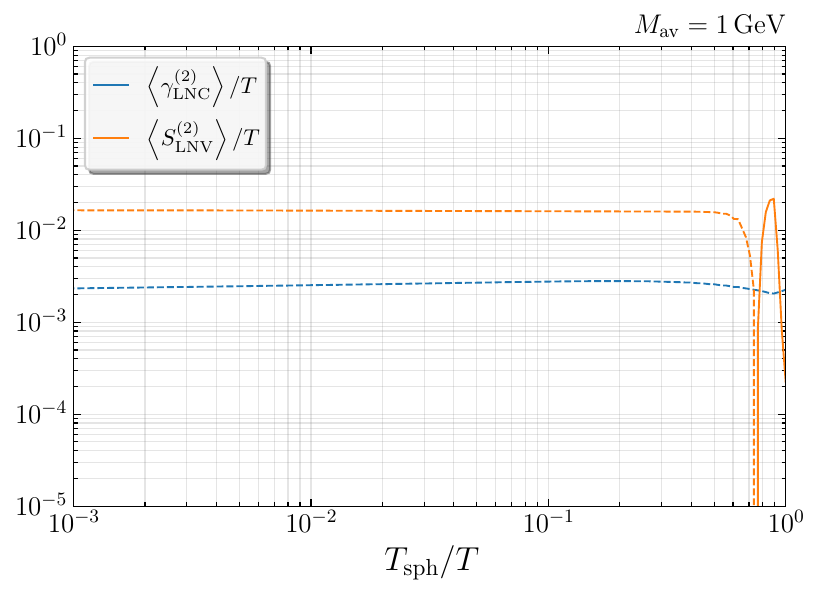}
  \includegraphics[width=0.32\textwidth]{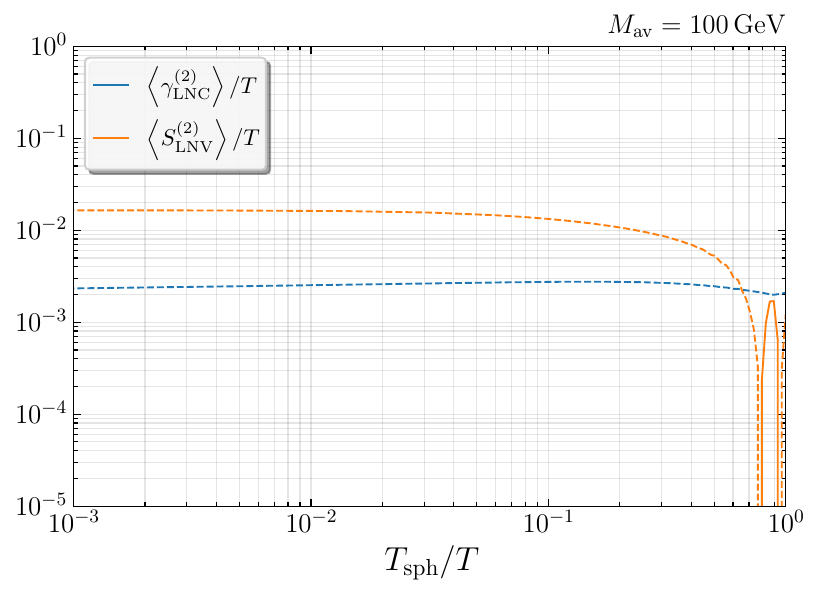}
  \includegraphics[width=0.32\textwidth]{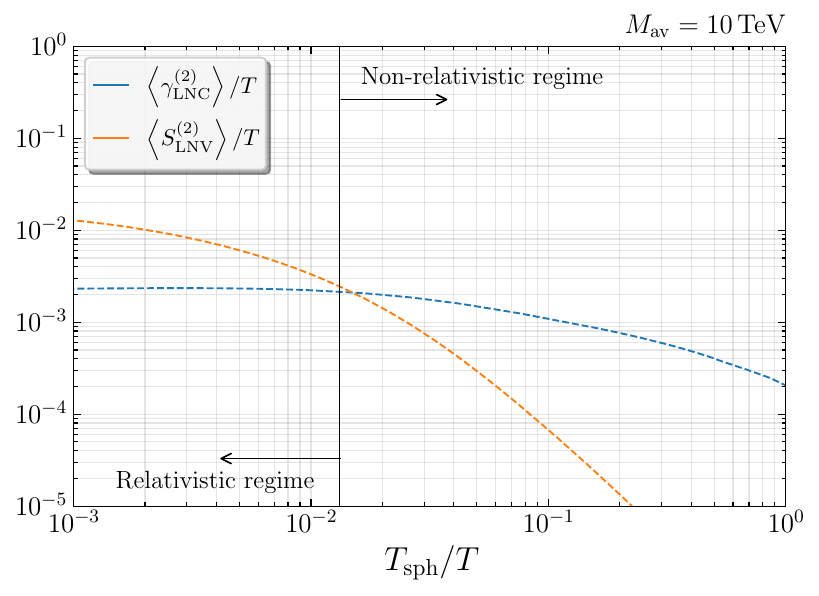}
 \caption{The rates $\langle \gamma_{\LNC}^{(2)}\rangle/T$ (blue) and $\langle S_{\LNV}^{(2)}\rangle/T$ (orange) as functions of $x = T_{\sph}/T$ for three benchmark masses $\Mav = 1\,\text{GeV}$ (left panel), $100\,\text{GeV}$ (middle panel) and $10\,\text{TeV}$ (right panel). The solid (dashed) style corresponds to positive (negative) contributions.}
 \label{fig:G2_and_S2}
\end{figure}
Our calculation instead treats these contributions separately, and our treatment of the indirect contribution remains valid also in the non-relativistic regime. For $\gamma_{\LNC}^{(0)}$ this distinction is not numerically significant, since the rate is dominated by direct contributions. In contrast, for $S_{\LNV}^{(0)}$ the indirect contributions become dominant in the relativistic regime. To avoid double counting, we therefore do not include the $Q_-$ contribution to $S_{\LNV}^{(0)}$ after the electroweak crossover.

We obtain the full rates $\gamma^{(0)}_{\LNC}$ and $S^{(0)}_{\LNV}$ by summing all the contributions in the regimes where they are kinematically allowed. Then, the rates $\tilde{\gamma}^{(1)}_{\LNC}$ and $\tilde{S}^{(1)}_{\LNV}$ are obtained as 
\begin{equation}\label{eq:washout_rates}
 \tilde{\gamma}^{(1)}_{\LNC} =- \frac{f_F'(y_0)}{f_F(y_0)} \gamma^{(0)}_{\LNC}\quad \text{and}\quad \tilde{S}^{(1)}_{\LNV} = - \frac{f_F'(y_0)}{f_F(y_0)} S^{(0)}_{\LNV}
\end{equation}
where $f_F'(y_0) = f_F(y_0) (f_F(y_0)-1)$ is the derivative of the Fermi-Dirac phase-space distribution.\footnote{The washout rates defined in Eq.~\eqref{eq:washout_rates}, entering the DMEs linearly in the chemical potential, are defined according to the adopted notation of Eqs.~\eqref{eq:DME_N} and \eqref{eq:DME_mu}, where, in particular, the non-linear terms proportional to $\gamma_{\LNC}^{(2)}$ and $S_{\LNV}^{(2)}$ are written in terms of $r_N-\mathds{1}$. In other works, e.g. \cite{Abada:2018oly, Hernandez:2022ivz}, the non-linear terms are written in terms of $r_N$ only, but the wash-out rates are defined accordingly as $\gamma^{(1)}_{\LNC} = \gamma_{\LNC}^{(2)} - \gamma_{\LNC}^{(0)}f_F'(y_0) /f_F(y_0)$ and $S^{(1)}_{\LNV} = S_{\LNV}^{(2)} - S_{\LNV}^{(0)}f_F'(y_0) /f_F(y_0)$.}

The relativistic contribution to $ \gamma^{(2)}_{\LNC}$ and $S^{(2)}_{\LNV}$ have been evaluated as functions of $|\vec{k}|/T$ and temperature in \cite{Ghiglieri:2017gjz, Ghiglieri:2017csp, Ghiglieri:2018wbs}, corresponding to $R_+$ and $R_-$ defined therein. We implement the numerical results for $R_+$ and $R_-$ as tabulated in \href{http://www.laine.itp.unibe.ch/leptogenesis/}{\ttfamily{this site}} and relate them to $\gamma_{\LNC}^{(2)}$ and $S_{\LNV}^{(2)}$ via
\begin{equation}
 \gamma_{\LNC}^{(2)} = \frac{R_+}{2}\left( 1 + \frac{y}{y_0}\right) \quad \text{and} \quad  S_{\LNV}^{(2)} = 2R_-y^2\left( 1 - \frac{y}{y_0}\right)
\end{equation}
We highlight, however, that by default in the \texttt{etabARS$\_$3RHN.py} module the non-linear terms are neglected, but we do provide the infrastructure to take them into account.

We finally average the rates $\langle \gamma_{\LNC}^{(0)}\rangle, \,\langle \tilde{\gamma}_{\LNC}^{(1)}\rangle,\, \langle \gamma_{\LNC}^{(2)}\rangle,\,\langle S_{\LNV}^{(0)}\rangle,\,\langle \tilde{S}_{\LNV}^{(1)}\rangle$ and $\langle S_{\LNV}^{(2)}\rangle$ according to Eq.~\eqref{eq:therm_average} for different values of the $\Mav$ and $x$ and store the results in tables which we provide within the code files. These tables are automatically interpolated before solving the set of DMEs, allowing for a more rapid evaluation of the BAU. We show our results for $\langle \gamma_{\LNC}^{(0)}\rangle/T$ and $\langle S_{\LNV}^{(0)}\rangle/T$ in Fig.~\ref{fig:G0_and_S0}, for $\langle \tilde{\gamma}_{\LNC}^{(1)}\rangle/T$ and $\langle \tilde{S}_{\LNV}^{(1)}\rangle/T$ in Fig.~\ref{fig:G1_and_S1} and for $\langle \gamma_{\LNC}^{(2)}\rangle/T$ and $\langle S_{\LNV}^{(2)}\rangle/T$ in Fig.~\ref{fig:G2_and_S2}. In these figures, the rates are given as functions of $x = T_{\sph}/T$ and the results are shown for the three benchmark masses $\Mav = 1\,\text{GeV}$, $100\,\text{GeV}$ and $10\,\text{TeV}$. \footnote{In non-relativistic limit for large $z$, the averaged production and wash-out rates becomes similar $\langle \gamma_{\LNC}^{(0)}\rangle \simeq \langle \tilde{\gamma}_{\LNC}^{(1)}\rangle \simeq z^2 \langle S_{\LNV}^{(0)}\rangle \simeq z^2\langle \tilde{S}_{\LNV}^{(1)}\rangle \simeq z/(16\pi)$, but evaluating this numerically with sufficient precision gets increasingly difficult as $z$ increases. To avoid potential numerical instabilities, we manually set all the rates equal to $z/(16\pi)$ when $z>10$ instead of performing the numerical integration.}

To allow for a better comparison with other works, we also list the results for $\langle {\gamma}_{\LNC}^{(1)}\rangle/T$ and $\langle {S}_{\LNV}^{(1)}\rangle/T$, with $\gamma^{(1)}_{\LNC} = \gamma_{\LNC}^{(2)} - \gamma_{\LNC}^{(0)}f_F'(y_0) /f_F(y_0)$ and $S^{(1)}_{\LNV} = S_{\LNV}^{(2)} - S_{\LNV}^{(0)}f_F'(y_0) /f_F(y_0)$. The values we obtain for the rates in the relativistic limit show a good agreement with other results in the literature. In particular, for $\Mav = 100\,\text{MeV}$ and $T = 10^6\,\text{GeV}$, we get $\langle\gamma_{\LNC}^{(a)}\rangle/T \times 10^3 = 9.15, 5.10, -2.19$ and $\langle S_{\LNV}^{(a)}\rangle/T \times 10^2 = 4.337, 0.855, -1.651$ respectively for $a = 0,1,2$, in agreement with, e.g., Table 1 of \cite{Hernandez:2022ivz}.
\newpage

\section{Predefined variables and methods in \texttt{ULSBase}}\label{app:ulsbase}

This appendix documents the variables and methods that are predefined in the \texttt{ULSBase}
base class (file \texttt{ulysses/ulsbase.py}) and are therefore available in any custom module
that inherits from it. Users writing extended models should not redefine these quantities.

\subsection*{Physical constants (keyword-overridable at construction)}

\begin{table}[h!]
\centering\small
\renewcommand{\arraystretch}{1.25}
\begin{tabular}{p{4.0cm} p{2.8cm} p{1.4cm} p{5.8cm}}
\hline
\textbf{Attribute} & \textbf{Default} & \textbf{Unit} & \textbf{Description} \\
\hline
\texttt{vev}                    & $174.0$               & GeV     & Higgs vacuum expectation value \\
\texttt{mhiggs}                   & $125.35$              & GeV     & Higgs boson mass \\
\texttt{mz}                   & $91.1876$             & GeV     & $Z$-boson mass \\
\texttt{mplanck}                   & $1.22\times10^{19}$   & GeV     & Planck mass \\
\texttt{gstar}                & $106.75$              & --      & Effective relativistic d.o.f.\ at high $T$ \\
\texttt{mstar}                & $1.0\times10^{-12}$   & GeV     & Neutrino equilibrium mass $m_*$ \\
\texttt{m2solar}       & $7.537\times10^{-23}$ & GeV$^2$ & Solar mass-squared splitting \\
\texttt{m2atm}& $2.521\times10^{-21}$ & GeV$^2$ & Atmospheric splitting (NO) \\
\texttt{matminvert}& $2.500\times10^{-21}$ & GeV$^2$ & Atmospheric splitting (IO) \\
\hline
\end{tabular}
\caption{Physical constants available in every \texttt{ULSBase} subclass. Neutrino oscillation
parameters are from \texttt{NuFit-6.1}~\cite{Esteban:2024eli}. Each can be overridden by passing
the corresponding keyword argument to the constructor.}
\label{tab:ulsbase_constants}
\end{table}

\subsection*{Solver and model configuration}

\begin{table}[H]
\centering\small
\renewcommand{\arraystretch}{1.25}
\begin{tabular}{p{4.2cm} p{3.8cm} p{6.0cm}}
\hline
\textbf{Attribute} & \textbf{Default} & \textbf{Description} \\
\hline
\texttt{zmin}           & $0.001$                               & Start of integration range ($z=M_1/T$) \\
\texttt{zmax}           & $1000$                                & End of integration range \\
\texttt{zsteps}         & $1000$                                & Number of $z$ steps (log-spaced) \\
\texttt{zs}             & -                                 & Log-spaced $z$ grid of length \texttt{zsteps} \\
\texttt{normfact}       & $0.013$                               & Conversion factor from $N_{B-L}$ to $\eta_B$ \\
\texttt{normfact\_ars}  & $(28/79)\,\pi^2/(27{\cdot}6{\cdot}\zeta(3))$ & Conversion factor from $N_{B-L}$ to $\eta_B$ for ARS models \\
\texttt{xmin}           & \texttt{1e-6}                         & Start of ARS $x=T_{sph}/T$-grid; auto-derived from $M_1$ if \texttt{None} \\
\texttt{xmax}           & min([$1, 20 \times T_{EW}/M1$])                       & End of ARS $x=T_{sph}/T$-grid; auto-derived from $M_1$ if \texttt{None} \\
\texttt{xsteps}         & $500$                                 & Number of ARS integration steps \\
\texttt{zcut}           & $1.0$                                 & Matching scale $z_{\rm cut}$ used in the ARS 2RHN model \\
\texttt{Lambda}         & \texttt{None}                         & ARS scale $\Lambda$ controlling the fast-mode quasi-static transition; overridable via \texttt{pdict} \\
\texttt{ordering}       & $0$                                   & Mass ordering: $0$\,=\,NO, $1$\,=\,IO \\
\texttt{loop}           & \texttt{False}                        & Use loop-corrected Yukawa matrix \\
\texttt{initial\_abundance} & $0$                               & Initial RHN abundance ($0$\,=\,vanishing, $1$\,=\,thermal; non-integer allowed) \\
\texttt{extended\_mode} & \texttt{False}                        & Allow extra keys in \texttt{setParams} beyond \texttt{pnames} \\
\texttt{which\_param}   & \texttt{'euler'}                      & Yukawa parametrisation: \texttt{'euler'}, \texttt{'single\_imaginary'}, or \texttt{'manual'} \\
\texttt{debug}          & \texttt{False}                        & Print diagnostic output \\
\texttt{pnames}         & - & List of expected parameter-dictionary keys \\
\texttt{pdict}          & \texttt{\{\}}                         & Full parameter dictionary passed to \texttt{setParams} \\
\texttt{evolname}       & \texttt{'z'}                          & Name of the evolution variable (used by plotter) \\
\texttt{path}           & \texttt{'./'}                          & Output directory for model results \\
\hline
\end{tabular}
\caption{Solver and model-switch attributes in \texttt{ULSBase}.}
\label{tab:ulsbase_solver}
\end{table}

\subsection*{Model parameters (set via \texttt{setParams})}

The following attributes are populated by \texttt{setParams} when a parameter dictionary is
passed. Mass inputs are interpreted as $\log_{10}$ values; angular inputs are in degrees and
internally converted to radians.

\begin{table}[h!]
\centering\small
\renewcommand{\arraystretch}{1.25}
\begin{tabular}{p{5.0cm} p{1.4cm} p{7.4cm}}
\hline
\textbf{Attribute} & \textbf{Unit} & \textbf{Description} \\
\hline
\texttt{m}  & GeV & Lightest active neutrino mass \\
\texttt{M1}, \texttt{M2}, \texttt{M3} & GeV & Heavy Majorana neutrino masses \\
\texttt{delta}  & rad & Dirac CP phase \\
\texttt{a21}, \texttt{a31} & rad & Majorana CP phases \\
\texttt{t12}, \texttt{t13}, \texttt{t23} & rad
  & PMNS mixing angles; optional --- if absent, \texttt{NuFit-6.1} best-fit values are used
    (NO: $33.76^\circ$, $8.62^\circ$, $43.27^\circ$;
     IO: $33.76^\circ$, $8.65^\circ$, $48.15^\circ$) \\
\texttt{x1}\ldots\texttt{x3}, \texttt{y1}\ldots\texttt{y3} & rad
  & Complex Casas--Ibarra rotation angles (\texttt{euler} param.) \\
\texttt{xN1}, \texttt{xN2}, \texttt{xnu1}, \texttt{xnu2},
  \texttt{x}, \texttt{y} & rad
  & Casas--Ibarra angles (\texttt{single\_imaginary} param., arXiv:2106.16226) \\
\texttt{Y11\_mag}, \texttt{Y12\_mag}, \ldots & --
  & Magnitudes of $h_{ij}$ (\texttt{manual} param.) \\
\texttt{Y11\_phs}, \texttt{Y12\_phs}, \ldots & rad
  & Phases of $h_{ij}$ (\texttt{manual} param.) \\
\hline
\end{tabular}
\caption{Model parameters populated by \texttt{setParams}. In the \texttt{manual}
parametrisation the nine magnitude--phase pairs replace all Casas--Ibarra angles.}
\label{tab:ulsbase_params}
\end{table}

\subsection*{Derived matrix properties}

\begin{table}[H]
\centering\small
\renewcommand{\arraystretch}{1.25}
\begin{tabular}{p{5.0cm} p{9.0cm}}
\hline
\textbf{Property} & \textbf{Description} \\
\hline
\multicolumn{2}{l}{\textit{Yukawa matrix variants}} \\
\texttt{h}               & Yukawa coupling matrix $\lambda_\nu$ ($3\times3$, complex);
                            dispatches to \texttt{h\_tree} or \texttt{h\_loop} \\
\texttt{h\_tree}         & Tree-level Yukawa:
                            $(i/v)\,U\,\sqrt{D_m}\,R^\top\!\sqrt{D_M}$ \\
\texttt{h\_loop}         & Loop-corrected Yukawa:
                            $(i/v)\,U\,\sqrt{D_m}\,R^\top\mathtt{fMR}$ \\
\texttt{h\_non\_unitary} & Yukawa with leading non-unitarity correction via $\eta$ \\
\hline
\multicolumn{2}{l}{\textit{Mixing and mass matrices}} \\
\texttt{U}      & PMNS matrix in PDG convention ($3\times3$, complex) \\
\texttt{R}      & Casas--Ibarra orthogonal matrix; dispatches to
                  \texttt{R\_ord} or \texttt{R\_alt} \\
\texttt{R\_ord} & $R=R_{23}R_{13}R_{12}$ (Euler-angle parametrisation) \\
\texttt{R\_alt} & $R=(O_{\nu,13}\,O_{\nu,23}\,R^C_{12}\,O_{N,23}\,O_{N,13})^\top$
                  (single-imaginary parametrisation) \\
\texttt{DM}     & Diagonal heavy mass matrix $\mathrm{diag}(M_1,M_2,M_3)$ \\
\texttt{SqrtDM} & $\sqrt{\mathrm{diag}(M_1,M_2,M_3)}$ \\
\texttt{SqrtDm} & $\sqrt{\mathrm{diag}(m_1,m_2,m_3)}$ (active neutrinos) \\
\texttt{Theta}  & Active-sterile mixing $\Theta_{ai}=v\,h_{ai}/M_i$ \\
\hline
\multicolumn{2}{l}{\textit{Non-unitarity}} \\
\texttt{RV}  & Non-unitary mixing $R_V=i\,U\,\sqrt{D_m}\,R^\top D_M^{-1/2}$ \\
\texttt{eta} & Non-unitarity parameter $\eta=-\tfrac{1}{2}R_V R_V^\dagger$ \\
\hline
\multicolumn{2}{l}{\textit{Effective and loop-corrected masses}} \\
\texttt{meff1}, \texttt{meff2}, \texttt{meff3}
                      & Effective neutrino masses
                        $\tilde{m}_i=(h^\dagger h)_{ii}v^2/M_i$ \\
\texttt{m\_tree}      & Tree-level light neutrino mass matrix \\
\texttt{m\_loop}      & One-loop light neutrino mass matrix \\
\texttt{fMR}          & Loop-corrected analogue of $\sqrt{D_M}$;
                        replaces \texttt{SqrtDM} when \texttt{loop=True} \\
\texttt{fMLoop(x)}    & Scalar loop function
                        $x\!\left[\ln r_H^2/(r_H^2{-}1)+3\ln r_Z^2/(r_Z^2{-}1)\right]$,\;
                        $r_{H,Z}=x/M_{H,Z}$ \\
\texttt{fMLoopHelper} & Diagonal matrix
                        $\tfrac{1}{32\pi^2v^2}
                        \mathrm{diag}(f_{ML}(M_1),f_{ML}(M_2),f_{ML}(M_3))$ \\
\hline
\multicolumn{2}{l}{\textit{ARS oscillation scale}} \\
\texttt{zosc\_mat} & $3\times3$ matrix
                    $z_{ij}^{\rm osc}=\bigl(12T_{\rm EW}^3/(\Delta M^2_{ji}M_0)\bigr)^{1/3}$
                    ($T_{\rm EW}=131.7\,\mathrm{GeV}$); used by ARS models \\
\hline
\end{tabular}
\caption{Key matrix properties in \texttt{ULSBase}. All are \texttt{@property} descriptors
recomputed on access, except \texttt{fMLoop} which is a regular method.}
\label{tab:ulsbase_matrices}
\end{table}

\subsection*{Kinetic functions}

\begin{table}[H]
\centering\small
\renewcommand{\arraystretch}{1.25}
\begin{tabular}{p{5.0cm} p{9.0cm}}
\hline
\textbf{Method / Property} & \textbf{Description} \\
\hline
\multicolumn{2}{l}{\textit{Decay parameters and widths}} \\
\texttt{k1}, \texttt{k2}, \texttt{k3}
  & Decay parameters $K_i=\tilde{m}_i/m_*$ \\
\texttt{Gamma1}, \texttt{Gamma2}, \texttt{Gamma3}
  & Total decay widths $\Gamma_i=(h^\dagger h)_{ii}M_i/(8\pi)$ [GeV] \\
\hline
\multicolumn{2}{l}{\textit{Boltzmann equation terms}} \\
\texttt{D1(k,z)}, \texttt{D2(k,z)}, \texttt{D3(k,z)}
  & Decay term $D_i\propto K_i\,z\,\mathcal{K}_1(M_iz/M_1)/\mathcal{K}_2(M_iz/M_1)$ \\
\texttt{W1(k,z)}, \texttt{W2(k,z)}, \texttt{W3(k,z)}
  & Washout term $W_i\propto K_i\,(M_iz/M_1)^3\mathcal{K}_1(M_iz/M_1)$ \\
\texttt{N1Eq(z)}, \texttt{N2Eq(z)}, \texttt{N3Eq(z)}
  & Equilibrium abundance $N_i^{\rm eq}(z)=\tfrac{3}{8}(M_iz/M_1)^2\mathcal{K}_2(M_iz/M_1)$ \\
\texttt{DS(k,z)}   & Decay $+$ $\Delta L=1$ scattering combined rate \\
\texttt{scat(z)}   & Multiplicative washout correction incorporating $\Delta L=1$
                     scatterings \\
\texttt{NDW1(k,z)} & Returns the triplet
                     $[N_1^{\rm eq}(z),\;D_1(k,z),\;W_1(k,z)]$ \\
\hline
\multicolumn{2}{l}{\textit{Loop functions}} \\
\texttt{f1(x)}
  & Vertex correction:
    $\tfrac{2}{3}x^2\!\left[(1+x^2)\ln\tfrac{1+x^2}{x^2}-\tfrac{2-x^2}{1-x^2}\right]$;\;
    returns $1$ for $x>10^4$ \\
\texttt{f2(x)}
  & Self-energy correction: $\tfrac{2}{3}/(x^2-1)$ \\
\hline
\end{tabular}
\caption{Kinetic functions available in \texttt{ULSBase} (part~1).
$\mathcal{K}_n$ denotes the modified Bessel function of the second kind of order $n$.}
\label{tab:ulsbase_kinetic1}
\end{table}

\begin{table}[H]
\centering\small
\renewcommand{\arraystretch}{1.25}
\begin{tabular}{p{5.2cm} p{8.8cm}}
\hline
\textbf{Method / Property} & \textbf{Description} \\
\hline
\multicolumn{2}{l}{\textit{Standard flavoured CP asymmetries
  ($a,b\in\{0,1,2\}\equiv\{e,\mu,\tau\}$)}} \\
\texttt{epsilon(i,j,k,m)}
  & General unflavoured CP asymmetry; $i,j,k$ are RHN indices, $m$ is the lepton index \\
\texttt{epsilon1ab(a,b)}
  & Off-diagonal CP asymmetry element $\epsilon^{(1)}_{\alpha\beta}$ from $N_1$ decays \\
\texttt{epsilon2ab(a,b)} & Flavoured CP asymmetry from $N_2$ decays \\
\texttt{epsilon3ab(a,b)} & Flavoured CP asymmetry from $N_3$ decays \\
\hline
\multicolumn{2}{l}{\textit{Regularised CP asymmetries (Breit--Wigner resonance treatment)}} \\
\texttt{epsilon1ab\_reg(a,b)}
  & As \texttt{epsilon1ab} with self-energy poles between $N_1$--$N_2$
    and $N_1$--$N_3$ regularised \\
\texttt{epsilon2ab\_reg(a,b)}
  & As \texttt{epsilon2ab} with $N_2$--$N_1$ and $N_2$--$N_3$ poles regularised \\
\texttt{epsilon3ab\_reg(a,b)}
  & As \texttt{epsilon3ab} with $N_3$--$N_1$ and $N_3$--$N_2$ poles regularised \\
\hline
\multicolumn{2}{l}{\textit{Resonant leptogenesis CP asymmetries}} \\
\texttt{epsilonaaRES(a)}
  & Flavour-diagonal resonant CP asymmetry for the quasi-degenerate $N_1$--$N_2$ case;
    combines mixing ($f_{\rm mix}$) and oscillation ($f_{\rm osc}$) contributions \\
\texttt{epsiloniaaRES(a,i,j)}
  & Generalised resonant CP asymmetry for pair $(N_i,N_j)$;
    retains both $f_{\rm mix}$ and $f_{\rm osc}$ \\
\texttt{epsiloniaaRESmix(a,i,j)}
  & As \texttt{epsiloniaaRES} retaining only $f_{\rm mix}$ \\
\hline
\multicolumn{2}{l}{\textit{Diagonal CP asymmetry shorthands (properties)}} \\
\texttt{eps100}, \texttt{eps111}, \texttt{eps122}
  & $\epsilon_1^{ee}$,\; $\epsilon_1^{\mu\mu}$,\; $\epsilon_1^{\tau\tau}$ \\
\texttt{eps200}, \texttt{eps211}, \texttt{eps222}
  & $\epsilon_2^{ee}$,\; $\epsilon_2^{\mu\mu}$,\; $\epsilon_2^{\tau\tau}$ \\
\texttt{eps300}, \texttt{eps311}, \texttt{eps322}
  & $\epsilon_3^{ee}$,\; $\epsilon_3^{\mu\mu}$,\; $\epsilon_3^{\tau\tau}$ \\
\hline
\multicolumn{2}{l}{\textit{Flavour projection helpers}} \\
\texttt{hterm(a,b)}
  & Projection probability $|h_{ab}|^2/(h^\dagger h)_{00}$ \\
\texttt{c1a(a)}, \texttt{c2a(a)}, \texttt{c3a(a)}
  & Normalised coupling coefficients
    $h_{a,i-1}/\sqrt{(h^\dagger h)_{i-1,i-1}}$ for $i=1,2,3$ \\
\texttt{resonance(z)}
  & Ratio $\Gamma_1(z)/\Delta M_{21}$;
    diagnostic for proximity to the resonant regime \\
\hline
\end{tabular}
\caption{Kinetic functions available in \texttt{ULSBase} (part~2).}
\label{tab:ulsbase_kinetic2}
\end{table}

\subsection*{Evolution data interface}

The methods in Table~\ref{tab:ulsbase_evol} control which ODE columns are stored and passed to
the plotter. The base-class defaults return empty lists, so the plotter produces no output unless
the subclass overrides them. \texttt{setEvolData} and its variants should be called at the end of
\texttt{EtaB} after the ODE integration is complete.

\begin{table}[H]
\centering\small
\renewcommand{\arraystretch}{1.25}
\begin{tabular}{p{4.6cm} p{9.4cm}}
\hline
\textbf{Method / Property} & \textbf{Description} \\
\hline
\texttt{flavourindices()} & Returns a list of integer column indices from the ODE solution
  array \texttt{ys} to be shown in the upper plot panel. Default: \texttt{[]}. \\
\texttt{flavourlabels()} & Returns a list of legend label strings (one per entry in
  \texttt{flavourindices()}). Default: \texttt{[]}. \\
\texttt{extendedindices()} & Returns a list of column indices for additional quantities shown
  in the lower plot panel; activates the two-panel layout if non-empty. Default: \texttt{[]}. \\
\texttt{extendedlabels()} & Returns a list of legend label strings (one per entry in
  \texttt{extendedindices()}). Default: \texttt{[]}. \\
\texttt{setEvolData(ys)} & Stores the ODE trajectory for standard leptogenesis models;
  selects columns via \texttt{flavourindices()} and \texttt{extendedindices()}, then appends
  $N_1$ and $\eta_B$ (using \texttt{normfact}) as the final two columns. \\
\texttt{setEvolDataARS(ys)} & As \texttt{setEvolData} for ARS models: no $N_1$ column is
  appended and $\eta_B$ is computed using \texttt{normfact\_ars}. \\
\texttt{setEvolDataPBH(ys)} & Stores the raw ODE trajectory for PBH-sourced leptogenesis
  models without column reordering. \\
\texttt{evolData} & Property returning the stored trajectory as an $N_z\times N_{\rm col}$
  matrix; column~0 is $z$ and subsequent columns follow the order set by
  \texttt{setEvolData[ARS]}. \\
\hline
\end{tabular}
\caption{Evolution data interface methods in \texttt{ULSBase}. Subclasses override
\texttt{flavourindices}, \texttt{flavourlabels}, \texttt{extendedindices}, and
\texttt{extendedlabels}, and call the appropriate \texttt{setEvolData} variant at the end of
\texttt{EtaB}.}
\label{tab:ulsbase_evol}
\end{table}

\subsection*{Dark matter utilities}

The methods in Table~\ref{tab:ulsbase_dm} are available in any subclass and are used by
\texttt{etab1BE1F\_DM\_FreezeIn.py}:

\begin{table}[H]
\centering\small
\renewcommand{\arraystretch}{1.25}
\begin{tabular}{p{5.2cm} p{8.8cm}}
\hline
\textbf{Method / Property} & \textbf{Description} \\
\hline
\texttt{Y\_DM(N\_DM)}            & Convert final comoving DM number to yield
                                   $Y_{\rm DM}=n_{\rm DM}/s$ \\
\texttt{OmegaDMh2(Y\_DM, m\_dm)} & Compute $\Omega_{\rm DM}h^2$ from yield and DM mass \\
\texttt{dmYield}                 & Stored $Y_{\rm DM}$ after \texttt{EtaB} has been evaluated \\
\texttt{OmDMh2}                  & Stored $\Omega_{\rm DM}h^2$ after \texttt{EtaB} has been evaluated \\
\hline
\end{tabular}
\caption{Dark matter utility methods. These use the PDG values
$s_0=2891.2\,\mathrm{cm}^{-3}$ and $\rho_c/h^2=1.054\times10^{-5}\,\mathrm{GeV\,cm}^{-3}$~\cite{ParticleDataGroup:2024cfk}.}
\label{tab:ulsbase_dm}
\end{table}

\subsection*{Miscellaneous utilities}

\begin{table}[H]
\centering\small
\renewcommand{\arraystretch}{1.25}
\begin{tabular}{p{4.8cm} p{9.2cm}}
\hline
\textbf{Method / Property} & \textbf{Description} \\
\hline
\texttt{isPerturbative}      & Returns \texttt{True} if all Yukawa elements satisfy
                               $|h_{ij}|<\sqrt{4\pi}$ \\
\texttt{constants}           & Property returning a formatted string of all physical
                               constant values; printed automatically when
                               \texttt{debug=True} \\
\texttt{extended\_summary()} & Override to return a formatted string of additional
                               outputs computed inside \texttt{EtaB}
                               (e.g.\ DM yield); printed by \texttt{uls-calc} after
                               $\eta_B$. Default: \texttt{None} \\
\texttt{printParams()}       & Prints current parameter values with angles
                               back-converted to degrees and masses as $\log_{10}$ \\
\texttt{shortname()}         & Returns a short identifier string for the model;
                               override in subclass. Default: empty string \\
\texttt{setZMin(x)}, \texttt{setZMax(x)}, \texttt{setZSteps(x)}
                             & Update the corresponding $z$-grid bound or step count
                               and regenerate \texttt{zs} \\
\texttt{setZS()}             & Regenerate the log-spaced $z$ grid from the current
                               \texttt{zmin}, \texttt{zmax}, \texttt{zsteps} \\
\hline
\end{tabular}
\caption{Miscellaneous utility methods and properties in \texttt{ULSBase}.}
\label{tab:ulsbase_misc}
\end{table}

\bibliography{Biblio}
\bibliographystyle{JHEP}

\end{document}